\documentclass[12pt]{article}
\usepackage{amsfonts}
\usepackage{graphicx}
\usepackage{epsfig}
\usepackage{amssymb}
\usepackage[utf8x]{inputenc}
\usepackage{caption2}
\usepackage{appendix}

\newcommand{\mysection}{\setcounter{equation}{0}\section}
\newcommand{\text}{\rm}
\begin{document}
\begin{titlepage}
\vspace{-2cm}
\begin{flushright}
RECAPP-HRI-2010-013\\[2ex]
{\large \tt hep-ph/yymmnnn}\\
\end{flushright}
\vspace{+2cm}
\begin{center}
{\Large {\bf 
{Graviton plus vector boson production to NLO in QCD at the LHC}}}
\\[1cm]

M.\ C.\ Kumar$^{a}, $\footnote{mckumar@hri.res.in}
\hspace{.5cm} 
Prakash Mathews$^{b}, $\footnote{prakash.mathews@saha.ac.in}
\hspace{.5cm}
V.\ Ravindran$^{a}, $\footnote{ravindra@hri.res.in}
\hspace{.5cm}
Satyajit Seth$^{b}, $\footnote{satyajit.seth@saha.ac.in}
\\[1cm]

${}^a$ Regional Centre for Accelerator-based Particle Physics\\
Harish-Chandra Research Institute, Chhatnag Road, Jhunsi,\\
 Allahabad 211 019, India
\\[.7cm]

${}^b$ Saha Institute of Nuclear Physics, 1/AF Bidhan Nagar, 
Kolkata 700 064,
India
\end{center}
\vspace{1cm}

\begin{abstract}
\noindent
We present the next-to-leading order QCD corrections to the associated 
production of the vector gauge boson ($Z/W^\pm$) and the graviton in 
the large extra dimension model at the LHC. We estimate the impact of 
the QCD corrections on the total cross sections as well as the differential 
distributions of the gauge bosons and find that they are significant.  
We also study the dependence of the cross sections on the arbitrary 
factorization scale and show the reduction in the scale uncertainties at 
NLO level. Further, we discuss the ultraviolet sensitivity of the theoretical 
predictions.
\end{abstract}
%
%
\vspace{1cm}
\end{titlepage}

\section{Introduction}

    The standard model (SM) of particle physics has been very
successful in explaining the fundamental interactions of 
the elementary particles, and its predictions have been
verified experimentally to a very good accuracy except for the
discovery of the Higgs boson, the only elementary scalar particle 
in the SM. In spite of its merits, the SM has many open questions 
that are not addressed within its domain and a plenty of room is 
left open for some beyond SM physics scenarios to address them. 
Supersymmetry, extra dimensions, techni-color models are a few to name 
such beyond SM scenarios.  With the advent of the high energetic hadron 
colliders like Tevatron and the Large Hadron 
Collider (LHC),  it is quite feasible to probe these new scenarios in 
the laboratory experiments. The LHC with its unprecedented center mass energy 
of $14$ TeV and with luminosities as high as $10^{34}~\text{cm}^{-2}~s^{-1}$,  
offers the best possibility of discovering not only the elusive Higgs boson 
but also of the possible new physics that is hidden so far at lower energies.
On the other hand,  the Tevatron which is operating at a center of mass energy 
of $1.96$ TeV has been probing such new scenarios.
%

One of such beyond SM scenarios that have gained a lot of interest and have been 
studied well in the context of collider phenomenology is the large extra 
dimensional (LED) model proposed by Arkani-Hamed, Dimopoulos and Dvali \cite{ADD}. 
This model is theoretically well motivated and it addresses the hierarchy  problem
with the concept of extra spatial dimensions.  The size of the extra dimensions 
in this model can be of macroscopic size but still consistent with the data from 
the experiments to date.  A viable mechanism to hide  these extra $d$ spatial 
dimensions from the SM particles is to confine the latter to a $3$-brane and 
allow only the gravity to propagate the full $4+d$ dimensional space time.
For simplicity, the extra dimensions can be assumed to be flat, of the 
same size and compactified on a $d$-dimensional torus of radius $R/(2\pi)$.
After the compactification, the scale $M_s$ of the extra dimensional theory 
is related to the Planck scale $M_p$ as:
\begin{eqnarray}
M_p^2 = C_d~M_s^{2+d} ~R^d
\label{relation}
\end{eqnarray}
where $C_d= {2 ~ (4\pi)^{-{d \over 2}} / \Gamma(d/2)}$ and $R$ is the size of 
the extra dimensions.  This compactification implies that a massless graviton 
propagating in $4+d$ dimensions manifests itself as a tower of massive graviton 
modes in $4$-dimensions, with mass $ m_{\vec{n}}^2  = 4 \pi^2 \vec{n}^2/R^2$ 
where $\vec{n} = \{n_1, n_2, ...., n_d\}$ and $n_i = \{0, 1, 2, ...\}$.
Here, the zero mode corresponds to the $4$-dimensional massless graviton.
As the inverse square law of gravity has been tested down to only few 
$\mu m$ so far \cite{expt}, the size of the extra spatial dimensions in 
this model can be taken as large as this limit.  The hierarchy between 
the electroweak scale and the Planck scale can then be accounted for by this 
large volume of the extra dimensions, as can be seen from eqn.(\ref{relation}). 
For $M_s \sim {\cal O}(\text{TeV})$, the above limit on $R$ constrains the number 
of extra dimensions to $d \ge 2$.

   In the effective theory valid below the scale $M_s$, these gravitons couple 
to the SM fields through energy momentum tensor $T^{\mu\nu}$ of the latter with the 
coupling $\kappa = \sqrt{16\pi}/M_p$, as given by 
\cite{GRW, HLZ}
\begin{eqnarray}
{\cal L}_{int} &=& - \frac{\kappa}{2} \sum_{\vec n=0}^\infty
T^{\mu\nu} (x) ~h_{\mu\nu}^{(\vec n)} (x).
\label{int}
\end{eqnarray}
Since the coupling is through the energy momentum tensor, the gravitons can couple to 
all the SM fields with the same coupling strength $\kappa$ irrespective of their 
charge, color and flavor. The Feynman rules for the above interaction
lagrangian are given in \cite{GRW,HLZ,Mathews:2004pi} and 
in the first reference of \cite{us2}. 
To order $\kappa^2$, the above action allows processes
involving SM fields and virtual gravitons in the intermediate state 
or real gravitons in the final state.  
In the context of collider phenomenology, 
this gives rise to a very rich and interesting signals that 
can be seen at the LHC. The virtual exchange of 
the gravitons can lead to the deviations from the SM predictions 
whereas the real emission of the gravitons can lead to the missing 
energy signals.  Though the coupling of each graviton mode to the SM fields 
is $M_p$ suppressed, the large multiplicity of the available graviton modes 
can give rise to observable effects.  Hence, there will be a summation 
over the graviton modes at the amplitude level for the virtual graviton exchanges, 
and at the cross section level for the real graviton emissions. 
As the size of the extra dimensions could be large in this model, the mass 
splitting i.e. $2\pi/R$ is very small and hence this summation over the graviton 
modes can be approximated to be an integral in the continuum limit, with 
the density of the graviton modes given by \cite{HLZ}
\begin{eqnarray}
\rho(m_{\vec n}) = \frac{R^d ~ m_{\vec n}^{d-2}} 
{(4\pi)^{d/2} ~ \Gamma(d/2)}
\end{eqnarray}
For the real graviton production process at the collider experiments, 
the inclusive cross section is given by the following convolution:
\begin{eqnarray}
{d \sigma}&=& 
\int dm_{\vec{n}} ~ \rho(m_{\vec{n}}) ~ m_{\vec{n}}^{d-1} ~ 
{d \sigma_{m_{\vec{n}}}} ~,
\label{d_state2}
\end{eqnarray}
where $d\sigma_{m_{\vec{n}}}$ is the cross section for the production of a 
single graviton of mass $m_{\vec{n}}$.  
This collective contribution of the graviton modes results in their 
non-negligible interaction with the SM fields, and offers the best 
possibility of probing the low scale quantum gravity effects at the  
colliders experiments. Consequently, a very rich and interesting 
collider signals of some important processes have been reported in 
the literature, but most of them are available only at the leading 
order (LO) in the perturbation theory \cite{GRW,HLZ,Peskin,us,Wu}.  
At the hadron colliders like LHC or Tevatron, the QCD radiative 
corrections are very significant for they can enhance the LO 
predictions as well as decrease the arbitrary scale uncertainties in 
theoretical predictions. Further, the presence of a hard jet in the final 
state, due to these radiative corrections, has the potential to modify 
the shapes of the transverse momentum distributions of the particles that 
are under study at LO.  Obtaining such a modification to the shapes of 
the distributions is beyond the scope of the normalization of the corresponding 
LO distributions  by a constant K-factor, and it requires an explicit computation 
of the cross sections or distributions to next-to-leading order (NLO) in QCD. 
Owing to this importance of the radiative corrections, they have been computed 
for some important processes involving virtual or real graviton effects. 
The K-factors in some cases are found to be as high as a factor of two.  
Pair production processes are the best to exemplify the case of virtual graviton 
effects, where the NLO QCD corrections are computed for di-lepton \cite{us2}, 
di-photon \cite{us3}, di-Z and $W^+ ~W^-$ \cite{us4} production processes. 
In the context of missing energy signals in the large extra dimensional model,
the NLO QCD corrections are presented for the processes (i) jet plus graviton
production \cite{KKLZ} and (ii) photon plus graviton production \cite{GLGW}. 
In each of these two cases, it is shown that the K-factors can be as high as 
$1.5$ at the LHC.

In the present work,  we compute the NLO QCD corrections to the 
associated production of vector gauge boson and the graviton at the LHC and 
give a quantitative estimate of the impact of these radiative corrections.  
The paper is organized as follows. In sec.2, we discuss the importance of 
graviton plus vector gauge boson production process, outline the phase space 
slicing method for computing the NLO QCD corrections and present the analytical 
results. In sec.3, we give the numerical results for both the neutral gauge 
boson and the charged gauge boson cross sections. Finally, we present the 
conclusions in sec.4.

\section{Graviton plus vector boson production}

The gravitons when produced at the collider experiments escape 
the experimental detection due to their small couplings and negligible 
decays into SM particles.  The production of vector bosons ($V=Z,~W^{\pm}$) 
together with such an {\it invisible} gravitons ($G$) can give rise to a very large 
missing transverse momentum signals at the collider experiments.
The study of graviton plus gauge boson production, hence, in general will be a 
useful one in probing the new physics at the LHC. This process has been studied
at leading order (LO) in the context of lepton colliders \cite{Kingman, Giudice}
as well as at the hadron colliders \cite{Ask}, and also has been implemented 
in Pythia8 \cite{Ask2}. The process is an important one and stands complementary
to the more conventional ones involving the graviton production, like jet plus
graviton or photon plus graviton productions, that are generally useful in
the search of the extra dimensions at collider experiments.

   It is important to note that there is a Standard Model (SM) background 
which gives signals similar to those of associated production of $Z$ and
$G$.  This SM background receives a dominant contribution coming from 
the $ZZ$ production process, where one of the $Z$-bosons in the final 
state decays into a pair of neutrinos ($Z \to \nu \bar{\nu}$) leading 
to $Z$-boson plus missing energy signals. The other $Z$-boson  can be 
identified via its decays to leptons, mostly electrons and muons, and 
then constraining the lepton invariant mass close to the mass of the 
$Z$-boson to consider only the on-shell $Z$-bosons. A detailed study of 
the event selection and the minimization of other SM contributions to
this process $ZZ \to l \bar{l}\nu\bar{\nu}$, using MC@NLO and Pythia, 
is taken up in the context of ATLAS detector simulation and is 
presented in \cite{ATLAS}.  Any deviation from this SM prediction will
hint some beyond SM scenario and hence a study of this process will be
useful in searching the new physics.

    In the context of extra dimensions, a study of the $Z$ plus graviton 
production at LO at the LHC is discussed in \cite{Ask}, where the $Z$-boson 
identification is done with the leptonic decay modes and using the cuts on 
the leptons as specified in \cite{ATLAS}.  At the LHC, a similar study is 
done where the signals of $Z$-boson plus missing energy in this model 
are compared against those coming from the SM $(Z\nu\bar{\nu})$ background 
and are presented in our recent study \cite{us1}.
It is worth noting here that a signal of $Z$-boson plus missing 
energy can also come from the production of $Z$ plus unparticle ${\cal U}$, 
where the unparticle leads to missing
energy signals.  A study of such process based on ATLAS detector 
simulation \cite{Ask} shows that the vector unparticles are difficult to 
be probed using this channel, whereas the tensor unparticles can give signals 
identical to that of the graviton. In view of the above, it is worth 
studying the $Z$ plus missing energy signals, in particular $ZG$ production,
which will be useful to confirm the extra dimensional signals once they 
are seen in the main channels like jet or photon plus missing energy.

In what follows, we describe the computation of NLO cross sections for the
process under study. Since our focus is on the QCD part in this work, we 
will confine our calculation to the production of on-shell $Z$-boson. 
A more detailed study involving the $Z$-boson decays into leptons 
requires a full detector level simulation with the appropriate cuts 
at NLO and is beyond the scope of the present paper.

\subsection{Leading Order Calculation}
\begin{figure}[tbh]
\centerline{
\epsfig{file=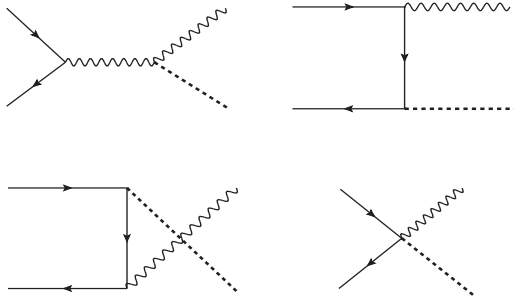,width=7cm,height=4cm,angle=0}}
\caption{Feynman diagrams that contribute to the associate production
of the vector boson and the graviton at the leading order.}
\label{lo}
\end{figure}

At the lowest order in the perturbation theory, the associated production
of the vector gauge boson and the graviton takes place via the quark 
anti-quark initiated subprocess, given by
\begin{eqnarray}
q_a(p_1) + {\bar{q}_b}(p_2) \rightarrow V(p_3) + G(p_4),
\end{eqnarray}
where $V = Z, W^\pm$ and $a,b$ are flavor indices. The corresponding Feynman diagrams are shown 
in Fig.\ (\ref{lo}). These diagrams are obtained by considering 
the tree level $q\bar{q}V$ diagram and by attaching the graviton 
line to all the possible external legs and the $q\bar{q}V$ vertex.
The Feynman rules and the summation of polarization tensor of 
the graviton are given in  \cite{GRW,HLZ}. 
The couplings of the fermions
to the $Z$ and $W$ bosons are given by 
\begin{eqnarray}
-i \frac{eT_Z}{2}~ \gamma^\mu (C_v -C_a \gamma^5)~, \quad \quad
-i \frac{eT_W}{2}~ \gamma^\mu (1 - \gamma^5)
\label{coupling}
\end{eqnarray}
where
\begin{eqnarray}
T_Z = \frac{1}{\hbox{cos} \theta_W~ \hbox{sin} \theta_W}, 
\quad \quad \nonumber
T_W = \frac{1}{\sqrt{2}~ \hbox{sin}~\theta_W}
\end{eqnarray}
and the coefficients $C_v$ and $C_a$ are 
\begin{eqnarray}
C_v = T_3^f - 2~\hbox{sin}^2 \theta_W~Q_f,
 \quad \qquad C_a = T_3^f
\end{eqnarray}
Here, $Q_f$ and $T_3^f$ denote the electric charge and 
the third component of the isospin of the quarks respectively,
and $\theta_W$ is the weak mixing angle.  For the vector gauge 
boson, the propagator in the unitary gauge $(\xi \rightarrow \infty)$ 
has been used throughout. 
This choice of the unitary gauge in the electro weak sector has the advantage 
of having vanishing goldstone and ghost contributions. The leading order matrix 
elements for the associated production of $Z$-boson and the graviton are 
computed using the algebraic manipulation program FORM \cite{FORM} and are 
given by (in $n$-dimensions)
%

\begin{eqnarray}
  \sum_{spin}~\overline {|M|}^2 &= &  \frac{1}{4} \frac{1}{3} \frac{1}{96}~
(C_v^2 + C_a^2) ~\frac{\kappa^2 T_Z^2}{(D^2 t^2 u^2)} ~\Big[
         12m^{10}(n-2)tu +  m^2tu\{3(n-2)^2t^4  
\nonumber \\[1.2ex]
{} &-& 2[-68 + n(104 + (-31 + n)n)]t^3u + 
              2[284 + n(-264 + (63 - 2n)n)]t^2u^2  
\nonumber \\ [1.2ex]
{} &-& 2[-68 + n(104 + (n-31)n)]tu^3 + 3(n-2)^2u^4 - 48m_z^6(n-2)(t + u)
\nonumber \\ [1.2ex]
{} &-&  4m_z^4[3(n-9)(n-2)t^2 + 2(124 + 3(n-21)n)tu + 3(n-9)(n-2)u^2] 
\nonumber \\[1.2ex]
{} &+&  4m_z^2(t + u) [3(n-5)(n-2)t^2 - 
                 2(-90 + n(n+35))tu + 3(n-5)(n-2)u^2]\} 
\nonumber \\ [1.2ex]
{} &-&  3m^8\{12(n-2)tu(t + u) + m_z^2[(n-2)^2t^2 + 2(16 + (n-14)n)tu 
\nonumber \\ [1.2ex]
{} &+&            (n-2)^2u^2]\} -
      3m^6\{4m_z^4(18 + (n-13)n)tu - 2m_z^2(t + u) [(n-2)^2t^2 
\nonumber \\ [1.2ex]
{} &+&  2(26 + n(2n-21))tu + (n-2)^2u^2] - tu[(n-2)(10 + n)t^2 
\nonumber \\ [1.2ex]
{} &+&         2(-32 + 3n(2 + n))tu + (n-2)(10 + n)u^2]\} 
\nonumber \\ [1.2ex]
{} &+&      2t^2u^2\{32m_z^6(n-2) - 8m_z^4(-6 + n + n^2)(t + u) 
\nonumber \\ [1.2ex]
{} &+&     (40 + (n-17)n)(t + u) [(n-2)t^2 + 2(n-4)tu + (n-2)u^2] 
\nonumber \\[1.2ex]
{} &-&     m_z^2[(n-2)(48 + (n-25)n)t^2  
\nonumber \\ [1.2ex] 
{} &+& 2(-156 + n(118 + (n-27)n))tu + (n-2)(48 + (n-25)n)u^2]\}  
\nonumber \\[1.2ex]
{} &+&  m^4\{48m_z^6(n-2)tu + 24m_z^4(18 + (n-12)n)tu(t + u)  
\nonumber \\ [1.2ex]
{} &-&  6tu(t + u) [(n-2)nt^2 + 2(-12 + n(3n-4))tu + (n-2)nu^2] 
\nonumber \\[1.2ex]
{} &-& m_z^2[3(n-2)^2t^4 + 12(n-6)(3n-5)t^3u +2(604 + n(25n-344))t^2u^2 
\nonumber \\[1.2ex]
{} &+& 12(n-6)(3n-5)tu^3 + 3(n-2)^2 u^4]\}\Big]
\label{Mlo}
\end{eqnarray}
%
where $D=(s-m_Z^2)$ and $s,t,u$ are the usual Mandelstam invariants. The over 
all bar in LHS of eq.\ (\ref{Mlo}) represents that the matrix elements have 
been averaged over the spins and the colors of the initial state particles and 
summed over those of the final state ones.

\subsection{Next-to-Leading Order Calculation}
At the NLO in the perturbation theory, the cross sections
receive ${\cal O}(\alpha_s)$ contributions from real emission as well as
virtual diagrams.  The integration over the phase space of the real emission diagrams will 
give rise to infra-red (IR) (soft and collinear) divergences in the limit 
where the additional parton at NLO is either soft and/or collinear to the initial 
state partons. On the other hand, the integration over the loop momenta in 
the virtual diagrams will also give rise to infrared divergences, in addition
to the ultraviolet (UV) divergences. In our calculation, we regulate all these 
divergences using dimensional regularization with $n = 4 + \epsilon$ being the 
number of space time dimensions.  
Completely anti-commuting $\gamma_5$ prescription \cite{CFH} is used to handle 
$\gamma_5$ in $n$ dimensions.
Here, it should be noted that as the gravitons couple to the energy momentum tensor 
of the SM fields, which is a conserved quantity, there won't be any UV divergences 
coming from the loop diagrams.  

There are several methods available in the literature to compute NLO QCD corrections.
Standard methods based on fully analytical computation deal with the phase space 
and loop integrals in $n$-dimensions and give a finite ${\cal O}(\alpha_s)$ 
contribution to the cross sections, after the real and the virtual contributions 
are added together and the initial state collinear singularities are absorbed into
the bare parton distribution functions.
However, these methods are not useful whenever the 
particles in the final state are subjected to either histogramming or experimental 
cuts or some isolation algorithms. In such cases,  semi analytical methods like 
{\it phase space slicing method} or {\it dipole subtraction method} are extremely 
useful. In the present work, we have resorted to the former 
with two cut offs to compute the radiative corrections. In this method, the IR 
divergences appearing in the real diagrams can be handled in a convenient way by 
slicing the soft and collinear divergent regions from the full three body phase space. 
The advantage of this method is that the integration over the remaining 
phase space can be carried out in $4$-dimensions, rather than in $n$-dimensions, 
using standard Monte-Carlo techniques.  In what follows, we give some of the details 
about the implementation of this phase space slicing method in our NLO computation.

\subsubsection{Real Emission Processes}
\begin{figure}[tbh]
\centerline{
\epsfig{file=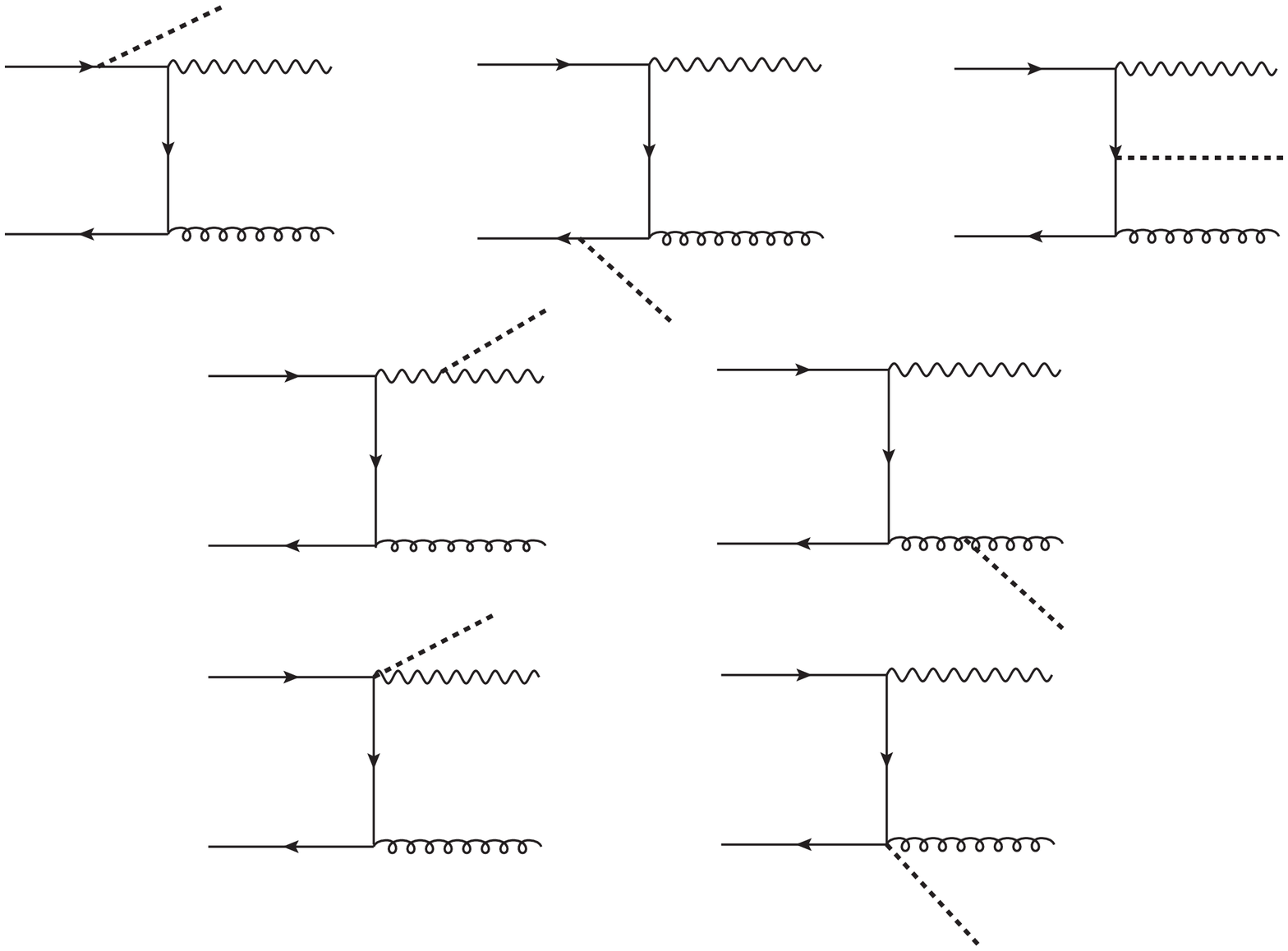,width=10cm,height=8cm,angle=0}}
\caption{Real gluon emission diagrams}
\label{real}
\end{figure}

There are two types of subprocesses that contribute to the associated production 
of the vector gauge boson and the graviton at NLO in QCD. They proceed by 
$q\bar{q}$ and $qg$ initial states.   At parton level, the $2 \to 3$ quark 
anti-quark initiated subprocess is given by
\begin{eqnarray}
\nonumber 
q_a(p_1)~ + ~\bar{q}_b (p_2)  \rightarrow  V(p_3)~  + ~ G(p_4) ~ + ~ g (p_5).
\end{eqnarray}
We find that $14$ diagrams contribute to this subprocess and a few of them are 
depicted in Fig.\ \ref{real}.  These diagrams are obtained by taking the $t$-channel 
$q\bar{q} \to Vg$ diagram at tree level and by attaching the graviton line to all 
the possible external as well as internal lines and to the vertices. The remaining 
diagrams are obtained by interchanging the vector boson and the graviton 
lines in Fig.\ \ref{real}.
In general, diagrams such as these involving gluons and massless quarks are prone 
to be singular in the soft and collinear regions of the $3$-body phase space 
integration.
In the phase space slicing method that we adopted here,
these soft and collinear regions are separated from the full $3$-body phase 
space using two small cut-off parameters, namely $\delta_s$ and $\delta_c$,
that define these singular regions. In the center of mass frame of the partons, 
the soft region is defined as: $0 \leq E_5 \leq {\frac{1}{2}}\delta_s\sqrt{s}$, 
where $E_5$ is the gluon energy and $\sqrt{s}$ is the parton center of mass energy.  
Integration of the eikonal approximated $2 \rightarrow 3$
matrix elements over the soft region of the phase space
gives the ${\cal{O}}(\alpha_{s})$ $2$-body contribution,
\begin{eqnarray}
 d\hat\sigma_S = a_s ~C_F~
F(\epsilon, \mu_R,s)
\left(\frac{16}{\epsilon^2} + \frac{16}{\epsilon}\ 
\ln\ \delta_s + 8\ \ln^2\ \delta_s\right)~d\hat\sigma_0 
\label{sigmaS}
\end{eqnarray}
where 
\begin{eqnarray*}
F(\epsilon, \mu_R, s) = 
\left[ 
\frac{\Gamma(1 + \frac{\epsilon}{2})}{\Gamma(1 + \epsilon)} 
\left(\frac{4\pi\mu_R^2}{s}\right)^{-\frac{\epsilon}{2}}\right],
~~ C_F = \frac{N^2-1}{2N} ~~~{\rm and}~~~ a_s ={ \alpha_s(\mu_R) \over 4\pi}.
\end{eqnarray*}
Here, $\alpha_s(\mu_R)=g_s^2(\mu_R)/4\pi$ with $g_s$ being the running 
strong coupling constant, $\mu_R$ 
is the renormalization scale and $N$ is the number of colors.
The region complementary to that of the soft $``S"$, i.e.  
$E_5 > {\frac{1}{2}}\delta_s\sqrt{s}$, is defined as the hard region ($H$) of 
the phase space. Within this hard region $``H"$, the emitted gluon can be collinear
to the incoming massless quark or anti-quark and hence can give rise to hard collinear 
divergences. By introducing another small cut-off parameter($\delta_c$), we 
separate these collinear divergences from the hard region.  The hard collinear
region($HC$) can be defined as :$0 \leq -t_{ij} \leq \delta_c s$ 
$(i=1,2~ {\rm and }~ j =5)$, where $t_{ij} = (p_i - p_j)^2$.
In the collinear limit, both the $2 \to 3$ matrix elements and the $3$-body
phase space get simplified to be expressed in terms of the born cross section
as: 
\begin{eqnarray}
 d\sigma_{HC}^{q\bar{q}}~ & = &~4a_s ~d\hat\sigma_0 ~ F(\epsilon, \mu_R, s)
\left(\frac{1}{\epsilon}\right)\!\!
\Big\{\!\!\left[P_{qq}(z,\epsilon)f_{q{/}P}(x_1/z)~f_{\bar q/P}(x_2) +
(q \leftrightarrow \bar q)\right] 
\nonumber \\[1.2ex]
&& + (x_1 \leftrightarrow x_2)\Big\}\frac{dz}{z}\ 
\left(\delta_c\frac{1-z}{z}\right)^{\frac{\epsilon}{2}}dx_1~dx_2
\label{qqbar}
\end{eqnarray}
where $f_{a/P}(x)$s' are the bare parton distribution functions (PDF) and
$P_{ab}(z,\epsilon)$ are the unregulated splitting functions in $n$-dimensions
and are related to the usual Altarelli-Parisi splitting kernels as
$P_{ab}(z,\epsilon) = P_{ab}(z) - {\epsilon \over 2} P'_{ab}(z)$ \cite{Harris}.  
Here $z$ denotes the fraction of the incoming parton $b$'s momentum carried 
by the parton $a$. Note that for $P_{qq}$ splitting in the hard region, since a 
fraction of the parton momentum i.e. $\delta_s$ is already carried  away by the 
gluon, the effective limits of the integration for $z$ will be $0 < z < 1-\delta_s$. 

Apart from the $q\bar{q}$ initiated subprocess at NLO, there will 
also be a $q(\bar{q})g$ initiated subprocess given by
\begin{eqnarray}
\nonumber 
q_a(p_1)~ + ~  g(p_2)  & \rightarrow & V(p_3)~ +  G(p_4)~ + ~ q_b(p_5).
\end{eqnarray}
Here the emitted parton, being a quark or an anti-quark instead of a gluon, 
won't give rise to soft singularity.  However, there will be hard 
collinear singularities whenever the emitted quark (anti-quark) becomes 
collinear to the incoming partons. These collinear singularities are 
separated using the cut-off $\delta_c$ in the same way as in the case 
of $q\bar{q}$ initiated subprocess. The cross section in this collinear 
region turns out to be
\begin{eqnarray}
d\sigma_{HC}^{qg,\bar{q}g} ~ &=& ~ 4a_s~d\hat\sigma_0~F(\epsilon, \mu_R, s)~
\!\!\left(\frac{1}{\epsilon}\right)\!\!
\Big\{\!\!\left[P_{\bar{q}g}(z,\epsilon)~f_{q{/}P}(x_1)~f_{g/P}(x_2/z) + 
(q \leftrightarrow \bar q)\right]
\nonumber \\[1.2ex]
&& + (x_1 \leftrightarrow x_2)\Big\}\frac{dz}{z}\
\left(\delta_c\frac{1-z}{z}\right)^{\frac{\epsilon}{2}}dx_1dx_2
\label{qg}
\end{eqnarray}
These initial state collinear divergences appearing in eqns. (\ref{qqbar} \& \ref{qg})
as poles in $\epsilon$ are purely due to the massless nature of the partons 
involved in the scattering process.  These divergences can be factored out 
from the parton level cross sections and be absorbed into the bare parton 
distribution functions at an arbitrary factorization scale $\mu_F$,  a process 
called mass factorization.  In the $\overline{MS}$ scheme, the scale dependent 
parton distribution functions, $f_{a/P}(x,\mu_F)$, can be expressed in terms of the 
bare parton distribution functions as given by
\begin{eqnarray}
f_{a/P}(x,\mu_F)   =   f_{a/P}(x) + ~
2a_s~\sum\limits_{b} \left(\frac{1}{\epsilon}\right)
F(\epsilon, \mu_R,\mu_F)
~\int\limits_x^1\frac{dz}{z} ~P_{ab}(z)~f_{b/P}(x/z),
\label{pdf}
\end{eqnarray}
where $a,b= q(\bar{q}),g$.
Substituting these parton densities in $d\hat\sigma_0$ produces collinear singular 
counter terms which 
when added with the hard collinear contributions results 
in the following ${\cal{O}}(a_s)$ contribution \cite{us3,us4}:
\begin{eqnarray}
d\sigma_{coll}~ & = &~ 2a_s~d\hat\sigma_0~F(\epsilon, \mu_R, s)
\Big(\Big\{f_{\bar{q}/p}(x_2,\mu_F)
[\widetilde{f}_{q/p}(x_1,\mu_F) + f_{q/p}(x_1,\mu_F)
\nonumber \\[1.2ex]
&&\left(-\frac{2}{\epsilon} + \ln \frac{s}{\mu_F^2}\right)A_{q \rightarrow q+g}] + 
(q \leftrightarrow \bar{q})\Big\} + (x_1 \leftrightarrow x_2) \Big)dx_1dx_2
\label{sigmaHC}
\end{eqnarray}
where $A_{q \rightarrow q+g}= C_F \left(2 \ln \delta_s + \frac{3}{2}\right)$.
The tilde parton distribution functions are given by \cite{GLGW,Harris}
\begin{eqnarray}
\widetilde{f}_{q/P}(x,\mu_F)  = \sum\limits_{b = q,g} 
\int_x^{1-\delta_s \delta_{qb}} \frac{dy}{y} f_{b/P}(x/y,\mu_F)
\times \widetilde{P}_{qb}(y)
\end{eqnarray}
\begin{eqnarray}
{\rm with} \quad \widetilde{P}_{ab}(y)   =  P_{ab}(y) ~ \hbox{ln}
\Big(\delta_c~\frac{1-y}{y}\frac{s}{\mu_F^2} \Big) - P'_{ab}(y).
\end{eqnarray}
Note that there is an additional factor of two, as the parton in the
final state can be collinear to either of the incoming partons, which
is implicit from $(q \leftrightarrow \bar{q})$ in eqn. (\ref{sigmaHC}).
At this stage, one can observe that the divergent pieces that
are proportional to $\ln\delta_s$ cancel among themselves.  However, there
are singularities still remaining that will get cancelled only with those 
coming from the loop integrals in the virtual diagrams. In what follows, 
we present the details of the virtual corrections to our process.
\subsubsection{Virtual Corrections}
\begin{figure}[tbh]
\centerline{
\epsfig{file=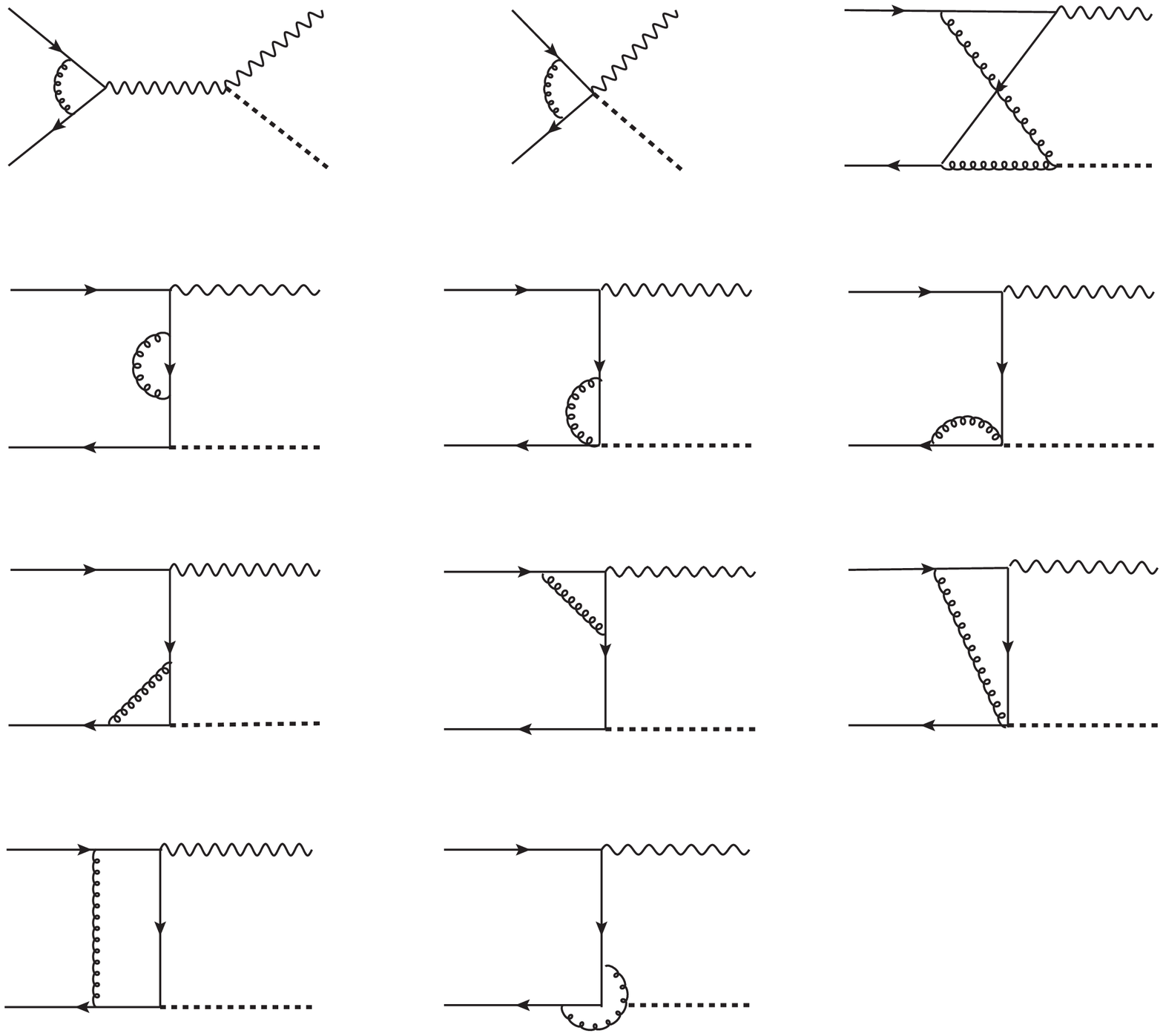,width=12cm,height=10cm,angle=0}}
\caption{Virtual gluon emission diagrams}
\label{virtual}
\end{figure}

The NLO cross sections also receive the contributions coming from the virtual 
corrections as well as the wave function renormalization to the $2 \rightarrow 2$ 
leading order processes. The corresponding Feynman diagrams are obtained by 
considering possible one loop virtual gluonic corrections to the tree level Feynman 
diagram for $q \bar{q} \to Z$ and then by attaching the graviton line to all possible 
internal as well as external lines and to vertices, as allowed by the Feynman rules. 
This way we find $27$ diagrams, out of which  $8$ diagrams correspond 
to external leg corrections and can be omitted as they vanish in the massless quark 
limit. Out of the remaining $19$ diagrams, $11$ are shown in Fig.\ \ref{virtual}. 
The rest of the diagrams can easily be obtained by inverting the charge flow direction 
of the quark lines in the last eight diagrams shown in the Fig.~(\ref{virtual}).
Interference of these one loop diagrams with the born diagrams  
gives ${\cal O}(a_s)$ contributions.  
Due to tensorial interaction of gravitons with the SM fields
the loop integrals involve higher powers of loop momenta
in their numerators and hence the reduction of tensorial integrals
to scalar ones becomes complicated.  We have written a symbolic program
using FORM \cite{FORM} to perform this reduction in $n$ dimensions.  The resulting
scalar integrals are then evaluated exactly (see \cite{SCALAR}) and they are listed
in the Appendix.  
Substituting these scalar integrals, we can express  
the ${\cal{O}}(a_s)$ contribution resulting from the
virtual processes as
\begin{eqnarray}
d\hat\sigma_V &=& a_s ~d\hat\sigma_0 ~ F(\epsilon, \mu_R, s)
C_F
\left(-{16 \over \epsilon^2} + {12 \over \epsilon } \right)
\nonumber\\[1.2ex]
&&+C \Bigg[V_1~ \ln^2\left({-t \over \mu^2}\right)
+V_2~ D_0^{fin}(p_1,k,q) 
+V_3~ D_0^{fin}(p_2,k,q) 
\nonumber\\[1.2ex]
&&+V_4~ \ln^2\left({-u \over \mu^2}\right)
+V_5~ \ln^2\left({-m^2 \over \mu^2}\right)
+V_6~ \ln^2\left({-m_Z^2 \over \mu^2}\right)
+V_7~ \ln^2\left({s \over \mu^2}\right)
\nonumber\\[1.2ex]
&&+V_8~ \ln\left({-t \over \mu^2}\right)
+V_9~ \ln\left({-u \over \mu^2}\right)
+V_{10}~i~C_0(k,q) 
+V_{11} 
+V_{12} 
+V_{13}~ \ln\left({m_Z \over \mu^2}\right)
\nonumber\\[1.2ex]
&&+V_{14}~ \ln\left({m \over \mu^2}\right)
+V_{15}~\zeta_2 \Bigg],
\label{sigmaV}
\end{eqnarray}
where $C = a_s \kappa^2~(C_v^2 + C_a^2)~T_Z^2~C_F /(4N)$,  
$C_0^{\rm fin}$ and 
the $D_0^{\rm fin}$s are the finite parts of the scalar integrals 
$C_0$ and $D_0$ respectively and are listed in the Appendix along with $V_i$s.  
It is clear from the above expression that the integration over the 
loop momenta in $4+\epsilon$ dimensions leads to 
soft and collinear singularities which appear as poles in $\epsilon$.
We found that the UV divergences that appear in the intermediate stages
cancel among various diagrams thanks to the conservation of SM energy momentum
tensor to this order in perturbation theory.
Now, when we add ${\cal{O}}(\alpha_s)$ contributions coming from 
eq.\ (\ref{sigmaS}), (\ref{sigmaHC}) and (\ref{sigmaV}), we observe that  
the remaining soft and collinear singularities cancel among themselves as expected, 
leaving a finite expression for the $2$-body contribution which can be 
computed using Monte-Carlo techniques.  In other words,  the $2$-body contribution, 
given by
\begin{eqnarray}
d\sigma^{\rm 2-body} = d\sigma_S ~ + ~ d\sigma_{coll} ~ + ~ d\sigma_V.
\label{2body}
\end{eqnarray}
is found to be free of both UV and IR singularities and hence suitable for further
numerical evaluation.

In addition to the above contribution, we also have the hard non-collinear region 
$\overline{HC}$ of the phase space which do not suffer from any IR singularities
by construction. 
The contributions from this region can be obtained by integrating the $2 \to 3$ 
matrix elements using standard Monte-Carlo 
integrations.  Owing to the divergence free nature of the integration, the 
$2 \to 3$ matrix elements computed in $4$-dimensions will suffice our purpose. 
These matrix elements are again computed using FORM.  
We have made several checks to ensure the correctness of our results, namely
the gauge invariances in QCD, electroweak and gravity sectors.
Since contributions from hard non-collinear regions involve three body phase space integrals
of final state particles having different masses, care is needed to 
parameterize as well as to determine the limits of various integrations.  We devote
our next sub section to discuss this.  

\subsection{Three body contribution}
In this section, we will present briefly how we have implemented 
various constraints imposed by the two cut off phase space slicing method 
and cuts on the phase space integrals for the $2 \rightarrow 3$ subprocesses.  
We are interested in the following cross section:
\begin{eqnarray}
d \sigma^{3-body} = \int\limits_{\overline{HC},cuts} d \Gamma_3 ~ |M_{q\bar{q},qg}^{2 \to 3}|^2.
\label{3body}
\end{eqnarray}

\noindent
where the three-body phase space measure is given by
\begin{eqnarray}
d\Gamma_3 = \Bigg(\prod\limits_{i=1}^3 \frac{d^3p_i}{(2\pi)^3 2E_i}\Bigg)
            (2\pi)^4 \delta^{(4)}(p_1 + p_2 -p_3 -p_4 -p_5).
\label{phase3}
\end{eqnarray}
It is easy to parameterize all the momenta in the center of mass frame of initial
state partons and then boost them to the lab frame or the center
of mass frame of the hadrons. The $4$-momenta of the massless partons in the 
initial state, moving along the $z$-axis, are given by
\begin{eqnarray}
p_1 = \frac{\sqrt{s}}{2}(1, 0, 0, 1), \quad p_2 = \frac{\sqrt{s}}{2}(1, 0, 0, -1)
\end{eqnarray}
where $\sqrt{s}$ is the parton center of mass energy. The corresponding $4$-momenta of the 
massive particles in the final state are given by
$p_i = (E_i, ~\vec{p}_i)$
with masses  $m_i^2 = E_i^2 - |\vec{p}_i|^2$, for $i = 3,4,5$.
For the three body case, it is easy to consider the momentum direction 
of one of the final state particles, say $\vec{p}_5$, as the reference 
direction and then parameterize the other two momenta $\vec{p}_3$ and $\vec{p}_4$
with respect to this direction: 
\begin{eqnarray}
\vec{p}_5   & = & |\vec{p}_5|~ (\hbox{sin}\theta, ~0, ~\hbox{cos}\theta)
\end{eqnarray}
where $\theta$ is the angle between $\vec{p}_5$ and the $z$-axis.
The momentum of $\vec{p}_3$ can now be parameterized with respect to the direction
of $\vec{p}_5$ and then followed by a rotation in the $xz$-plane by an angle of 
$\theta$ to get $\vec{p}_3 = (p_3^x, p_3^y, p_3^z)$ in the center of mass frame 
of the partons as given by
\begin{eqnarray}
\nonumber
p_3^x & = & |\vec{p}_3|~ \big(\hbox{cos}\theta~ \hbox{cos}\alpha~ \hbox{sin}\beta
            +\hbox{sin}\theta~ \hbox{cos}\beta\big) \nonumber \\[1.2ex]
\nonumber
p_3^y & = & |\vec{p}_3|~ \hbox{sin}\alpha~ \hbox{sin}\beta 
\nonumber \\[1.2ex]
p_3^z & = & |\vec{p}_3| \big(\hbox{cos}\theta~ \hbox{cos}\beta - 
                      \hbox{sin}\theta~ \hbox{cos}\alpha~ \hbox{sin}\beta\big)
\end{eqnarray}
where $\alpha$ and $\beta$ are the azimuthal and polar angles of $\vec{p}_3$ 
with respect to $\vec{p}_5$.  The 4-momentum of $p_4$ simply follows from 
the energy momentum conservation.
The three body phase space in eqn.(\ref{phase3}) can now be expressed in terms
of the angular variables, using
\begin{eqnarray}
\frac{d^3 p_i}{2E_i} = d^4p_i ~\delta(p_i^2 - m_i^2) = 
\frac{|\vec{p_i}|}{2} ~ dE_i ~ d\Omega_i
\end{eqnarray}
to get 
\begin{eqnarray}
d\Gamma_3 = \frac{|\vec{p_3}||\vec{p_5}|}{4(2\pi)^5}~dE_3 ~ d\Omega_3 ~dE_5 ~ d\Omega_5
~\delta(p_4^2 - m_4^2),
\end{eqnarray}
where $d\Omega_3 = d \hbox{cos}\beta~d\alpha$ and 
$d\Omega_5 = d \hbox{cos}\theta~d\phi$. 
Further, the angle $\beta$ can be eliminated using
\begin{eqnarray}
2~|\vec{p_3}| ~ |\vec{p}_5|~\hbox{cos}\beta = |\vec{p}_4|^2 - 
|\vec{p}_3|^2 - |\vec{p}_5|^2
\end{eqnarray}
Finally, out of the nine integration variables of the three body phase space, in 
eqn.(\ref{phase3}), we are 
left with four independent variables {\it viz.} $E_3$, $E_5$, $\theta$ and $\eta$,
(due to $4$-momentum conserving delta function and the rotational invariance of 
$\vec{p}_5$, the reference momentum direction). The three body phase space 
can then be written in terms of these four independent variables as
\begin{eqnarray}
d\Gamma_3 = \frac{1}{8(2\pi)^4}~dE_3~dE_5~d\hbox{cos}\theta~d\eta
\end{eqnarray}
The limits of integration of $E_3$ and $E_5$ can be obtained from the
constraint $|\hbox{cos}\beta| \le 1 $, and are given by
\begin{eqnarray}
E_5^{min} = m_5, \quad \qquad
E_5^{max} = \frac{1}{2\sqrt{s}}[s + m_5^2 - (m_3 + m_4)^2]
\end{eqnarray}
and
\begin{eqnarray}
E_3^{min, max} = \frac{1}{2\tau}
\Bigg[A(B + m_+ m_-) \pm \sqrt{(B - m_+^2)(B - m_-^2)} \Bigg]
\end{eqnarray}
where
\begin{eqnarray}
A = \sqrt{s} - E_5, \quad B = A^2 - |\vec{p}_5|^2 \quad \hbox{and} 
\quad m_\pm = m_3 \pm m_4
\end{eqnarray}
Finally, all the parton momenta can be boosted back to the lab frame or the center
of mass frame of the hadrons by a boost factor given, in the limit of the 
zero rest mass of the hadrons, by
\begin{eqnarray}
\beta = \frac{P_{cm}}{E_{cm}} = \frac{(x_1 - x_2)}{(x_1 + x_2)}
\end{eqnarray}
where $x_1$ and $x_2$ are the fractions of the incoming hadron momenta carried 
by the partons in the center of mass frame of the hadrons.  

We have implemented this phase space parameterization in our numerical code 
written in Fortran 77.  We set $p_i=k_i$ ($i=3,4,5$) and $m_3=m_V,m_4=m$ in our code.   
Here $m_V$ is mass of the gauge boson and $m$ is mass of the graviton.
The phase space integrations as well as 
various convolutions in the two and three body contributions
are done using VEGAS multi dimensional integration package. In what follows, 
we present the impact of our NLO corrections on various observables.

\section{Numerical Results}
\subsection{Neutral gauge bosons}
In this section, we present various kinematic distributions for the 
associate production of the graviton and the  vector gauge boson to 
NLO in QCD at the LHC. The results are presented for proton-proton 
collision energy of $\sqrt{S} = 14$ TeV.  As discussed before, the 
inclusive cross section for the graviton production involves the 
summation of all possible graviton modes. This summation in the 
continuum limit leads to an  integral over the graviton mass.
The limits of this integral are set by the kinematics from $0$ to 
$\sqrt{s} -m_V$, where $\sqrt{s}$  is the parton center of mass 
energy and $m_V = m_Z, m_W$.  
The masses of the gauge bosons and the weak mixing angle are given 
by \cite{PDG1}
\begin{eqnarray}
m_Z = 91.1876 ~ \rm{GeV},  \quad m_W = 80.398 ~ \rm{GeV},  \quad \text{sin}^2\theta_w = 0.2312 
\end{eqnarray} 
The fine structure constant is taken to be $\alpha = 1/128$. 
Throughout our study, we have used CTEQ6L1 and CTEQ6.6M 
parton density sets \cite{CTEQ}
for LO and NLO cross sections respectively.  The strong coupling constant is 
calculated at two loop order in the $\overline{MS}$ scheme with 
$\alpha_s(m_Z) = 0.118$ ($\Lambda_{\text{QCD}} = 0.226$ GeV). 
We have also set the number of light flavors $n_f = 5$.   
The following cuts are used for our numerical study,
\begin{eqnarray}
p_{T}^{Z, W} > p_T^{min},  \quad p_{T}^{miss} >  p_T^{min},
\quad |y^{Z,W}| \le 2.5.
\label{cuts1}
\end{eqnarray}
For the 2-body process, the missing transverse momentum is same as that
of the  gauge boson.  On the other hand, for the 3-body process, it need not be so 
due to the presence of an observable jet in the final state and hence 
it amounts purely to the graviton transverse momentum. The observable
jet is defined as the one that satisfies the following conditions:
\begin{eqnarray}
p_T^{jet} > 20 \text{GeV} \quad \text{and} \quad |\eta^{jet}| \le 2.5
\label{cuts2}
\end{eqnarray}
Whenever the jet does not satisfy the above conditions, the missing 
transverse momentum is approximated to be that of the gauge boson.

The LED model is an effective field theory valid below the UV cut-off scale $M_s$, 
which is expected to be of the order of a few TeV . At the LHC energies 
($\sqrt{S} = 14$ TeV), it is very well possible that the partonic center of mass 
energies can exceed this scale $M_s$ and lead to the signals that do not correspond 
to the compactified extra dimensions of the LED model.  This necessitates the need 
to quantify the UV sensitivity of the theory and this issue was already addressed in 
\cite{GRW}, according to which the cross sections can be computed in two different 
ways, one with {\it truncation} where the cross sections are set to zero whenever 
the hard scale $Q$ involved in the problem exceeds $M_s$, and the other with 
{\it un-truncation} where there is no such constraint imposed on the cross sections. 
As pointed out in \cite{GRW}, if these two results converge then the predictions are 
valid and the model is viable, otherwise the un-truncated cross sections can dominate 
the truncated ones, implying the calculations are not under control. In our calculation,
we choose the hard scale to be the invariant mass of the gauge boson and the 
graviton, $Q = M_{ZG}$, which at LO is the same as the center of mass 
energy of the partons $\sqrt{s}$.  We will consider both truncated as well 
as un-truncated cases, however, most of our distributions are obtained with our 
default choice of truncation scheme.

\begin{figure}[htb]
\centerline{
\epsfig{file=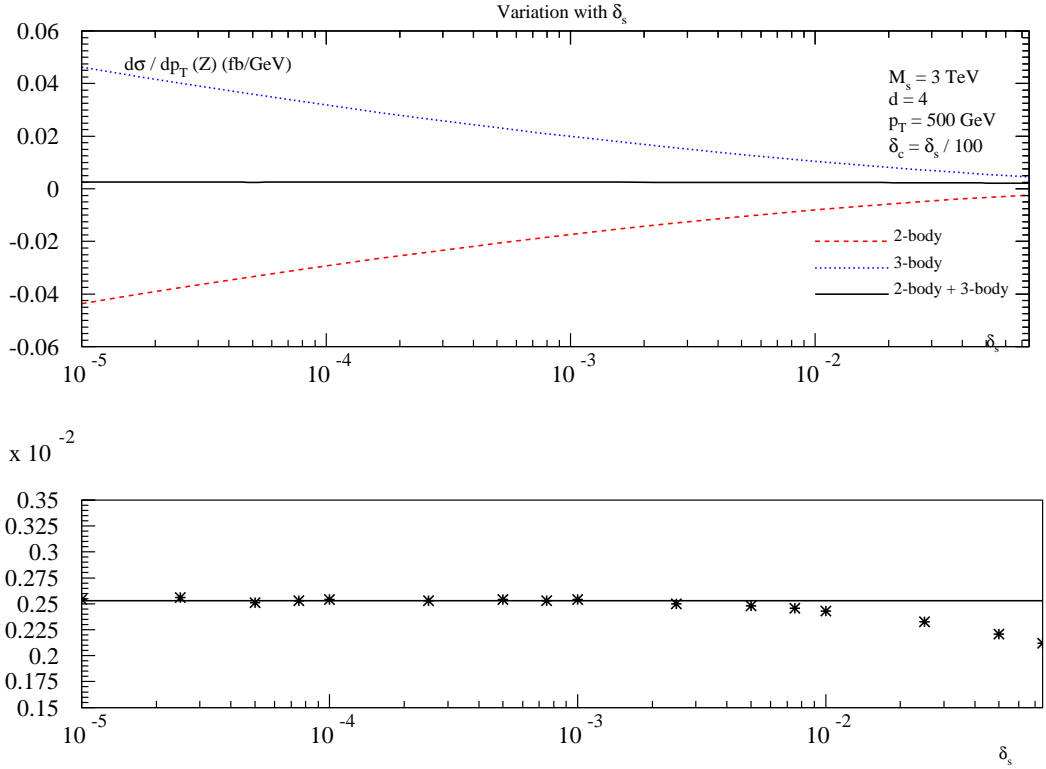,width=15cm,height=12cm,angle=0}}
\caption{Variation of the transverse momentum distribution of $Z$ boson
with $\delta_s$, keeping the ratio $\delta_s/\delta_c=100$ 
fixed, for $M_s=3$ TeV and $d=4$.}
\label{ds-z}
\end{figure}

Before proceeding to the kinematic distributions, we will do some consistency
checks on the calculation. First, we check for the stability of the cross 
sections against the variation of the slicing parameters, $\delta_s,\delta_c$. 
The sum of the 2-body 
and the 3-body contributions given in eqns. (\ref{2body}) and (\ref{3body}) 
is expected
to be independent of the choice of these slicing parameters that are introduced 
in the intermediate stages of the calculation.  In fig.(\ref{ds-z}), we show 
the dependency of the transverse momentum distribution $p_T^Z$ on the slicing 
parameter $\delta_s$ keeping the ratio of $\delta_s$ to $\delta_c$, fixed at 
a value of $100$.  This distribution is obtained using the hard truncation scheme
for a particular choice of the model parameters $M_s = 3$ TeV and $\delta =4$.
It can be seen from the fig.(\ref{ds-z}) that both the 2-body and the 3-body 
contributions vary with $\delta_s$ but their sum is fairly stable 
against the variation of $\delta_s$ over a wide range. This ensures the proper 
implementation of the slicing method in our NLO computation.

    Another useful check on the computation is to reproduce the cross sections 
for the associated production of the photon and the graviton at the LHC \cite{GLGW}. 
In order to do this, we recalculated both real emission as well as virtual contributions
for this process and the corresponding soft and collinear pieces.  
We found that the following replacements 
\begin{enumerate}
\item{${(C_V^2 + C_A^2) \over 4} \to Q_f^2$}
\item{$m_z \to 0$}
\item{$T_z \to e$}
\end{enumerate}
in the two body and three body real emission matrix elements of the $Z/W^\pm$ boson
with Graviton production processes correctly reproduce the corresponding matrix elements 
for photon with Graviton production processes.
Here, $Q_f$ is the charge of the fermion and $e$ is the electromagnetic coupling.
Using our symbolic program, the analytical expression 
for virtual contributions for this process agrees with one given in the appendix of \cite{GLGW}.
In addition, using these recalculated quantities, we reproduced 
all the numerical results in \cite{GLGW} after taking their choice of parameters, cuts etc.
It is important to note that the NLO cross sections, or the K-factors, 
are subject to the choice of the event 
selection or more precisely the cuts on the particles in the final state.  
In our calculation, however, the gauge bosons being  massive, we present our 
results according to the cuts given in eqns. (\ref{cuts1} \& \ref{cuts2}). 

\begin{figure}[htb]
\centerline{
\epsfig{file=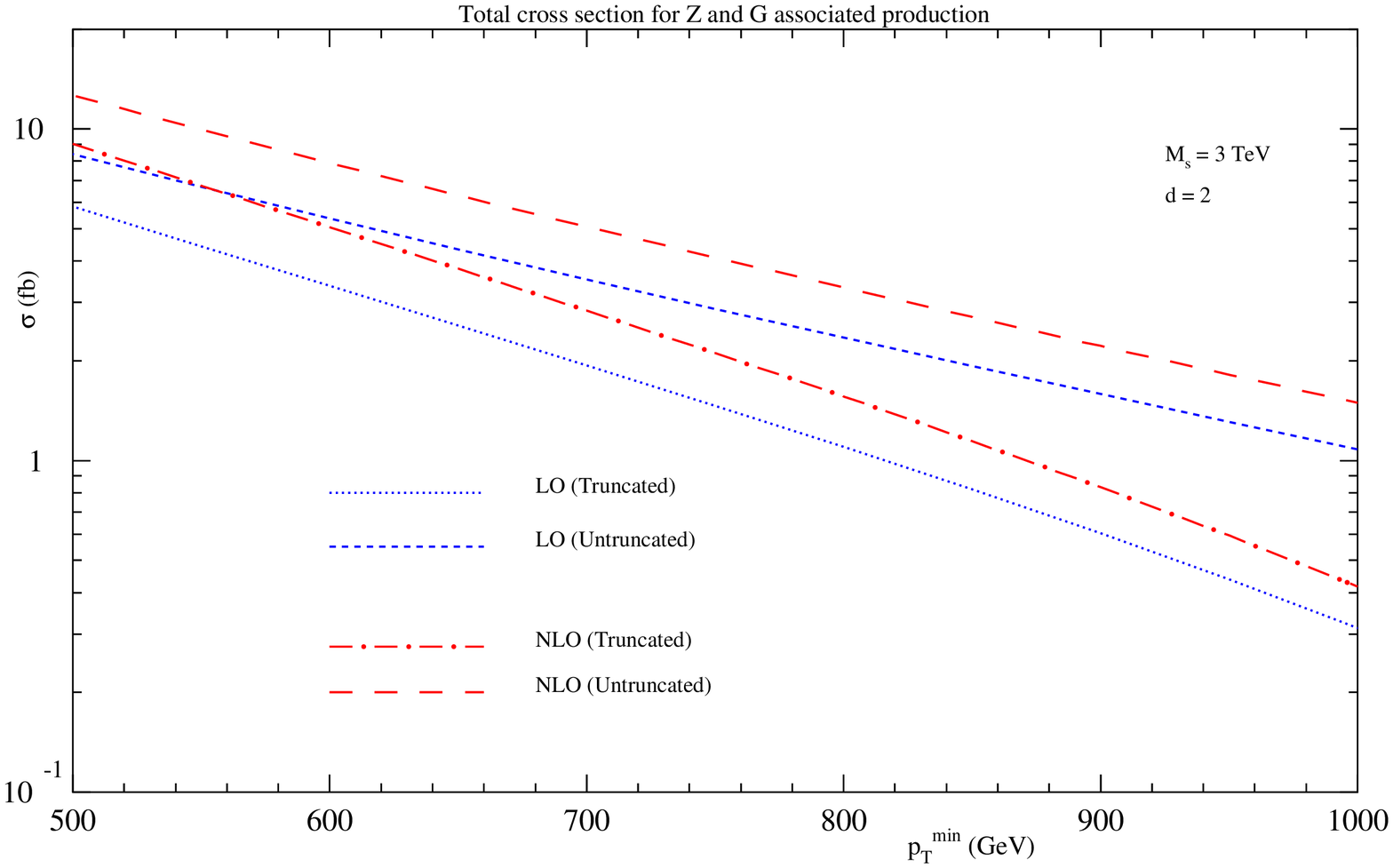,width=15cm,height=10cm,angle=0}}
\caption{Total cross section for the associated production of $Z$ and $G$
as a function of $P_T^{min}$ at the LHC, for $M_s = 3$ TeV and $d=2$.}
\label{tot-z-d2}
\end{figure}

\begin{figure}[htb]
\centerline{
\epsfig{file=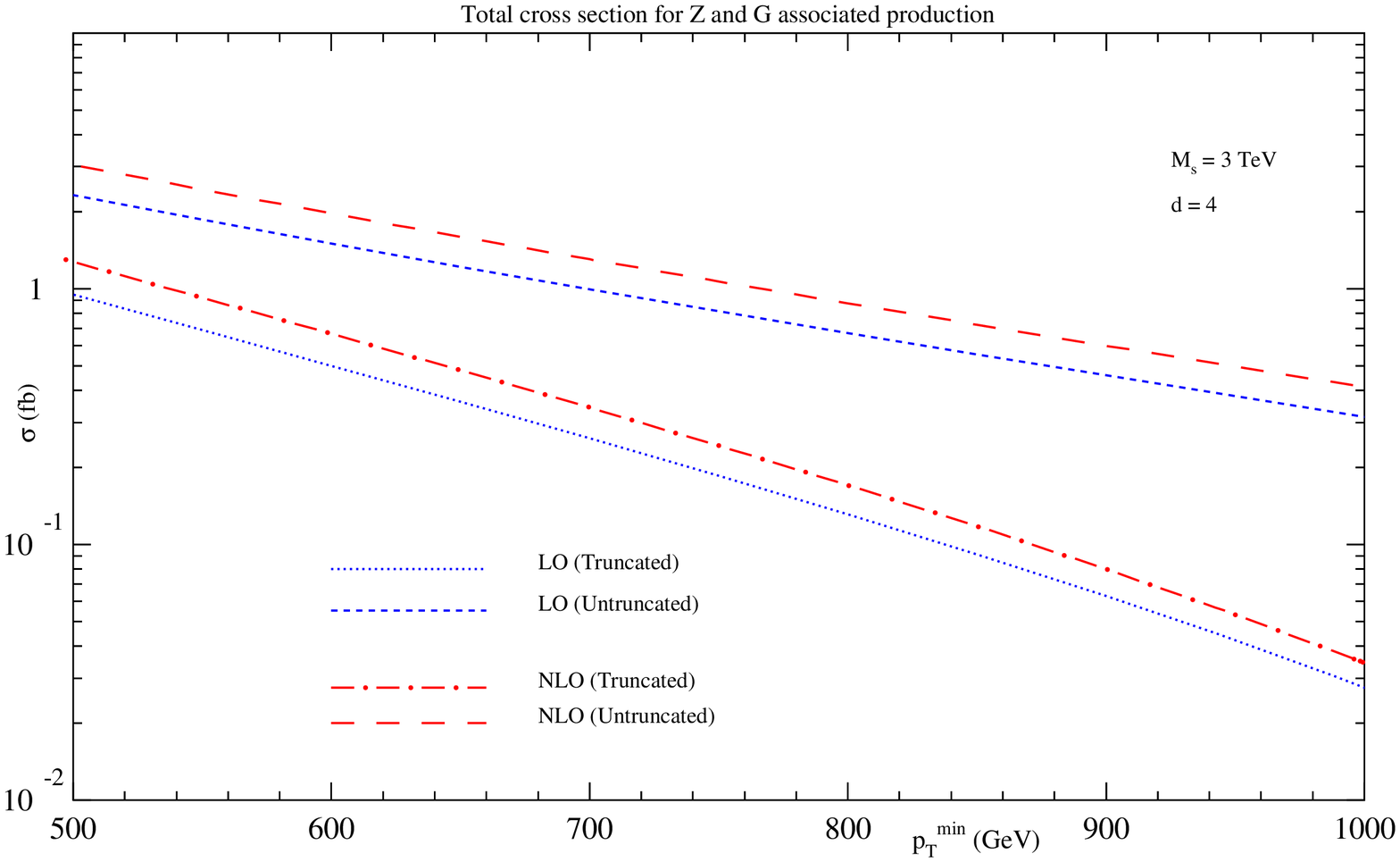,width=15cm,height=10cm,angle=0}}
\caption{Total cross section for the associated production of $Z$ and $G$
as a function of $P_T^{min}$ at the LHC, for $M_s = 3$ TeV and $d=4$.}
\label{tot-z-d4}
\end{figure}

\begin{figure}[htb]
\centerline{
\epsfig{file=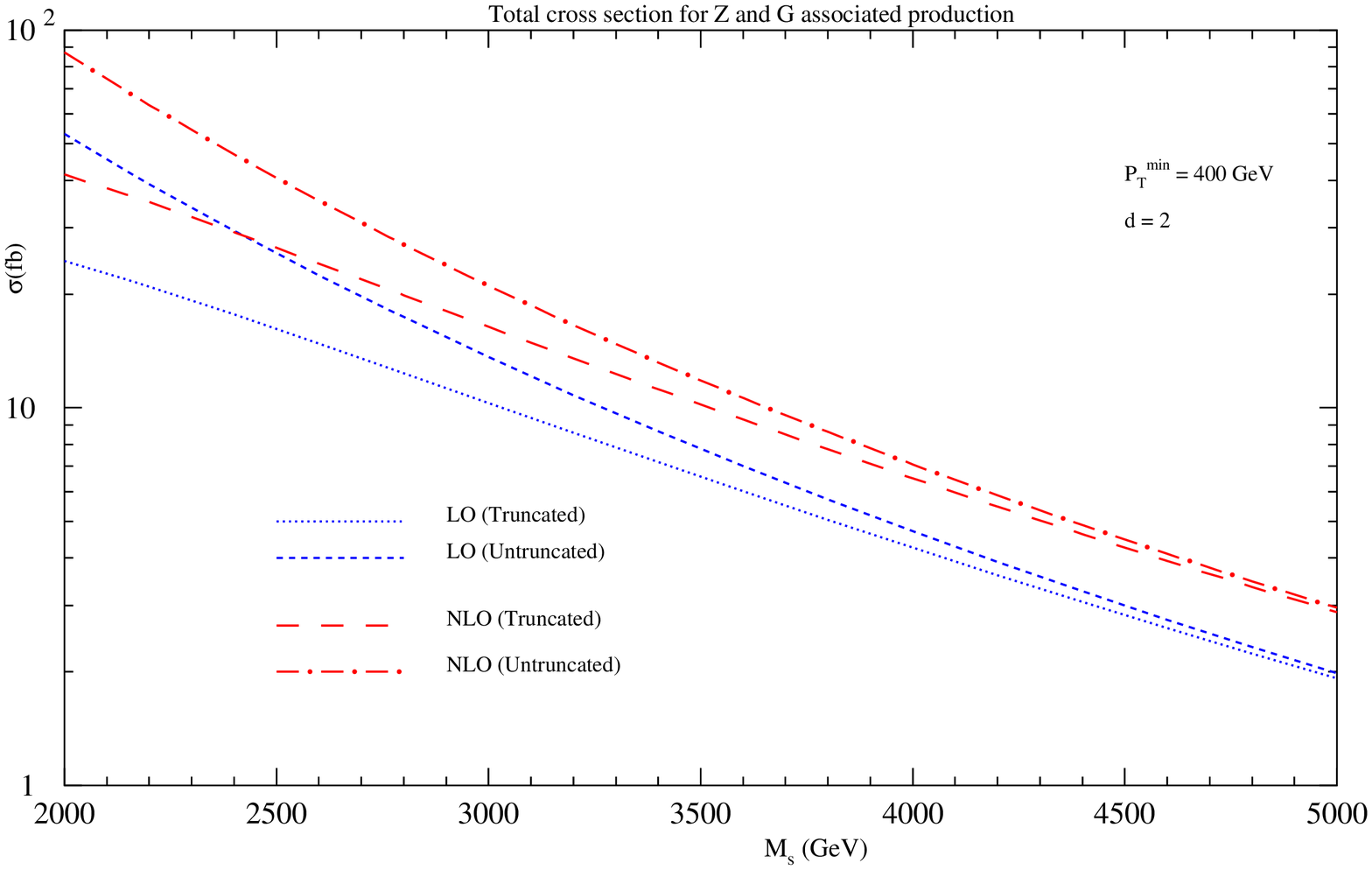,width=15cm,height=10cm,angle=0}}
\caption{Total cross section for the associated production of $Z$-boson
and the graviton at the LHC, shown as a function of $M_s$ for $d=2$.}
\label{totms-z-d2}
\end{figure}

\begin{figure}[htb]
\centerline{
\epsfig{file=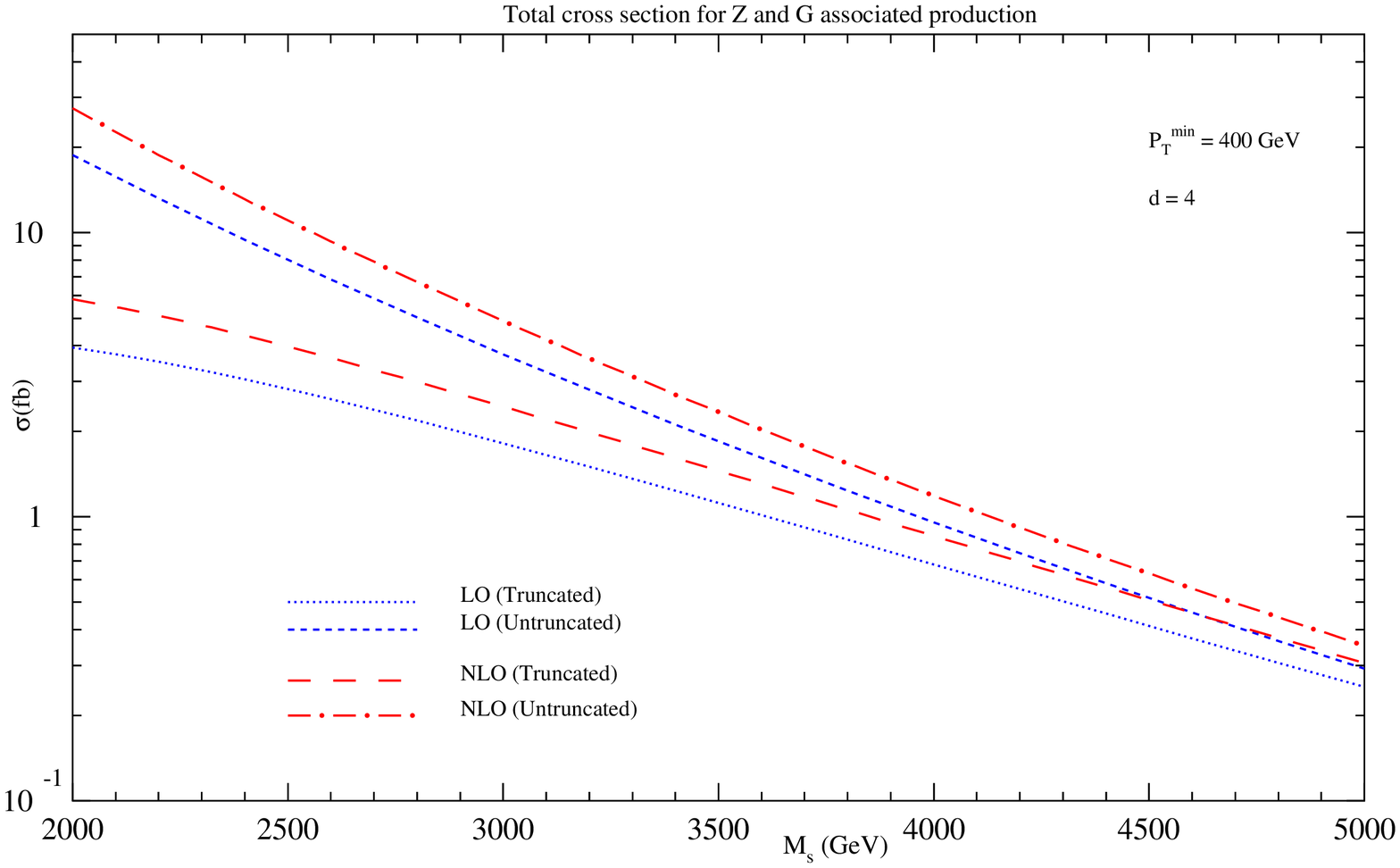,width=15cm,height=10cm,angle=0}}
\caption{Total cross section for the associated production of $Z$-boson and 
the graviton at the LHC, shown as a function of $M_s$ for $d=4$.}
\label{totms-z-d4}
\end{figure}

\begin{figure}[htb]
\centerline{
\epsfig{file=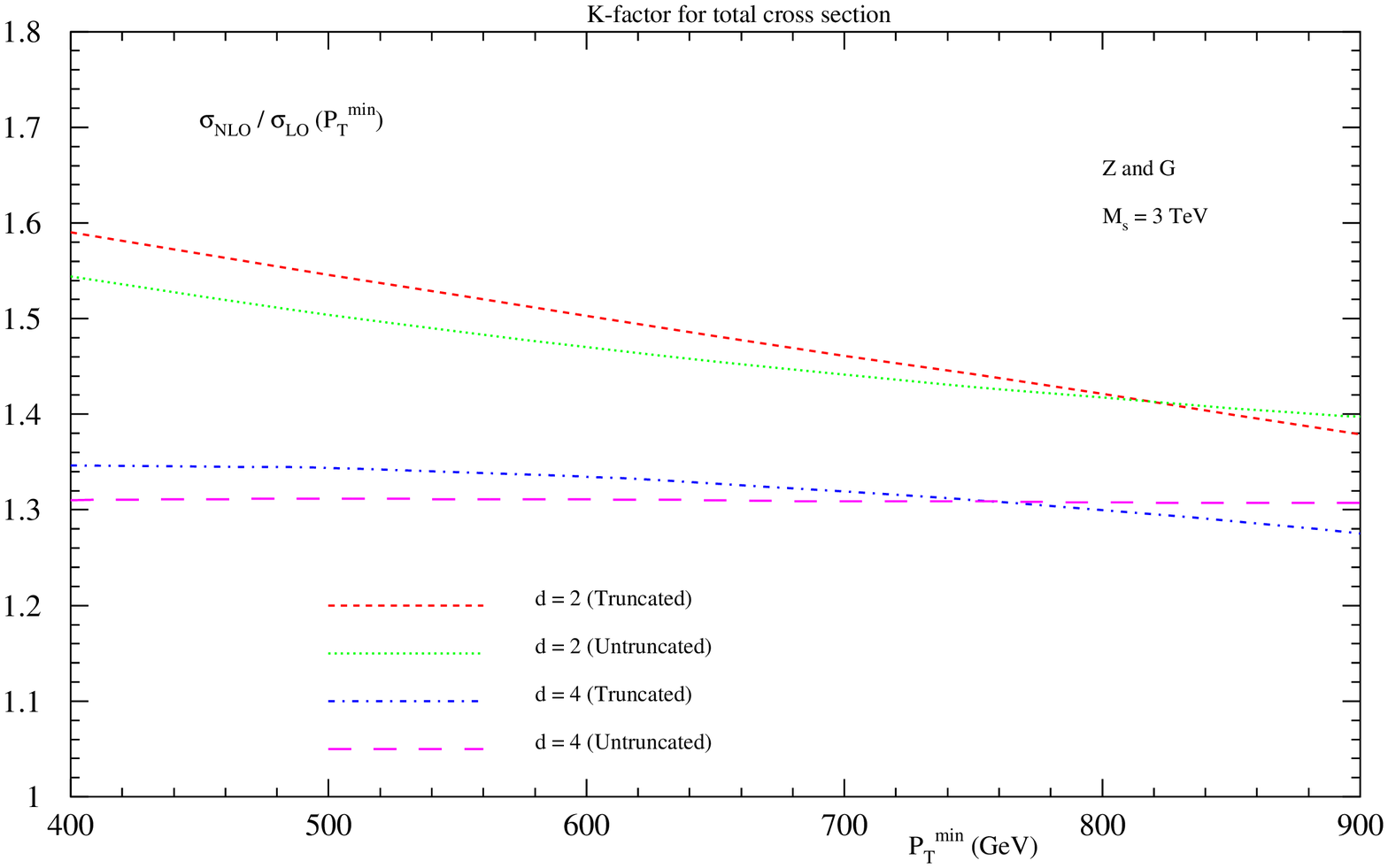,width=15cm,height=10cm,angle=0}}
\centerline{
\epsfig{file=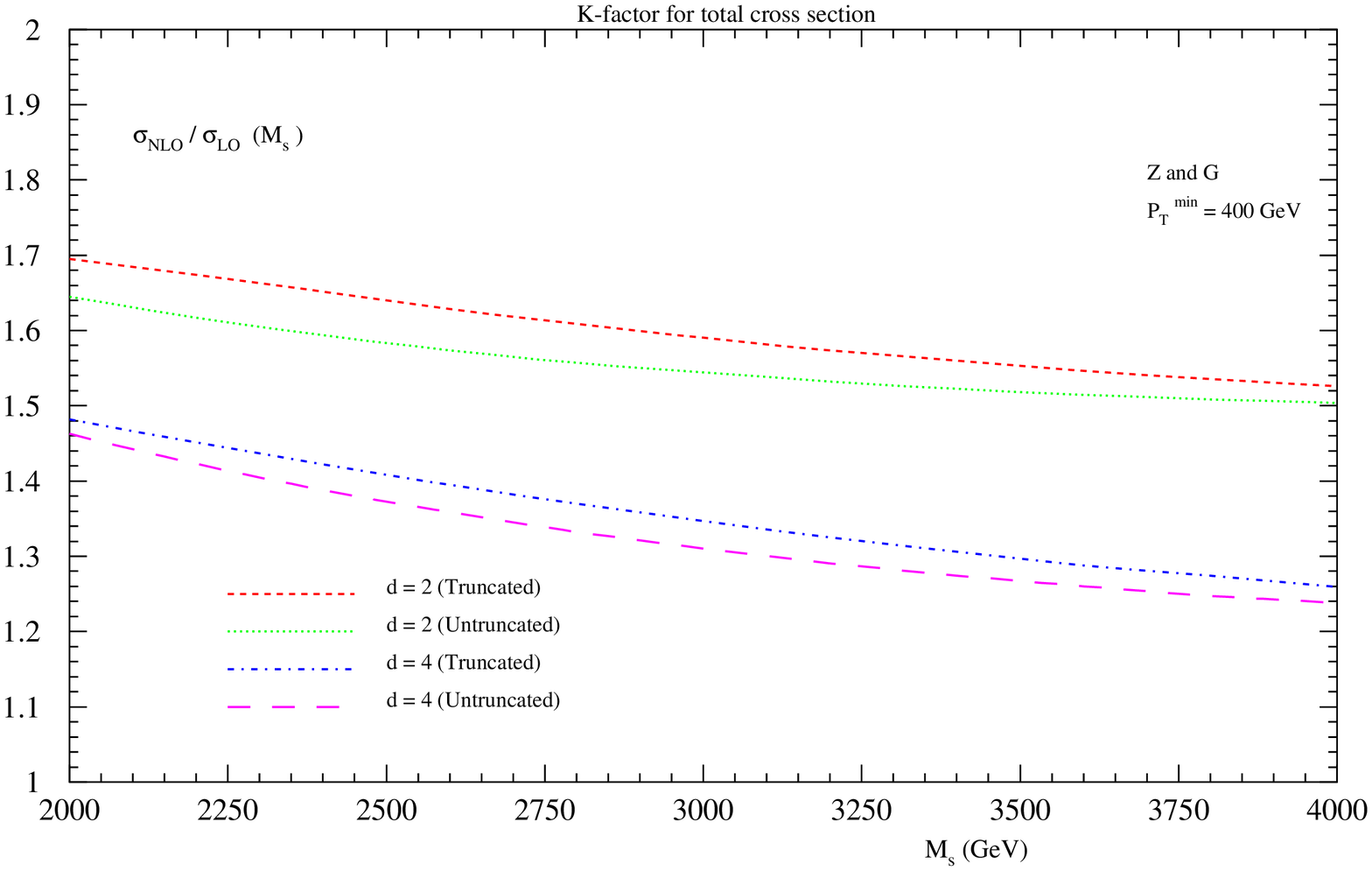,width=15cm,height=10cm,angle=0}}
\caption{K-factors of the total cross section for the associated production 
of the $Z$-boson  and the graviton at the LHC, given as a function of 
$P_T^{min}$ (top) and the scale $M_s$ (bottom).}
\label{tot-z-kf}
\end{figure}

In fig.(\ref{tot-z-d2}), the total cross section for the 
associated production of $Z$-boson and the graviton is shown as a function of 
$P_T^{min}$, to NLO in QCD at the LHC. 
The cross sections are given for both the truncated as 
well as the un-truncated cases and with the choice of model parameters 
$M_s = 3 $ TeV and $d =2$.  A similar plot is shown for $d = 4$ in 
fig.(\ref{tot-z-d4}). The K-factors are found to have a mild dependency on 
$P_T^{min}$, varying from $1.6$ to $1.4$.  
In fig.(\ref{totms-z-d2}), we have shown the 
variation of the truncated as well as un-truncated total cross sections 
with respect to the scale $M_s$, for the case $d = 2$. 
The difference between the 
truncated and the un-truncated cross sections is mainly due to the 
contributions coming from the region $Q > M_s$.  However, with 
increasing $M_s$ the  parton fluxes corresponding to $Q$ in this 
region rapidly fall down and hence the difference between the two cross 
sections decreases with increase in $M_s$.
Such a behavior is evident from the figures. (\ref{totms-z-d2}) 
and (\ref{totms-z-d4}), for $d =2$ and $d = 4$ respectively. 
The corresponding K-factors are also shown in  fig. (\ref{tot-z-kf}).  
In the rest of our calculation we choose $P_T^{min} = 400$ GeV 
and $M_s = 3$ TeV.

\begin{figure}[htb]
\centerline{
\epsfig{file=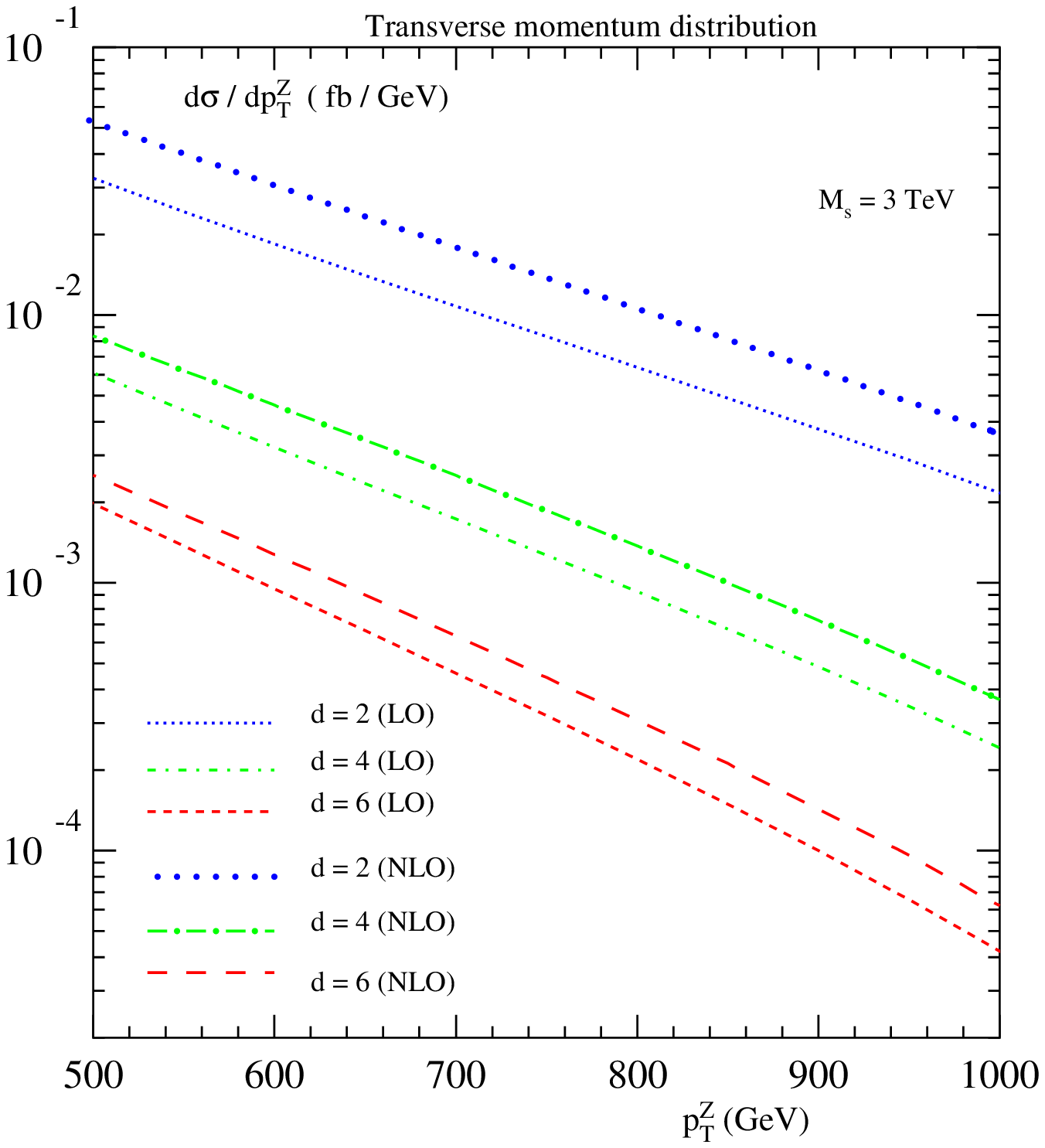,width=8cm,height=9cm,angle=0}}
\caption{Transverse momentum distribution of the $Z$-boson for $M_s = 3$ 
TeV is shown for different values of the number of extra dimensions $d$.}
\label{pt-z}
\end{figure}

\begin{figure}[htb]
\centerline{
\epsfig{file=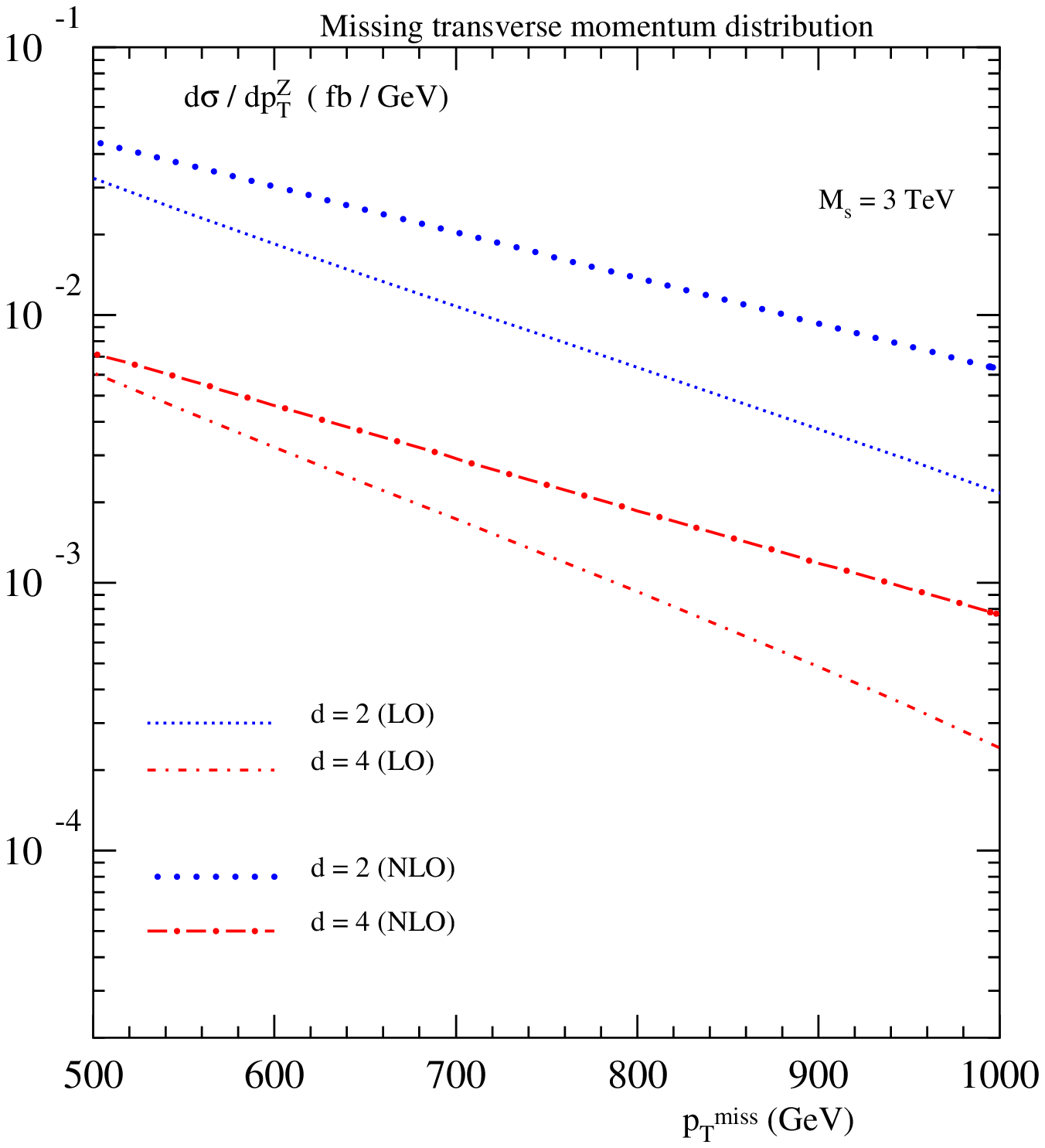,width=8cm,height=9cm,angle=0}
\epsfig{file=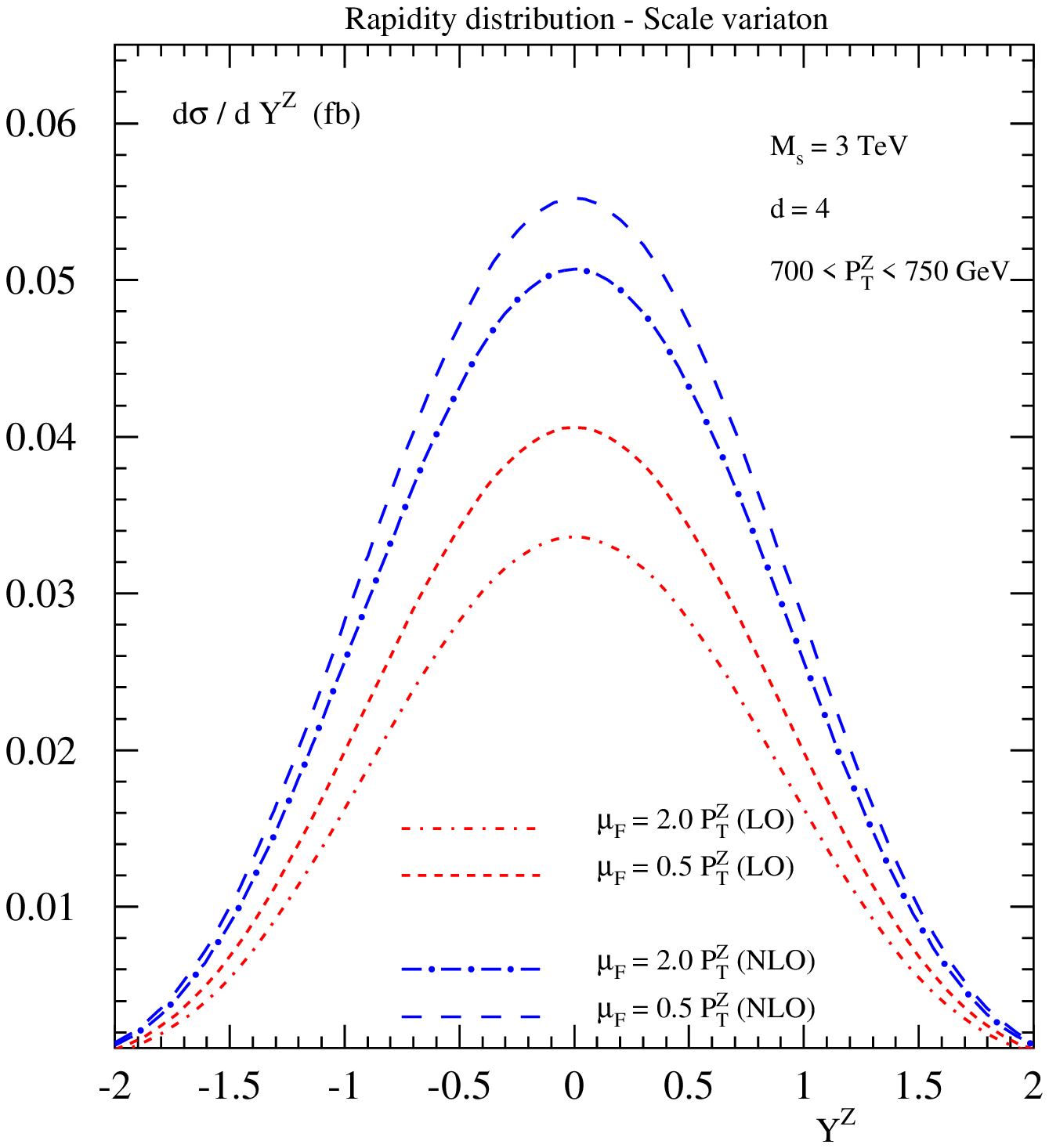,width=8cm,height=9cm,angle=0}}
\caption{Missing transverse momentum distribution of the graviton produced
in association with $Z$-boson at the LHC, for $M_s = 3$ TeV (left). The
scale uncertainties in the rapidity distribution of $Z$-boson for 
$M_s = 3$ TeV and $d=4$ (right).}
\label{ptmiss-z}
\end{figure}

Next, we present LO and NLO transverse momentum distributions 
of the $Z$-boson ($P_T^Z$) in fig.(\ref{pt-z}) for $d=2, 4, 6$ and the corresponding 
missing transverse momentum distributions ($P_T^{miss}$) in the left panel 
of fig.(\ref{ptmiss-z}) for $d=2, 4$.  
The QCD corrections enhance both $P_T^Z$ and $P_T^{miss}$ distributions.  Note
that the shape of the $P_T^Z$ distribution remains unaffected while this is
not the case for $P_T^{miss}$ distribution. 
Such a pattern can be understood from the definition of $P_T^{miss}$ 
mentioned before.  

Next, we present the rapidity distributions of $Z$ 
bosons.  The rapidity of massive gauge bosons is defined by
\begin{eqnarray}
Y = \frac{1}{2} \text{ log } \Big(\frac{E + p_z}{E - p_z} \Big),
\end{eqnarray}
where $E$ and $p_z$ are the energy and the longitudinal momentum components
of the gauge boson in the lab frame.  In the right panel of fig.(\ref{ptmiss-z}),
we have plotted the rapidity distribution of the $Z$-boson 
both at LO and at NLO for two different choices of the factorization scale: 
$\mu_F=P_T^Z/2~~ {\rm and}~~2P_T^Z$.  This distribution is obtained by  
integrating over the transverse momentum of the $Z$-boson from $700$ GeV 
to $750$ GeV, for $d=4$.  Note that the NLO corrections increase the
cross section.  As expected, the inclusion of order $a_s$ corrections reduces 
the dependence on the arbitrary factorization scale $\mu_F$. 
The percentage of uncertainty in the cross sections at the central rapidity region 
$Y=0$, due the variation of the scale from $\mu_F=P_T^Z/2$ to 
$\mu_F=2 P_T^Z$, is 18.9 at LO and it gets reduced to 8.6 at NLO.
\subsection{Charged gauge bosons}
    In this section we discuss the impact of NLO QCD corrections 
on the associated production of charged gauge bosons ($W^\pm$) 
and the graviton at the LHC.  The matrix elements for the $W^\pm$ case 
are identical to  those for the $Z$-boson case but for the masses of the gauge bosons 
and their couplings to the quarks as seen in eqn. (\ref{coupling}).  
Further, in the case of charged gauge 
bosons, the parton fluxes will also be different from those of the neutral 
gauge boson.  The parton fluxes for the quark anti-quark annihilation 
process in the case of $Z$-boson are of the form $q \bar{q}$
($q = {u, d, s, c, b}$), while they are of the form $u \bar{d} ~ (d \bar{u})$ 
for $W^+ (W^{-})$. For $W$ boson production cross sections, we will consider 
the mixing of quarks among different quark generations, as allowed by the 
CKM-matrix elements $V_{ij}$, with $(i = u, c, t)$ and $(j = d, s, b)$. 
In view of this, in the above parton fluxes, $u$ and $d$ correspond 
to any $up$-type and $down$-type quarks respectively.  The CKM matrix elements 
are given by \cite{PDG1}
\begin{eqnarray}
\nonumber
\noindent
|V_{ud}| = 0.97425 \quad |V_{us}| = 0.2252 \quad |V_{ub}| = 3.89 \times 10^{-3} 
\nonumber \\[1.2ex]
|V_{cd}| = 0.230   \quad |V_{cs}| = 1.023  \quad |V_{cb}| = 40.6 \times 10^{-3}
\end{eqnarray}
Since all our calculations are done in the massless limit of the partons, 
we have not included the top quark contribution in our calculation and 
set all $V_{tj}$'s to zero.

\begin{figure}[htb]
\centerline{
\epsfig{file=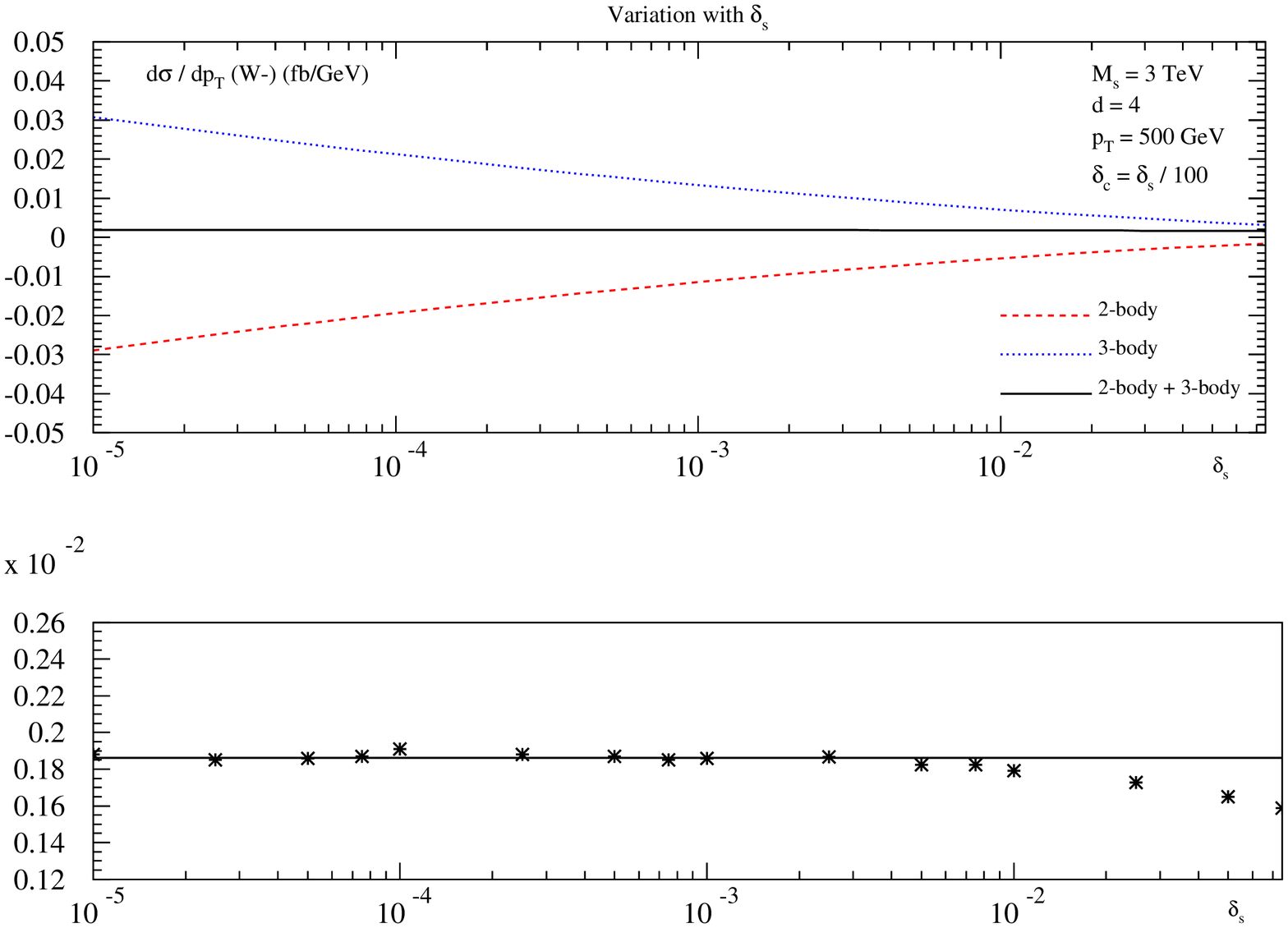,width=15cm,height=12cm,angle=0}}
\caption{Variation of the transverse momentum distribution of $W^-$ boson
with $\delta_s$, keeping the ratio $\delta_s/\delta_c=100$ 
fixed, for $M_s=3$ TeV and $d=4$.}
\label{ds-wm}
\end{figure}

\begin{figure}[htb]
\centerline{
\epsfig{file=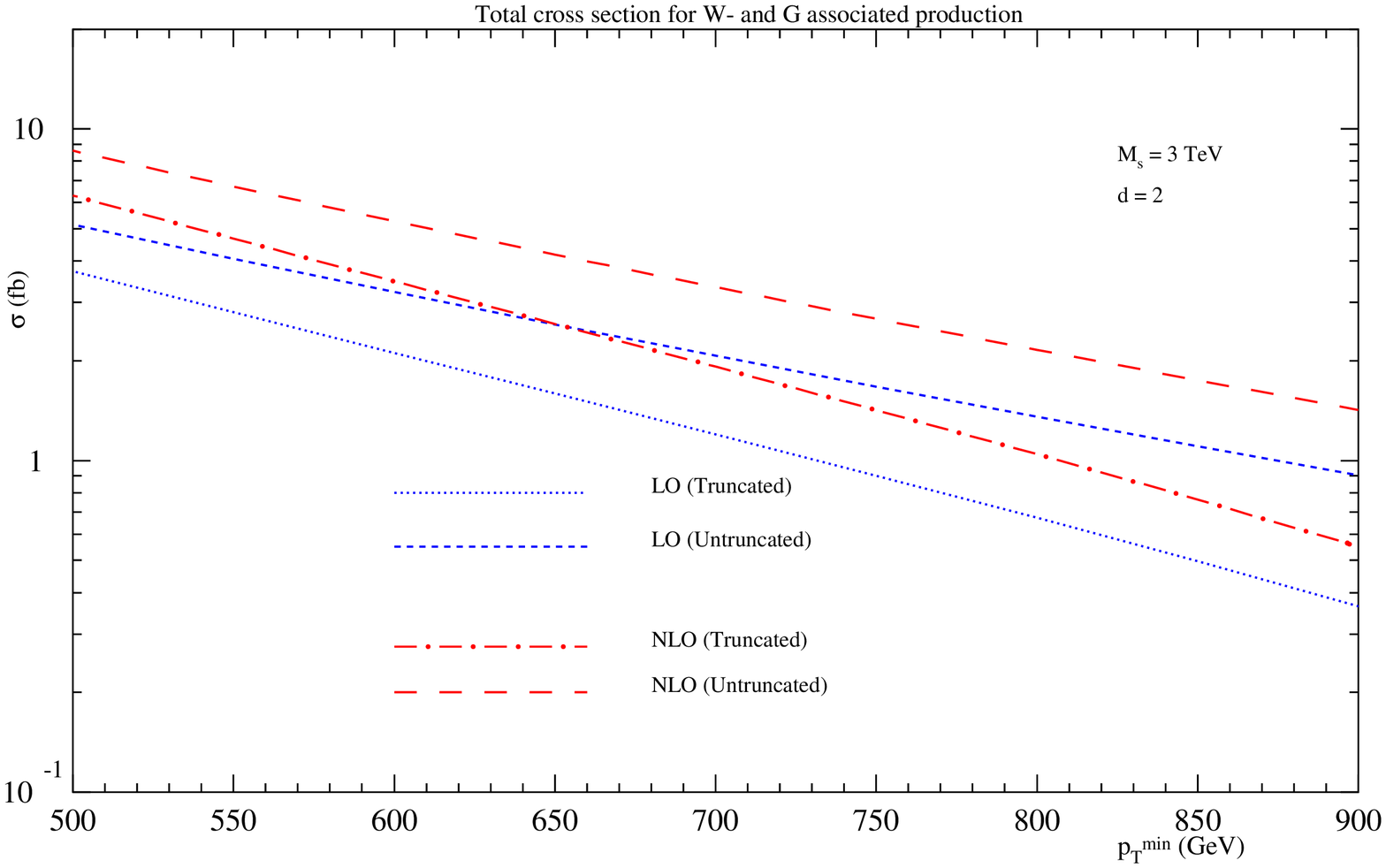,width=15cm,height=10cm,angle=0}}
\caption{Total cross section for the associated production of $W^-$ boson and 
the graviton at the LHC, as a function of $P_T^{min}$ for $M_s = 3$ TeV and $d=2$.}
\label{tot-wm-d2}
\end{figure}

\begin{figure}[htb]
\centerline{
\epsfig{file=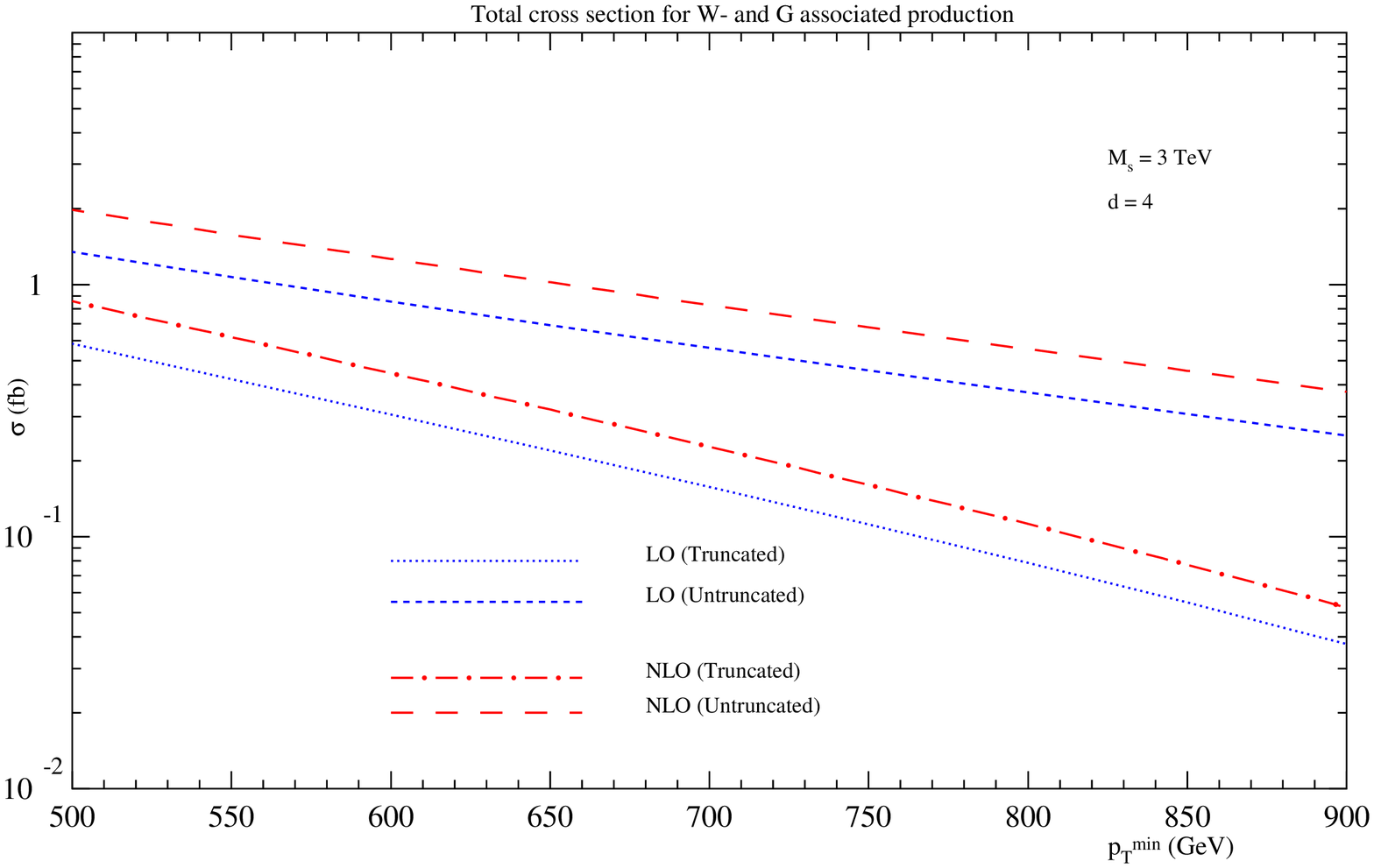,width=15cm,height=10cm,angle=0}}
\caption{Total cross section for the associated production of $W^-$ boson and 
the graviton at the LHC, shown as a function of $P_T^{min}$ for $M_s = 3$ TeV 
and $d=4$.}
\label{tot-wm-d4}
\end{figure}

\begin{figure}[htb]
\centerline{
\epsfig{file=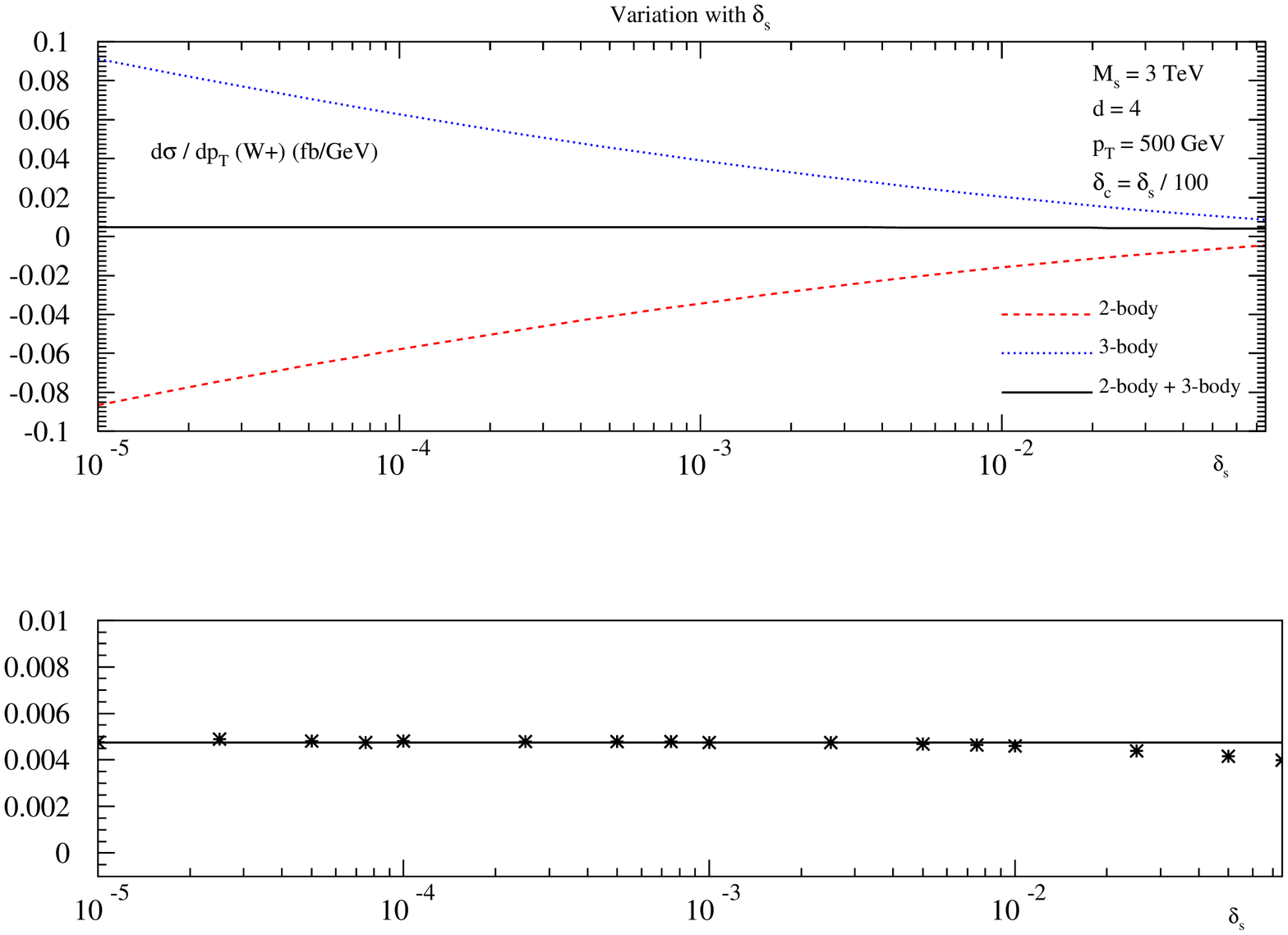,width=15cm,height=12cm,angle=0}}
\caption{Variation of the transverse momentum distribution of $W^+$ boson
with $\delta_s$, keeping the ratio $\delta_s/\delta_c=100$ 
fixed, for $M_s=3$ TeV and $d=4$.}
\label{ds-wp}
\end{figure}

\begin{figure}[htb]
\centerline{
\epsfig{file=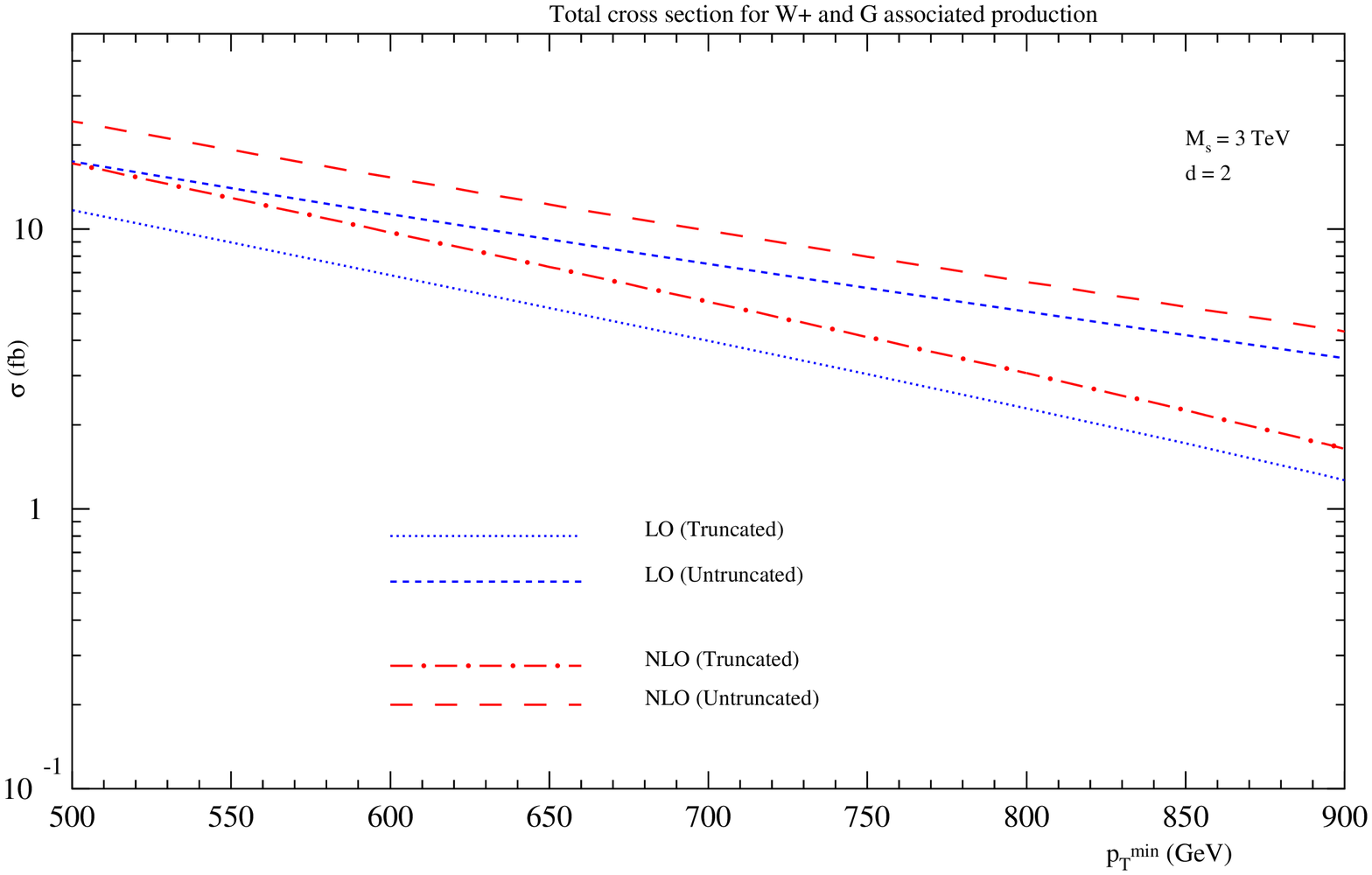,width=15cm,height=10cm,angle=0}}
\caption{Total cross section for the associated production of $W^+$ boson and 
the graviton at the LHC, shown as a function of $P_T^{min}$ for $M_s = 3$ TeV 
and $d=2$.}
\label{tot-wp-d2}
\end{figure}

\begin{figure}[htb]
\centerline{
\epsfig{file=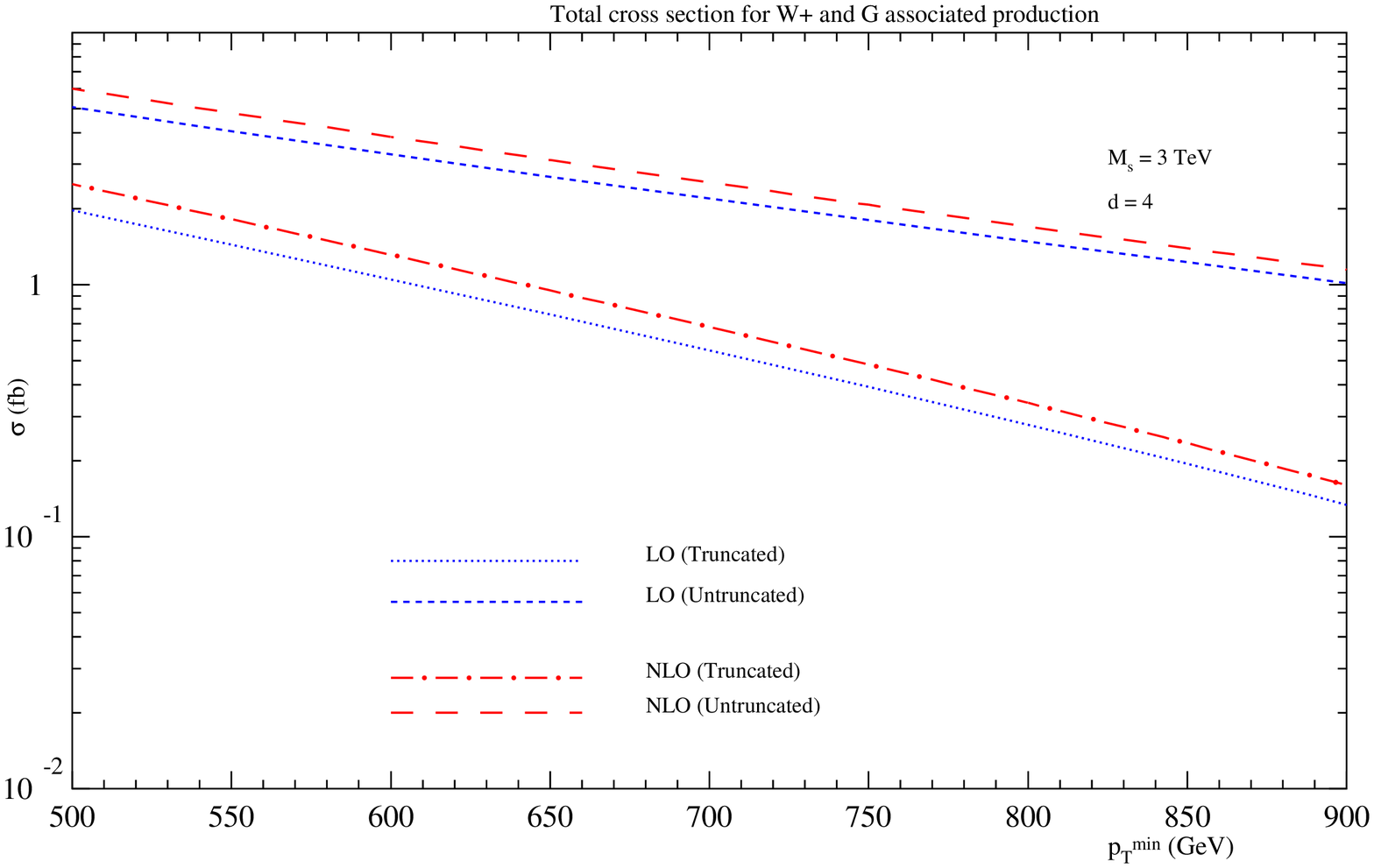,width=15cm,height=10cm,angle=0}}
\caption{Total cross section for the associated production of $W^+$ boson and 
the graviton at the LHC, shown as a function of $P_T^{min}$ for $M_s = 3$ TeV 
and $d=4$.}
\label{tot-wp-d4}
\end{figure}

    Similar to the case of $Z$ boson, we will present the total cross 
sections  as well as the differential distributions for the associated
production of $W^\pm$ boson and a graviton.   In fig.(\ref{ds-wm}) and fig.(\ref{ds-wp}), 
we have shown the stability of the 
transverse momentum distributions of $W^-$ and $W^+$ respectively, with 
the slicing parameter $\delta_s$.
These distributions are obtained for the choice of $P_T^{W} = 500$ GeV,
keeping the ratio $\delta_s/\delta_c$ fixed at 100.  It can be seen
from the figures that the sum of the $2$-body and $3$-body contributions
is fairly stable against the variation of the slicing parameters. 
This ensures the proper implementation of the slicing method in our
numerical code, taking into account the appropriate parton fluxes 
for $W^\pm$.  Next, we present the total cross sections as a 
function of $P_T^{min}$ as well as $M_s$. In fig.(\ref{tot-wm-d2})
and fig.(\ref{tot-wm-d4}), we show truncated 
as well as untruncated total cross sections for $W^-$ case,
as a function of $P_T^{min}$, for $d = 2$ and $d=4$ respectively.
It can be seen from these figures that the QCD corrections have enhanced 
the leading order cross sections considerably, but there is no significant
change in the shape of the cross sections.
Similar plots are shown for $W^+$ in  fig.(\ref{tot-wp-d2}) and fig.(\ref{tot-wp-d4}).

\begin{figure}[htb]
\centerline{
\epsfig{file=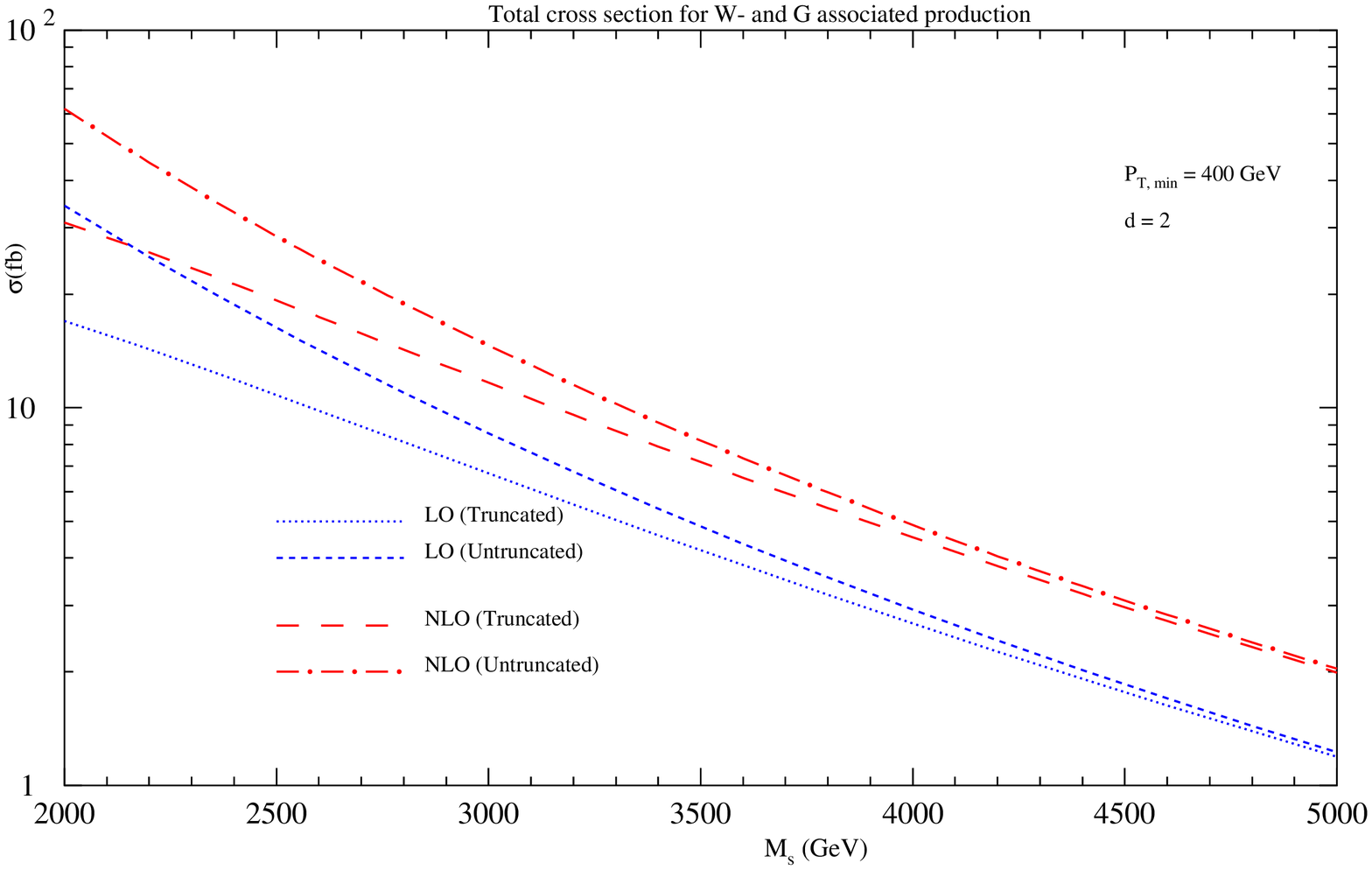,width=15cm,height=10cm,angle=0}}
\caption{Total cross section for the associated production of $W^-$ boson and 
the graviton at the LHC, given as a function of $M_s$ for $d=2$.}
\label{totms-wm-d2}
\end{figure}

\begin{figure}[htb]
\centerline{
\epsfig{file=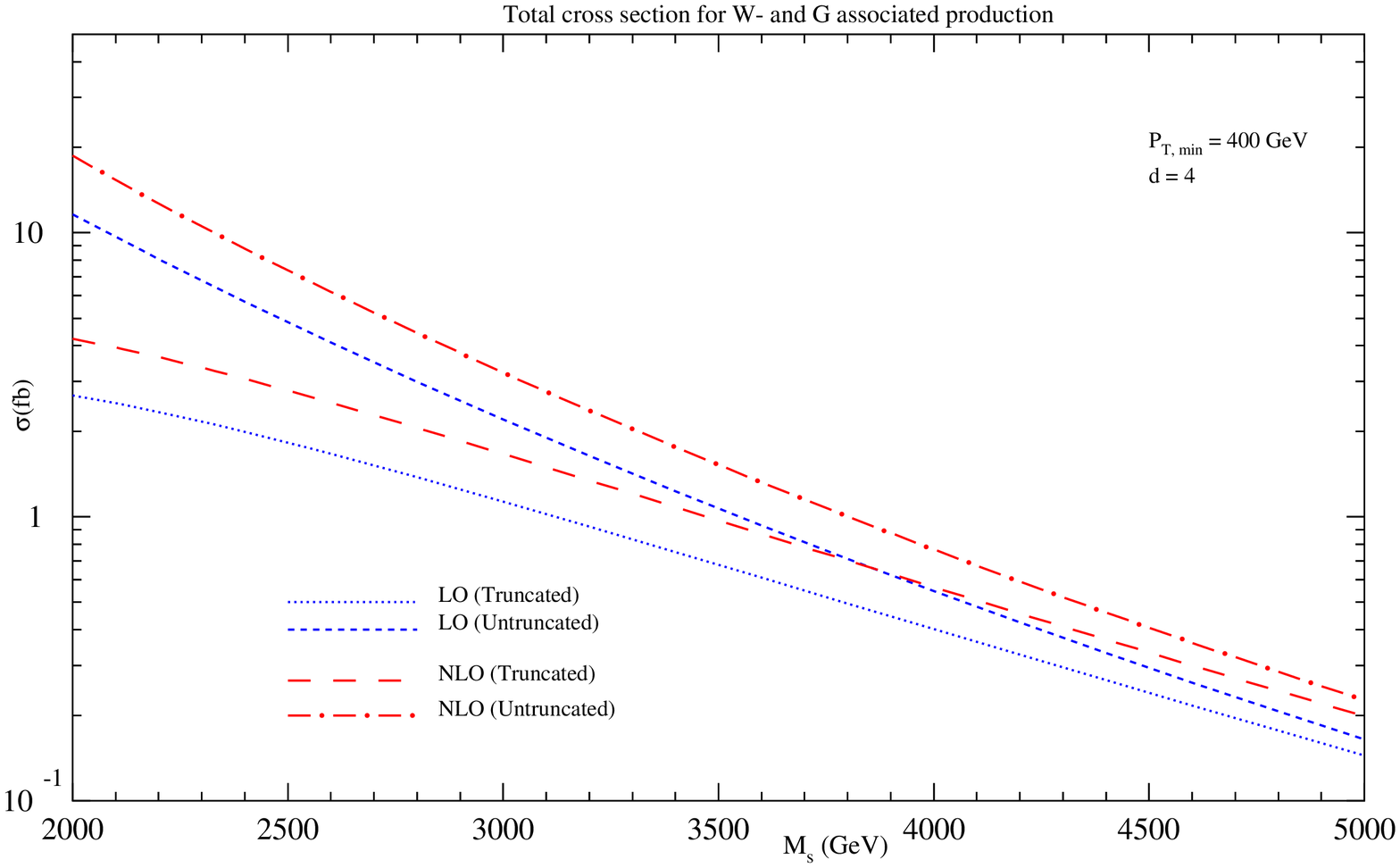,width=15cm,height=10cm,angle=0}}
\caption{Total cross section for the associated production of $W^-$ boson and 
the graviton at the LHC, shown as a function of $M_s$ for $d=4$.}
\label{totms-wm-d4}
\end{figure}

\begin{figure}[htb]
\centerline{
\epsfig{file=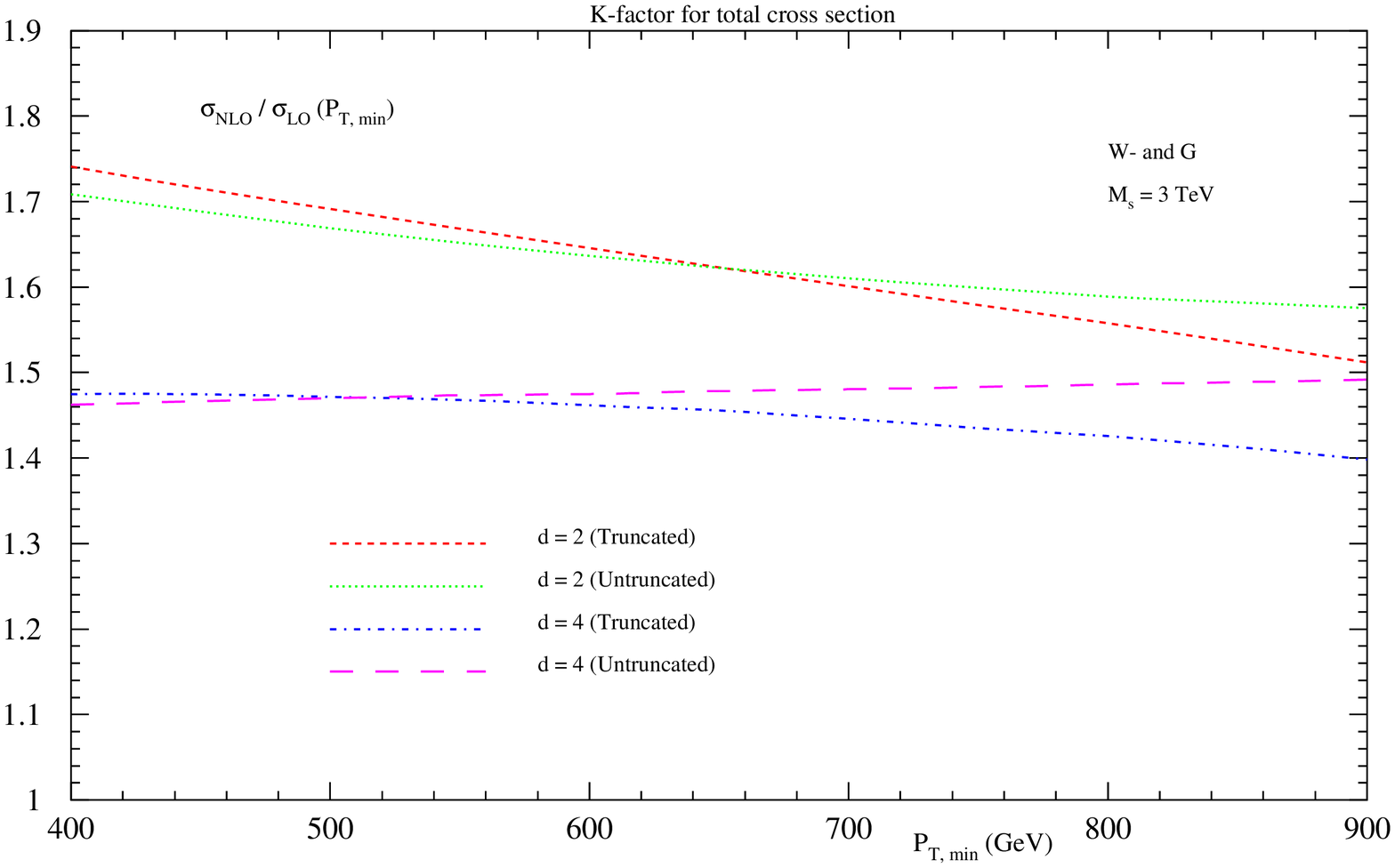,width=15cm,height=10cm,angle=0}}
\centerline{
\epsfig{file=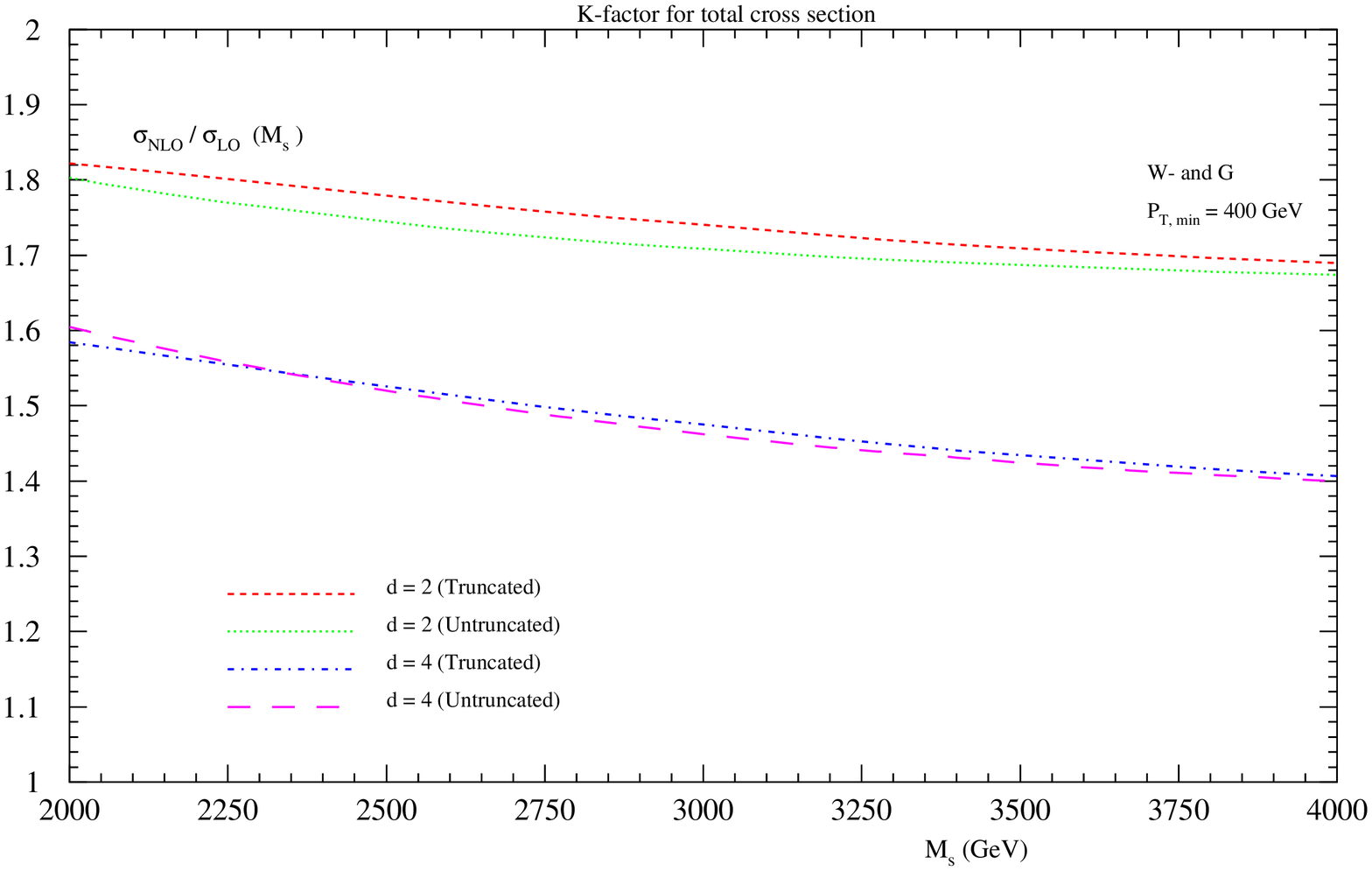,width=15cm,height=10cm,angle=0}}
\caption{K-factors of the total cross section for the associated production 
of $W^-$ boson and the graviton at the LHC, given as a function of $P_T^{min}$ 
(top) and the scale $M_s$ (bottom).}
\label{tot-wm-kf}
\end{figure}

\begin{figure}[htb]
\centerline{
\epsfig{file=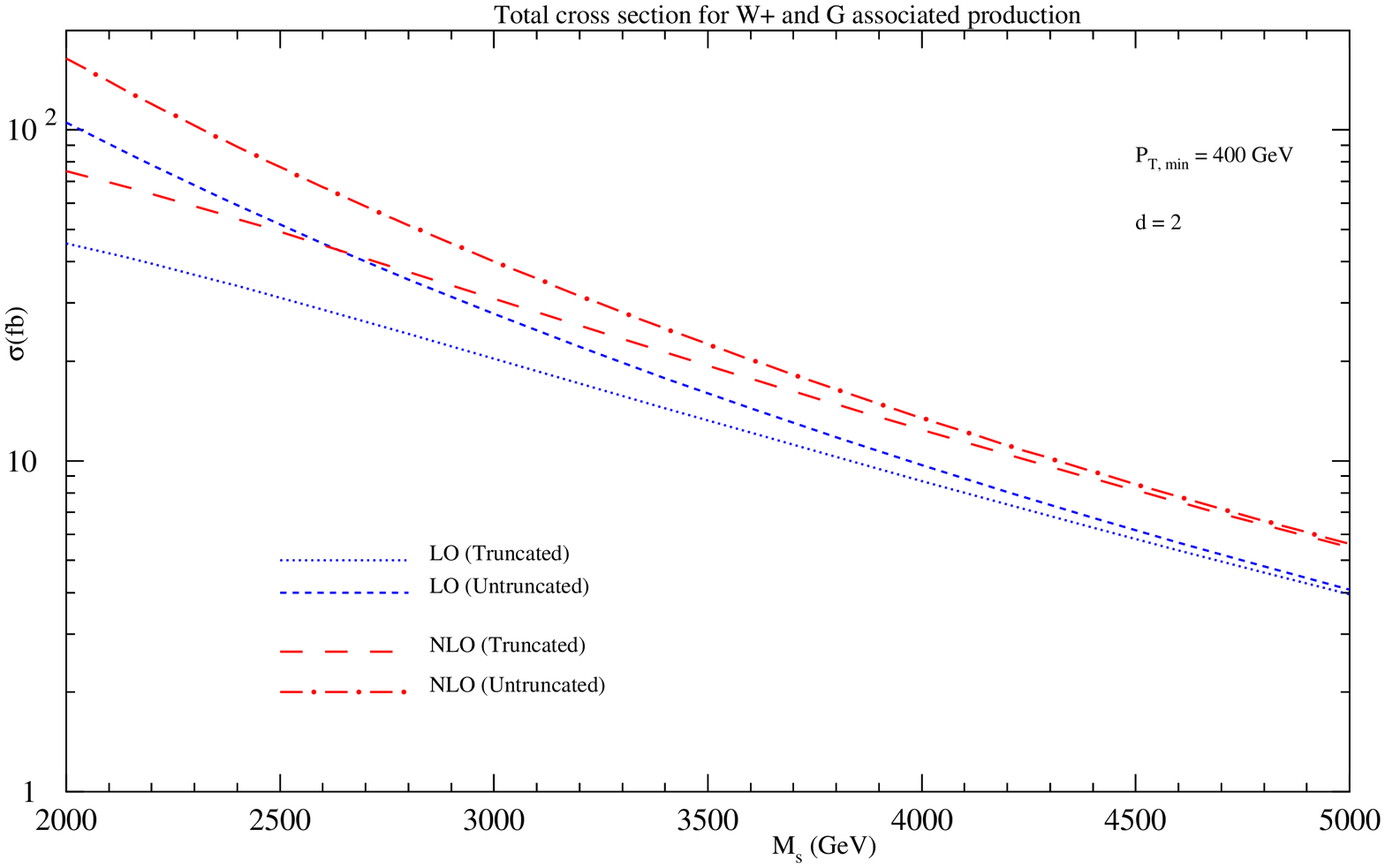,width=15cm,height=10cm,angle=0}}
\caption{Total cross section for the associated production of $W^+$ boson and 
the graviton at the LHC, shown as a function of $M_s$ for $d=2$.}
\label{totms-wp-d2}
\end{figure}

\begin{figure}[htb]
\centerline{
\epsfig{file=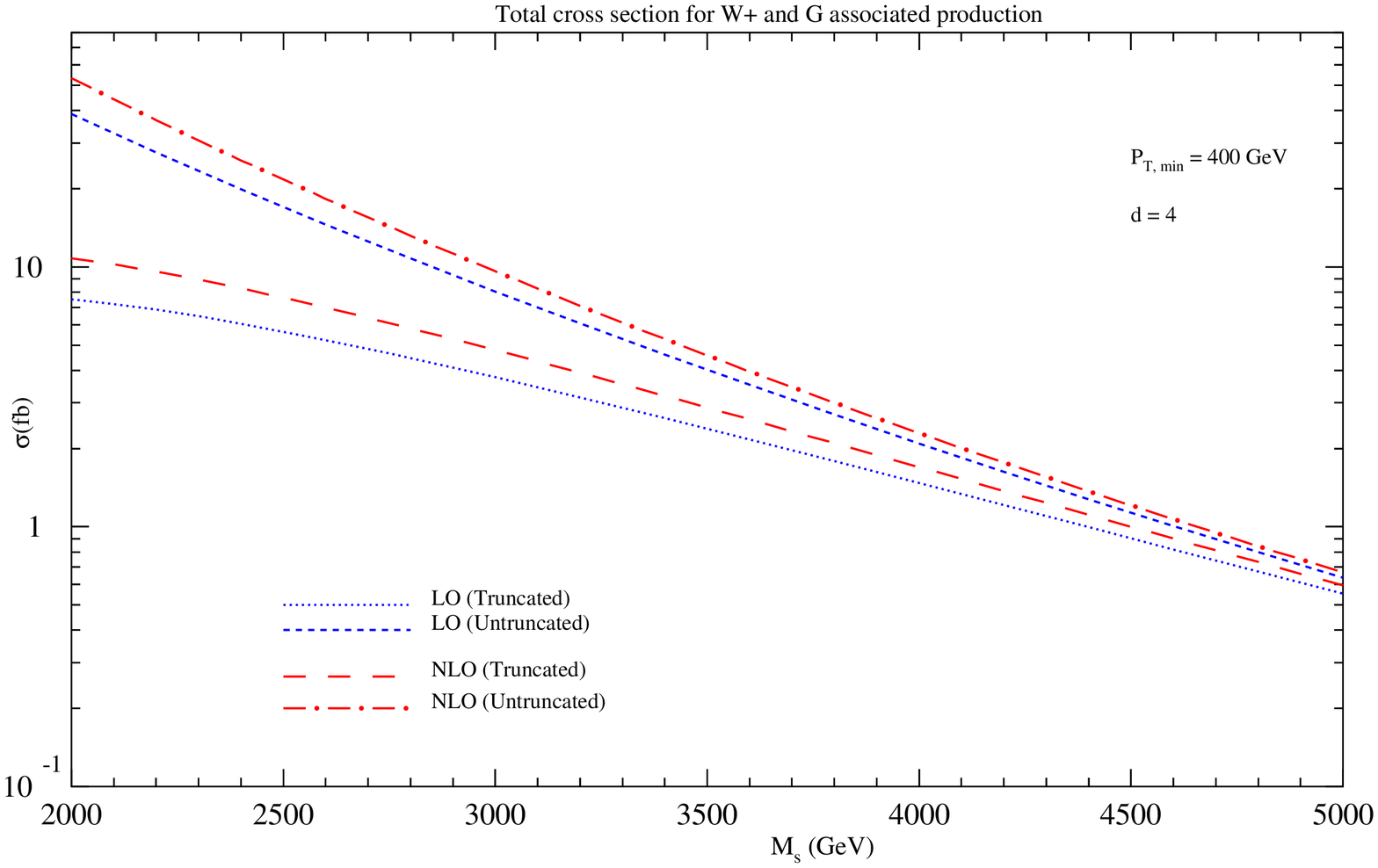,width=15cm,height=10cm,angle=0}}
\caption{Total cross section for the associated production of $W^+$ boson and 
the graviton at the LHC, shown as a function of $M_s$ for $d=4$.}
\label{totms-wp-d4}
\end{figure}

\begin{figure}[htb]
\centerline{
\epsfig{file=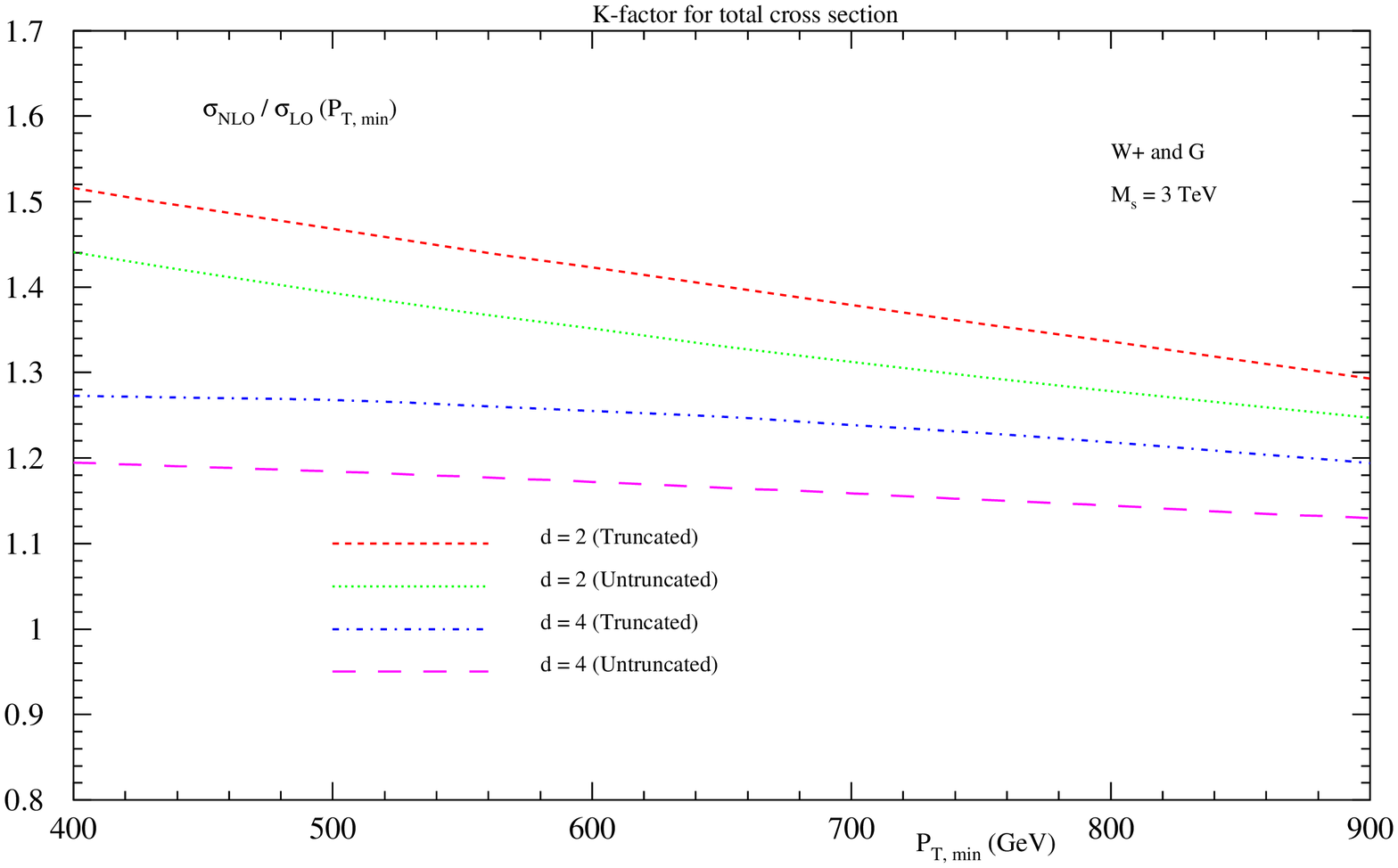,width=15cm,height=10cm,angle=0}}
\centerline{
\epsfig{file=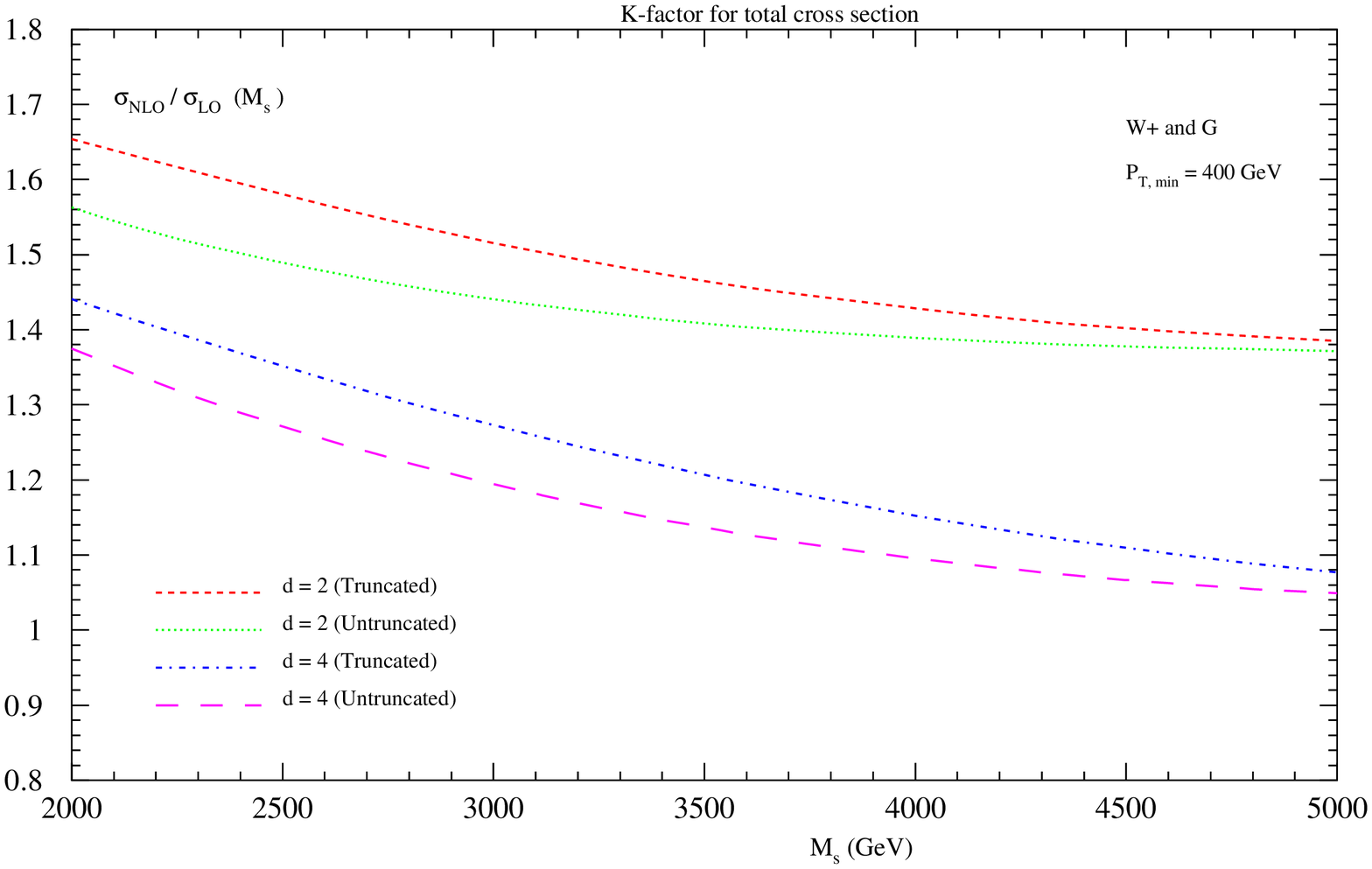,width=15cm,height=10cm,angle=0}}
\caption{K-factors of the total cross section for the associated production 
of $W^+$ boson and the graviton at the LHC, given as a function of $P_T^{min}$ 
(top) and the scale $M_s$ (bottom).}
\label{tot-wp-kf}
\end{figure}

In fig.(\ref{totms-wm-d2}) and fig.(\ref{totms-wm-d4}), we show the total 
cross sections for $W^-$ as a function of $M_s$ for $d=2$ and $d=4$, respectively. 
A set of similar plots for $W^+$ are shown in fig.(\ref{totms-wp-d2}) and 
fig.(\ref{totms-wp-d4}).  Note that, in each of the above, the cross sections for 
$W^+$ are somewhat higher than the corresponding ones for $W^-$. This difference in 
the total cross sections can be understood from the respective parton fluxes for $W^-$ 
and $W^+$ at the LHC. 
The corresponding K-factors are shown in fig.(\ref{tot-wm-kf}) for $W^-$ and in 
fig.(\ref{tot-wp-kf}) for $W^+$. For the choice of the parameters we have considered,
the K-factors are found to vary from $1.7$ to $1.4$ in the case of $W^-$ while they range
from $1.65$ to $1.05$ for $W^+$. Note that the K-factors for $W^-$ case are 
comparable but a little higher than those for $W^+$, which again can be accounted
for the differences in the parton fluxes.  The fact that the valence quark contributions 
are negligible and the parton fluxes at LO for $W^+$ are 
higher compared to those for $W^-$ explains the behavior the above factors.  

\begin{figure}[htb]
\centerline{
\epsfig{file=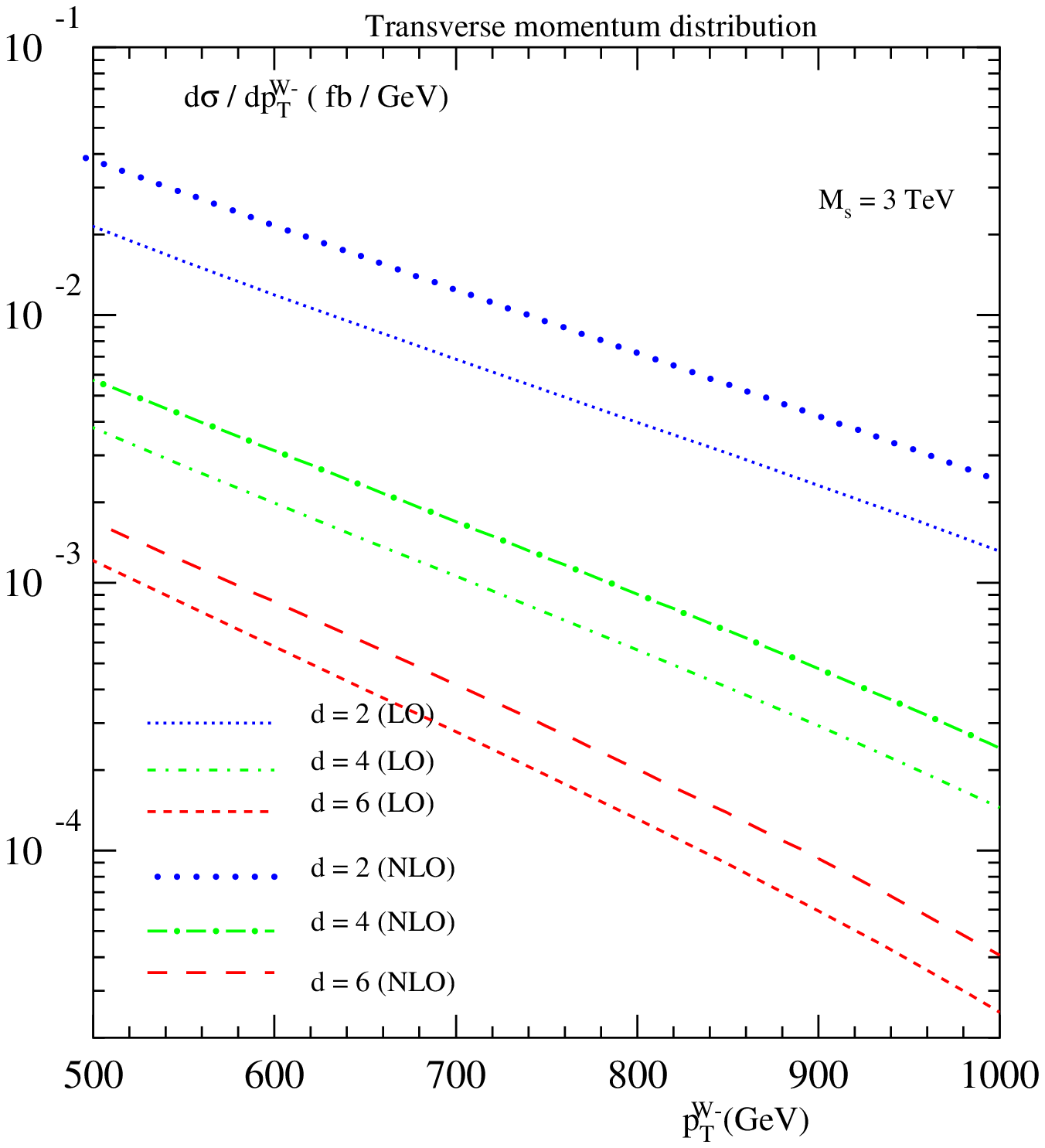,width=8cm,height=9cm,angle=0}}
\caption{Transverse momentum distribution of the $W^-$-boson for $M_s = 3$ 
TeV is shown for different values of the number of extra dimensions $d$.}
\label{pt-wm}
\end{figure}

\begin{figure}[htb]
\centerline{
\epsfig{file=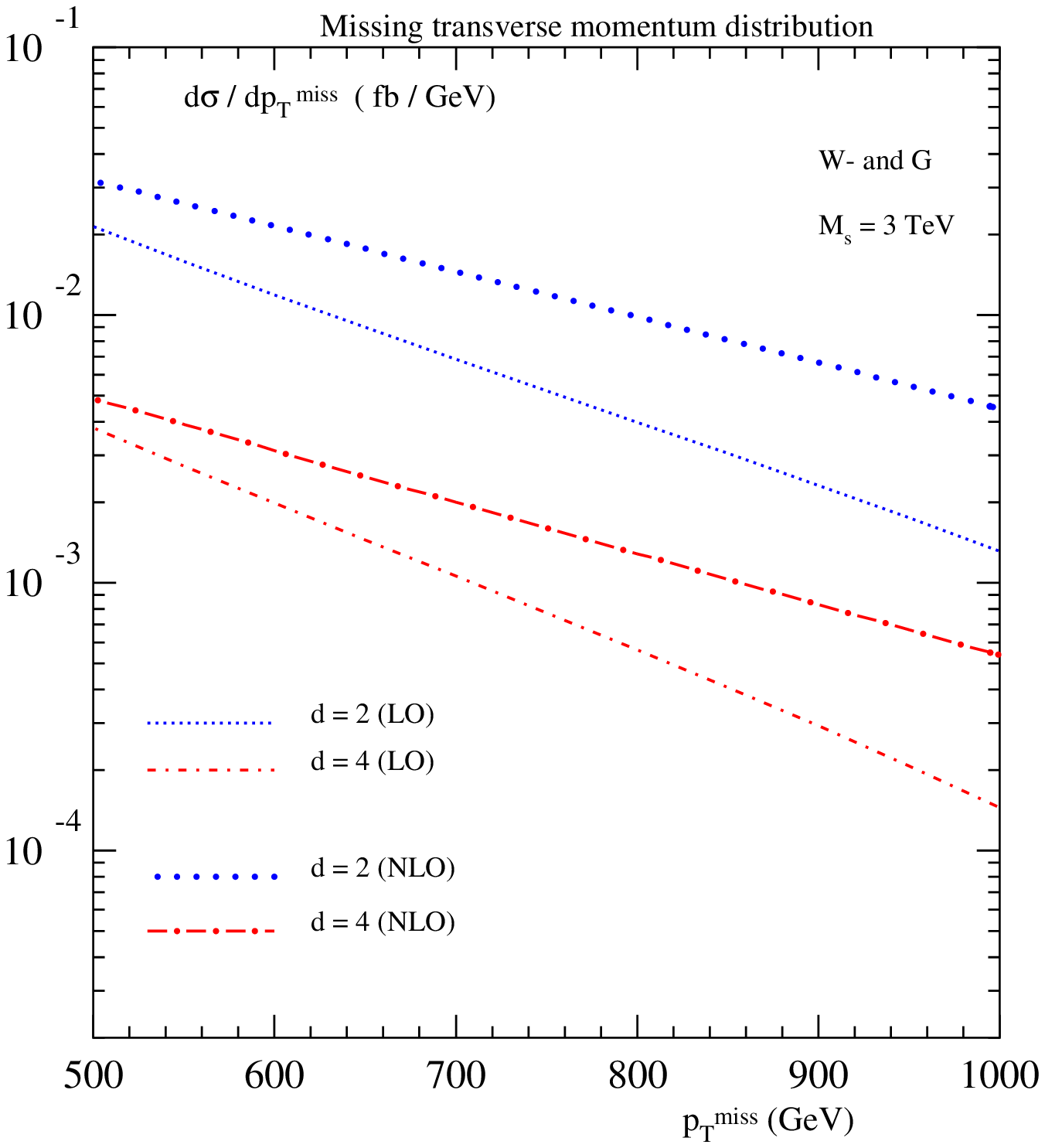,width=8cm,height=9cm,angle=0}
\epsfig{file=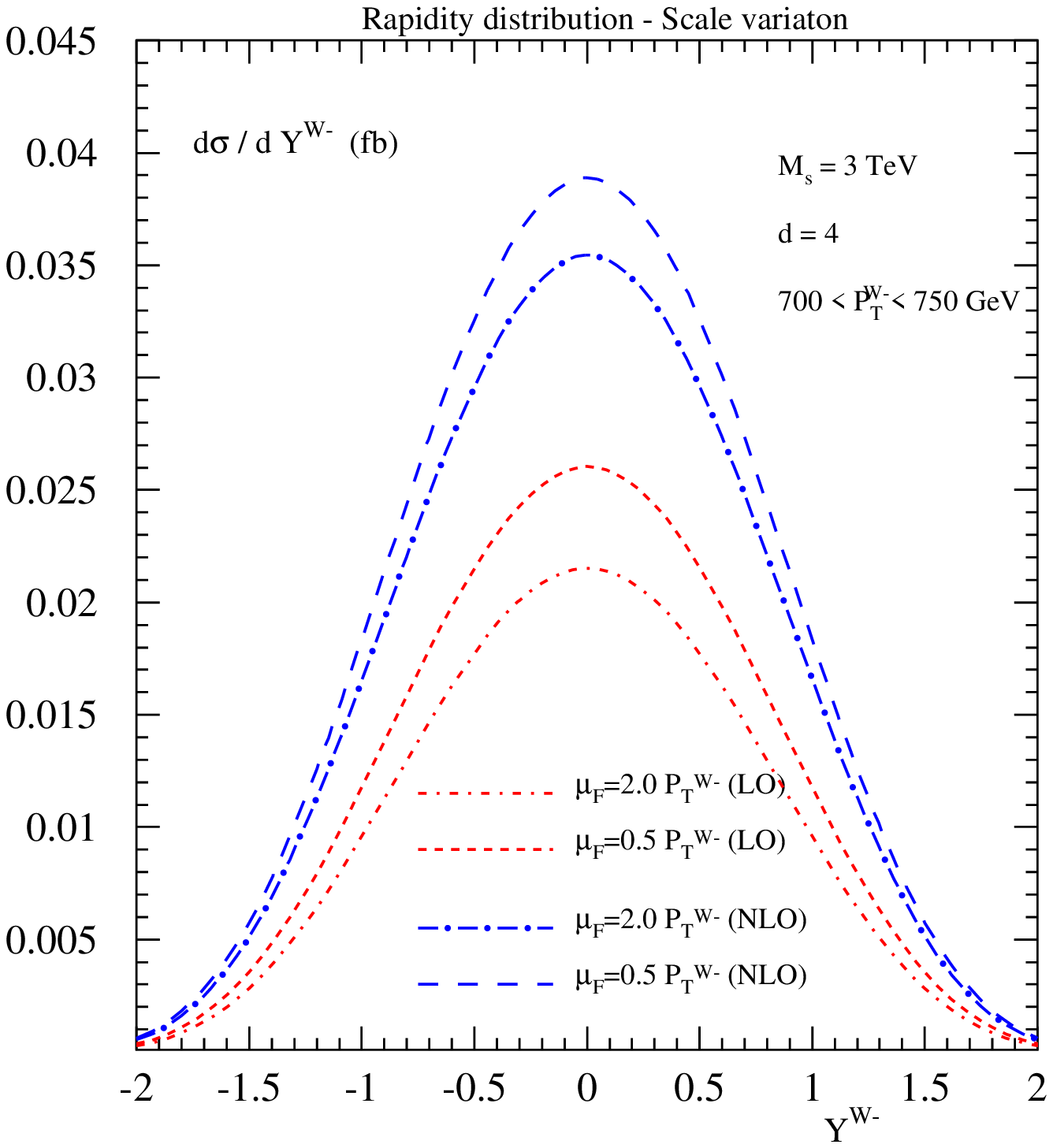,width=8cm,height=9cm,angle=0}}
\caption{Missing transverse momentum distribution of the graviton produced
in association with $W^-$-boson at the LHC, for $M_s = 3$ TeV (left). The
scale uncertainties in the rapidity distribution of $W^-$ boson for 
$M_s = 3$ TeV and $d=4$ (right).}
\label{ptmiss-wm}
\end{figure}

\begin{figure}[htb]
\centerline{
\epsfig{file=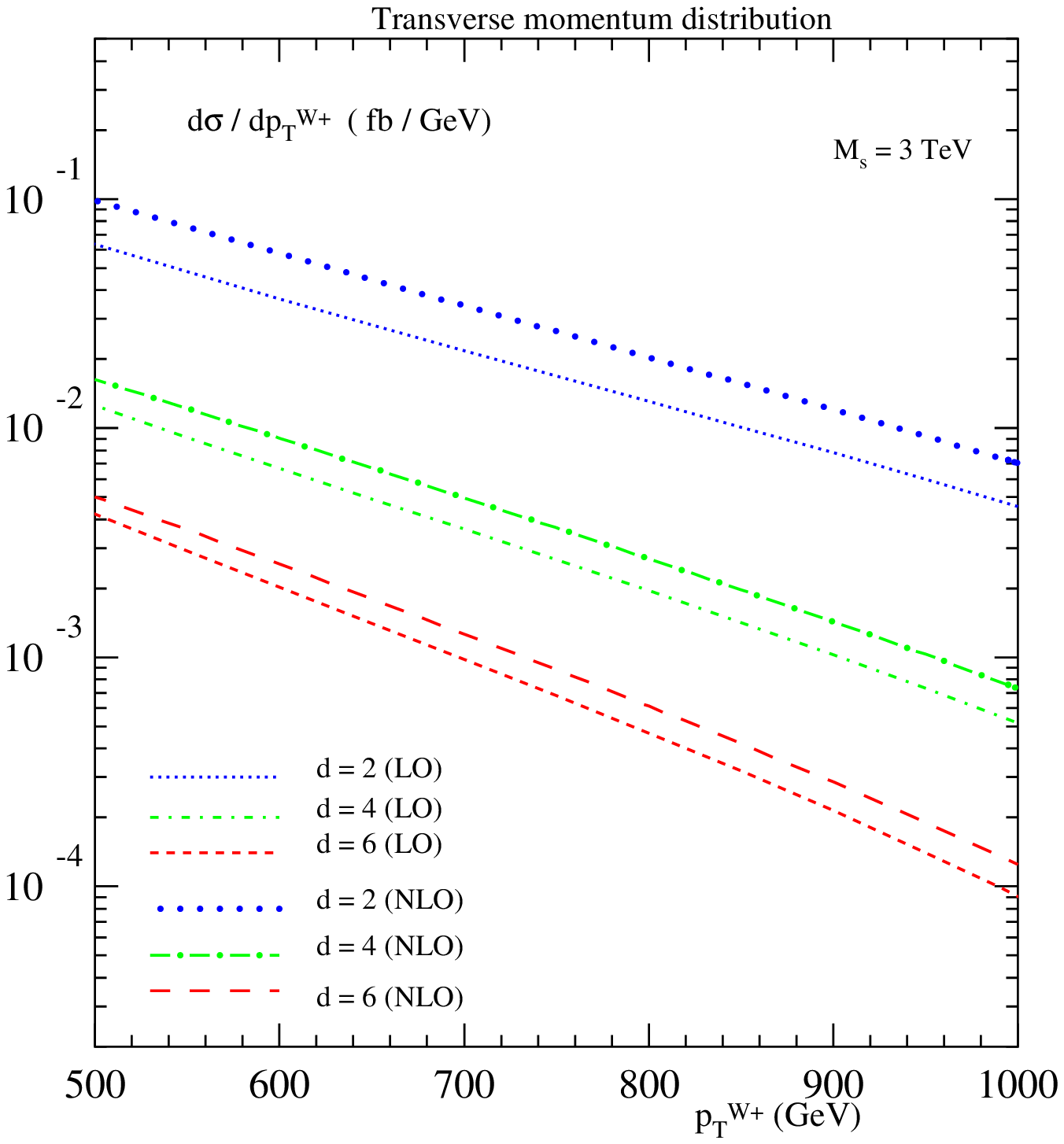,width=8cm,height=9cm,angle=0}}
\caption{Transverse momentum distribution of the $W^+$-boson for $M_s = 3$ 
TeV is shown for different values of the number of extra dimensions $d$.}
\label{pt-wp}
\end{figure}

\begin{figure}[htb]
\centerline{
\epsfig{file=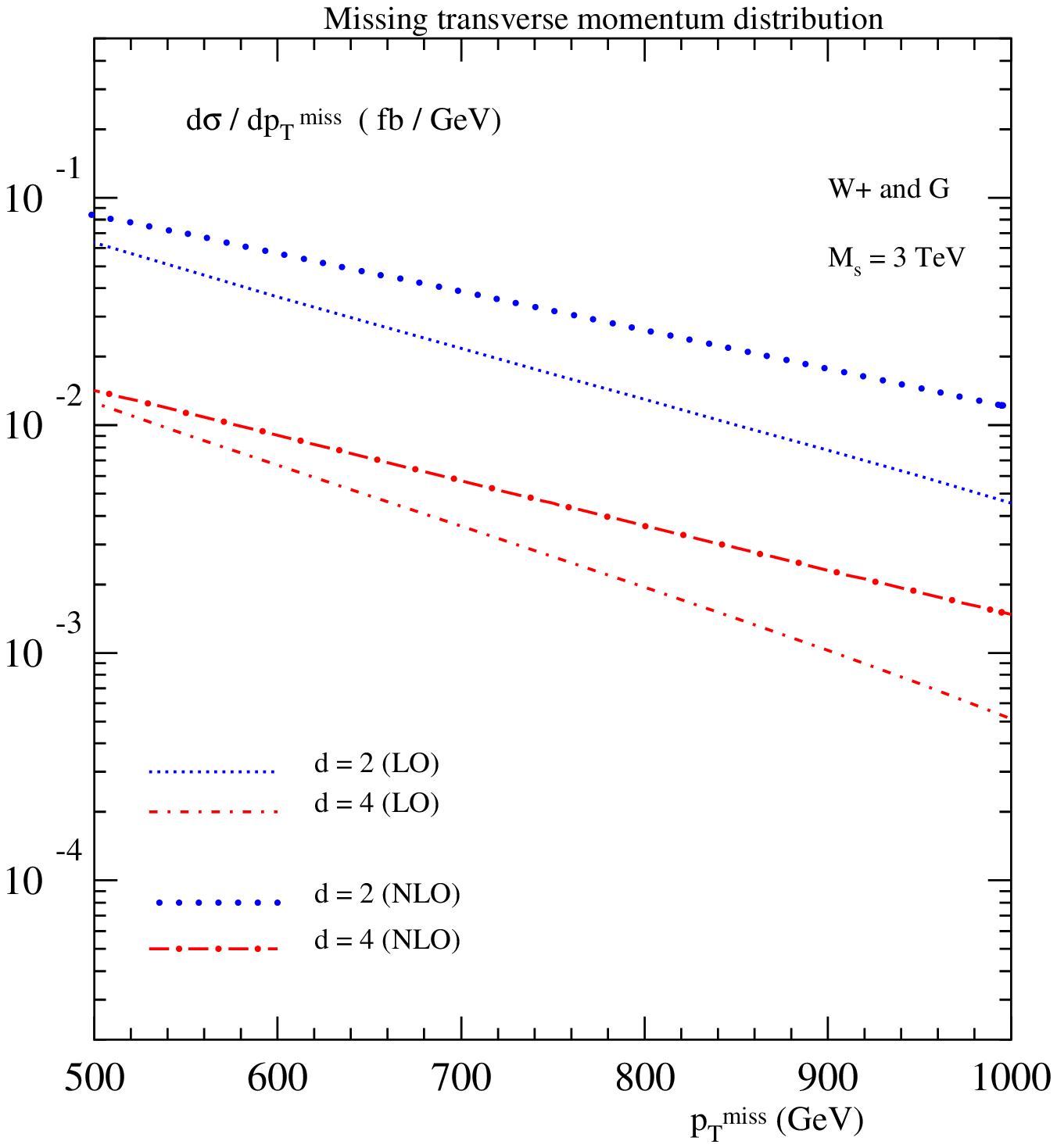,width=8cm,height=9cm,angle=0}
\epsfig{file=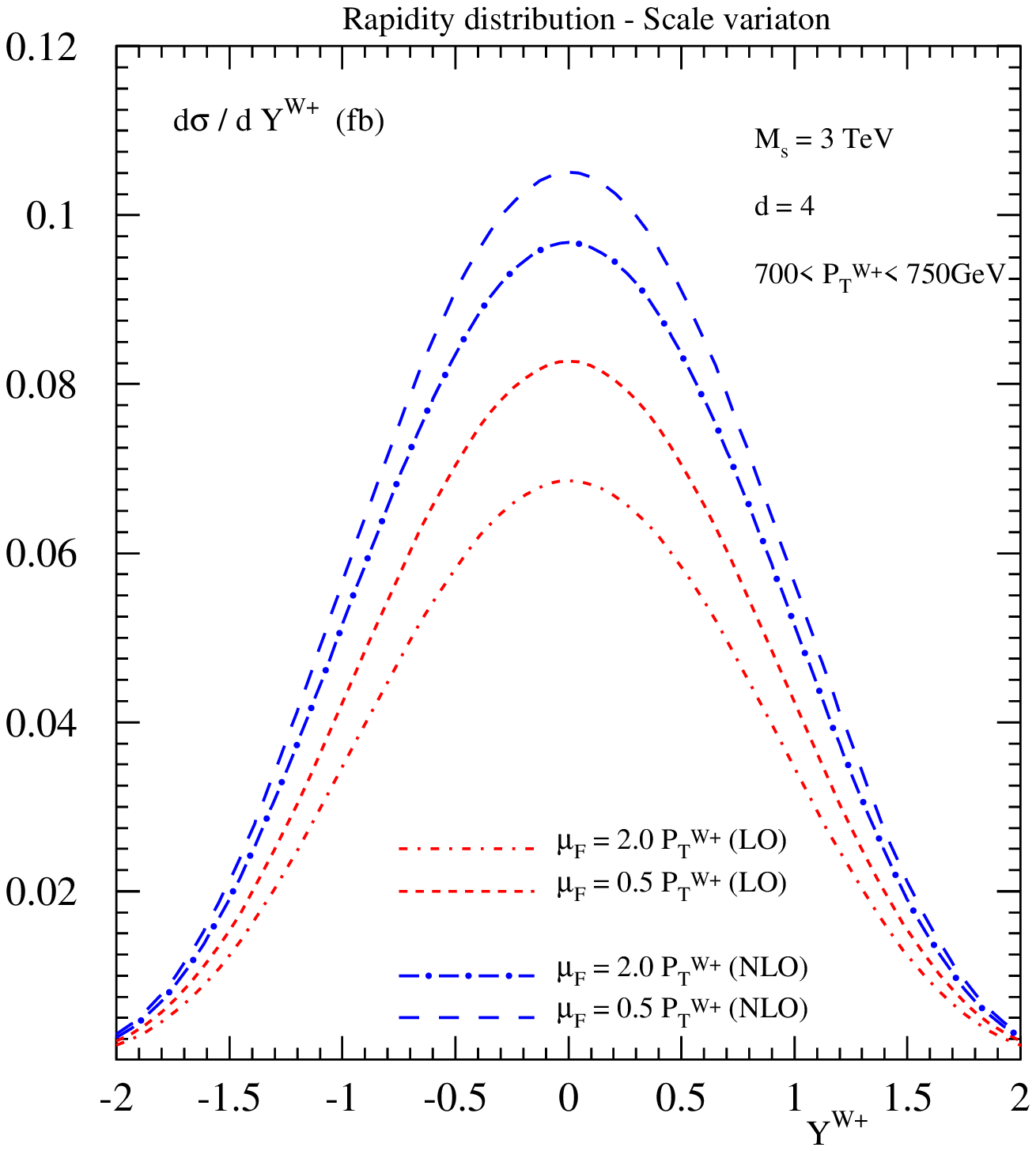,width=8cm,height=9cm,angle=0}}
\caption{Missing transverse momentum distribution of the graviton produced
in association with $W^+$-boson at the LHC, for $M_s = 3$ TeV (left). The
scale uncertainties in the rapidity distribution of $W+$-boson for 
$M_s = 3$ TeV and $d=4$ (right).}
\label{ptmiss-wp}
\end{figure}

Further, in figs.(\ref{pt-wm}) and (\ref{pt-wp}), we present the transverse momentum 
distribution of $W^-$ and $W^+$ respectively as a function of the number of extra 
dimensions $d$ and for $M_s = 3$ TeV.  Similarly, we show the missing transverse momentum 
distribution the graviton when produced in association with $W^-$ in the left panel
of fig.(\ref{ptmiss-wm}).  In the right panel, we present the scale uncertainties in
the rapidity distribution of $W^-$ by varying the factorization scale from 
$\mu_F = P_T^{W^-}/2$ to $\mu_F = 2P_T^{W^-}$. This rapidity distribution is obtained 
by integrating over the transverse momentum $P_T^{W^-}$ from $700$ GeV to $750$ GeV.
Similar plots are shown for $W^+$ in fig.(\ref{ptmiss-wp}). 
Note that the uncertainty resulting from the factorization scale $\mu_F$ get reduced
as we include order $a_s$ corrections.  The percentage of 
uncertainty at the central rapidity $Y^{W^\pm} = 0$ is decreased from $19.1$ to $9.3$ 
in the case of $W^-$, whereas it gets reduced from $18.8$ to $8.3$ in the case of $W^+$.

\section{Conclusions}
In this paper, we have systematically computed the full NLO QCD corrections 
to the associated production of the vector gauge boson and the graviton in theories 
with large extra dimensions at the LHC.  This process plays an 
important role in probing the extra dimensions at the collider experiments,
thanks to the large parton fluxes available at the LHC. 
We have used a semi-analytical two cut-off phase space slicing method
to compute these corrections.   We have quantified 
the ultraviolet sensitivity of the theoretical predictions by studying 
the cross sections in the truncated as well as the untruncated cases.
In both the cases, the radiative corrections are found to have enhanced 
the cross sections significantly but do not appreciably change their 
shapes. The K-factors for the neutral gauge boson are found to vary from 
$1.6$ to $1.2$ depending on the number of extra dimensions $d$, while they 
vary from $1.8$ to $1.3$ for the case of charged gauge bosons. 
Although, the choice of the model parameters has the potential to change 
the cross sections calculated in truncated or untruncated cases significantly, 
we notice that the K-factors remain almost the same in these two cases. 
In addition to the total cross sections, we have also 
studied the differential distributions of the vector gauge bosons and 
found that the radiative corrections are significant
and they do not affect their shapes except for the missing transverse 
momentum distribution.  At the hadron colliders, 
the leading order predictions often suffer from large uncertainties resulting from
the choice of factorization scale.   Reducing these uncertainties 
is one of the main motivations for doing NLO computation.  We have shown that 
this is indeed the case for the rapidity distributions of the gauge bosons 
by varying the factorization scale from $\mu_F = P_T/2$ to $\mu_F = 2P_T$, 
leading to reduction in the percentage of scale uncertainty to $9$\% from $19$\%.  
Hence, the results presented in this paper are more suitable for 
studies on associated production of vector boson and graviton 
in the context of extra dimension searches at the hadron colliders.
\section*{Acknowledgments}
The work of V.R. and M.C.K. has been partially supported by funds made 
available to the Regional Centre for Accelerator based Particle Physics 
(RECAPP) by the Department of Atomic Energy, Govt. of India.  We would 
like to thank the cluster computing facility at Harish-Chandra Research 
Institute where part of computational work for this study was carried 
out.  S.S. would like to thank UGC, New Delhi for financial support. 
S.S. would also like to thank RECAPP center for his visit, where part of
the work was done.


%


\mysection*{Appendix A}
\setcounter{section}{0}
\section{Finite part of the virtual contribution}

All the $V_i$'s appearing in eqn. (\ref{sigmaV}) are given below:
\begin{eqnarray}
V_1&=&  {1 \over (t^2 u (m_z^2 - s))} 
       ( (-2 m^8 t + 
            2 m^6 (m_Z^2 (-6 t + u) + t (7 t + 2 u)) 
       - m^4 (18 m_Z^4 t 
\nonumber\\[1.2ex]
&&        - 2 m_Z^2 (24 t^2 + 6 t u - u^2) 
           + 3 t (10 t^2 + 8 t u + u^2))  
          + m^2 t (-8 m_Z^6 + 26 t^3 
          + 36 t^2 u 
\nonumber\\[1.2ex]
&&           + 15 t u^2 + u^3 + 2 m_Z^4 (21 t + 5 u)  
              - 2 m_Z^2 (30 t^2 + 23 t u + u^2)) 
           - 4 t^2 (-2 m_Z^6 + 2 t^3 
\nonumber\\[1.2ex]
&&           + 4 t^2 u + 3 t u^2 + u^3 
            + m_Z^4 (6 t + 4 u) - m_Z^2 (6 t^2 + 8 t u + 3 u^2)))
        )
\nonumber \\[1.2ex]
V_2&=&       {1 \over (t^2 u (m_Z^2 -s))} 
            (16  (2 m^8 t + 
            2 m^6 (m_Z^2 (6 t - u) - t (7 t + 2 u)) 
          + m^4 (18 m_Z^4 t 
\nonumber\\[1.2ex]
&&       - 2 m_Z^2 (24 t^2 + 6 t u - u^2) + 
               3 t (10 t^2 + 8 t u + u^2)) - 
            m^2 t (-8 m_Z^6 + 26 t^3 + 36 t^2 u 
\nonumber\\[1.2ex]
&&           + 15 t u^2 + u^3 + 2 m_Z^4 (21 t + 5 u)  
              - 2 m_Z^2 (30 t^2 + 23 t u + u^2)) + 
            4 t^2 (-2 m_Z^6 + 2 t^3 
\nonumber\\[1.2ex]
&&           + 4 t^2 u + 3 t u^2 + u^3  
               +m_Z^4 (6 t + 4 u) - m_Z^2 (6 t^2 + 8 t u + 3 u^2))))
\nonumber\\
V_3&=&V_2|_{t \leftrightarrow u}
\nonumber\\
\nonumber\\
V_4&=&V_1|_{t \leftrightarrow u}
\nonumber\\
\nonumber\\[0.5ex]
V_5&=& {1 \over (2 t^2 u^2 (-m^2 + t + u))}
            ((4 m^8 t u - 
            2 m^6 (9 t u (t + u) + m_Z^2 (t^2 - 12 t u + u^2)) 
\nonumber\\[1.2ex]
&&            + m^4 (36 m_Z^4 t u + 3 t u (11 t^2 + 16 t u + 11 u^2) + 
               2 m_Z^2 (t^3 - 30 t^2 u - 30 t u^2 + u^3)) 
\nonumber\\[1.2ex]
&&          + 4 t u (-2 m_Z^6 (t + u) + 
               (t + u)^2 (2 t^2 + t u + 2 u^2) + 
               m_Z^4 (6 t^2 + 8 t u + 6 u^2) 
\nonumber\\[1.2ex]
&&            - m_Z^2 (6 t^3 + 11 t^2 u + 11 t u^2 + 6 u^3)) - 
            m^2 t u (-16 m_Z^6 + 52 m_Z^4 (t + u) 
\nonumber\\[1.2ex]
&&           - 2 m_Z^2 (31 t^2 + 46 t u + 31 u^2) + 
               3 (9 t^3 + 17 t^2 u + 17 t u^2 + 9 u^3))))
\nonumber\\[1.2ex]
V_6&=&         {1 \over (2 t^2 u^2 (-m^2 + t + u))} 
       ((4 m^8 t u - 
            2 m^6 (9 t u (t + u) + m_Z^2 (t^2 - 12 t u + u^2)) 
\nonumber\\[1.2ex]
&&          + m^4 (36 m_Z^4 t u + 3 t u (11 t^2 + 16 t u + 11 u^2) + 
               2 m_Z^2 (t^3 - 30 t^2 u - 30 t u^2 + u^3)) 
\nonumber\\[1.2ex]
&&            + 4 t u (-2 m_Z^6 (t + u) + 
               (t + u)^2 (2 t^2 + t u + 2 u^2) + 
               m_Z^4 (6 t^2 + 8 t u + 6 u^2) 
\nonumber\\[1.2ex]
&&           - m_Z^2 (6 t^3 + 11 t^2 u + 11 t u^2 + 6 u^3)) - 
            m^2 t u (-16 m_Z^6 + 52 m_Z^4 (t + u) 
\nonumber\\[1.2ex]
&&           - 2 m_Z^2 (31 t^2 + 46 t u + 31 u^2) + 
               3 (9 t^3 + 17 t^2 u + 17 t u^2 + 9 u^3))))
\nonumber\\[0.5ex]
V_7&=&V_6
\nonumber\\
V_8&=&{1 \over ((m^2 - t)^2 (m_Z^2 - t)^2 t^2 (m^2 - t - u) u)}
         ( (m^{12} m_Z^2 t (-3 m_Z^2 + 2 t) 
\nonumber\\[1.2ex]
&&         - 3 m^{10} m_Z^2 (m_Z^4 (6 t - u) + t^2 (6 t + u) - 
               m_Z^2 t (13 t + u)) + 
            m^8 (-27 m_Z^8 t 
\nonumber\\[1.2ex]
&&            + 12 t^4 u + 
               3 m_Z^6 (38 t^2 + 5 t u - u^2) 
              + m_Z^2 t^2 (48 t^2 + 11 t u + 3 u^2) 
\nonumber\\[1.2ex]
&&              - m_Z^4 t (138 t^2 + 50 t u + 3 u^2)) - 
            4 t^5 u (4 m_Z^6 - 2 m_Z^4 (5 t + 4 u) 
\nonumber\\[1.2ex]
&&             - t (2 t^2 + 5 t u + 3 u^2) + 
               m_Z^2 (8 t^2 + 13 t u + 4 u^2)) + 
            m^4 t^2 (12 m_Z^{10} 
\nonumber\\[1.2ex]
&&            - 2 m_Z^8 (39 t + 7 u) + 
               m_Z^6 (160 t^2 + 61 t u + 5 u^2) + 
               t^2 u (56 t^2 + 69 t u + 12 u^2) 
\nonumber\\[1.2ex]
&&             + m_Z^2 t (34 t^3 - 107 t^2 u - 99 t u^2 - 12 u^3) - 
               2 m_Z^4 (64 t^3 + t^2 u - 2 t u^2 + 3 u^3)) 
\nonumber\\[1.2ex]
&&              + m^6 t (-12 m_Z^{10} - 27 m_Z^6 t (8 t + 3 u) - 
               6 t^3 u (7 t + 4 u) + m_Z^8 (89 t + 15 u) 
\nonumber\\[1.2ex]
&&              + m_Z^2 t (-58 t^3 + 29 t^2 u + 16 t u^2 - 2 u^3) + 
               m_Z^4 (198 t^3 + 94 t^2 u + 23 t u^2 + 3 u^3)) 
\nonumber\\[1.2ex]
&&              + m^2 t^3 (4 m_Z^8 (4 t - u) + 
               4 m_Z^4 t (8 t^2 - 22 t u - 17 u^2) + 
               4 m_Z^6 (-10 t^2 + 6 t u + u^2) 
\nonumber\\[1.2ex]
&&              - t^2 u (34 t^2 + 65 t u + 27 u^2) + 
               2 m_Z^2 t (-4 t^3 + 51 t^2 u + 69 t u^2 + 18 u^3)))
          )
\nonumber\\[1.2ex]
V_9&=&V_8|_{t \leftrightarrow u}
\nonumber\\[1.2ex]
V_{10}&=&  - {1 \over (t u (-m^2 + t + u) (-4 m^2 m_Z^2 + (t + u)^2)^2)}
            (16 (104 m^{12} m_Z^4 (t + u) 
\nonumber\\[1.2ex]
&&            + 4 m^{10} m_Z^2 (96 m_Z^4 (t + u) - 
               2 m_Z^2 (73 t^2 + 74 t u + 73 u^2) - 
               3 (t^3 + 13 t^2 u 
\nonumber\\[1.2ex]
&&              + 13 t u^2 + u^3)) + 
            4 (t + u)^5 (-2 m_Z^6 (t + u) + 
               (t + u)^2 (2 t^2 + t u + 2 u^2) 
\nonumber\\[1.2ex]
&&              + m_Z^4 (6 t^2 + 8 t u + 6 u^2) - 
               m_Z^2 (6 t^3 + 11 t^2 u + 11 t u^2 + 6 u^3)) 
\nonumber\\[1.2ex]
&&              + m^4 (3 (t + u)^5 (11 t^2 + 16 t u + 11 u^2) - 
               16 m_Z^{10} (17 t^2 + 10 t u + 17 u^2) 
\nonumber\\[1.2ex]
&&              - 8 m_Z^4 (t + u)^3 (25 t^2 + 132 t u + 25 u^2) - 
               12 m_Z^6 (t + u)^2 (43 t^2 - 108 t u + 43 u^2) 
\nonumber\\[1.2ex]
&&               + 4 m_Z^2 (t + u)^4 (49 t^2 + 40 t u + 49 u^2) + 
               48 m_Z^8 (16 t^3 + 11 t^2 u + 11 t u^2 + 16 u^3))
\nonumber\\[1.2ex]
&&             - m^2 (t + u)^2 
             (-16 m_Z^8 (4 t^2 + 11 t u + 4 u^2) + 
               2 m_Z^2 (t + u)^3 (9 t^2 - 26 t u + 9 u^2) 
\nonumber\\[1.2ex]
&&              + 3 (t + u)^4 (9 t^2 + 8 t u + 9 u^2) - 
               4 m_Z^4 (t + u)^2 (45 t^2 + 26 t u + 45 u^2) 
\nonumber\\[1.2ex]
&&            + 8 m_Z^6 (25 t^3 + 64 t^2 u + 64 t u^2 + 25 u^3)) + 
            4 m^8 (146 m_Z^8 (t + u) + (t + u)^5 
\nonumber\\[1.2ex]
&&           + 26 m_Z^2 (t + u)^2 (t^2 + 5 t u + u^2) - 
              24 m_Z^6 (17 t^2 + 16 t u + 17 u^2) 
\nonumber\\[1.2ex]
&&            + 4 m_Z^4 (69 t^3 + 59 t^2 u + 59 t u^2 + 69 u^3)) + 
            2 m^6 (136 m_Z^{10} (t + u) - 9 (t + u)^6 
\nonumber\\[1.2ex]
&&            - 12 m_Z^8 (59 t^2 + 54 t u + 59 u^2) - 
               m_Z^2 (t + u)^3 (123 t^2 + 286 t u + 123 u^2) 
\nonumber\\[1.2ex]
&&             - 2 m_Z^4 (t + u)^2 (157 t^2 - 374 t u + 157 u^2) + 
               m_Z^6 (986 t^3 + 866 t^2 u 
\nonumber\\[1.2ex]
&&              + 866 t u^2 + 986 u^3))) \pi^2
          )
\nonumber\\[1.2ex]
V_{11}&=& -{1 \over  (t^2 u^2 (-m^2 + t + u) (-4 m^2 m_Z^2 + (t + u)^2)^2) }
       ( (-96 m^{12} m_Z^4 t u 
\nonumber\\[1.2ex]
&&           - 4 t^2 u^2 (t + u)^3 
             (-6 m_Z^6 + 16 m_Z^4 (t + u) - 15 m_Z^2 (t + u)^2 + 
               5 (t + u)^3) 
\nonumber\\[1.2ex]
&&              + 4 m^{10} m_Z^2 (120 m_Z^2 t u (t + u) + 
               t u (-3 t^2 + 34 t u - 3 u^2) + 
               12 m_Z^4 (t^2 
\nonumber\\[1.2ex]
&&             - 12 t u + u^2)) + 
            m^2 t u (2 m_Z^2 (t + u)^4 (t^2 + 18 t u + u^2) + 
               (t + u)^5 (3 t^2 
\nonumber\\[1.2ex]
&&            + 64 t u + 3 u^2) + 
               2 m_Z^4 (t + u)^3 (9 t^2 - 272 t u + 9 u^2) - 
              4 m_Z^6 (t + u)^2 (11 t^2 
\nonumber\\[1.2ex]
&&            - 196 t u + 11 u^2) + 
               24 m_Z^8 (t^3 - 13 t^2 u - 13 t u^2 + u^3)) + 
            m^4 (192 m_Z^{10} t u (t + u) 
\nonumber\\[1.2ex]
&&             - 9 t u (t + u)^4 (t^2 + 8 t u + u^2) - 
               8 m_Z^8 t u (71 t^2 + 30 t u + 71 u^2) 
\nonumber\\[1.2ex]
&&               - 2 m_Z^4 t u (t + u)^2 
                (167 t^2 - 1060 t u + 167 u^2) + 
               4 m_Z^6 t u (167 t^3 - 227 t^2 u 
\nonumber\\[1.2ex]
&&            - 227 t u^2 + 
                  167 u^3) - 
               m_Z^2 (t + u)^3 
                (3 t^4 - 20 t^3 u + 674 t^2 u^2 - 20 t u^3 + 
                  3 u^4)) 
\nonumber\\[1.2ex]
&&             - 2 m^8 (432 m_Z^8 t u + 
               t u (t + u)^2 (3 t^2 - 4 t u + 3 u^2) + 
               2 m_Z^2 t u (-15 t^3 + 179 t^2 u 
\nonumber\\[1.2ex]
&&                + 179 t u^2 - 
                  15 u^3) + 
               8 m_Z^6 (3 t^3 - 97 t^2 u - 97 t u^2 + 3 u^3) + 
               4 m_Z^4 (3 t^4 + 118 t^3 u 
\nonumber\\[1.2ex]
&&              + 66 t^2 u^2 + 
                  118 t u^3 + 3 u^4)) + 
            m^6 (-384 m_Z^{10} t u + 1312 m_Z^8 t u (t + u) 
\nonumber\\[1.2ex]
&&             + 4 t u (t + u)^3 (3 t^2 + 5 t u + 3 u^2) - 
               4 m_Z^6 t u (391 t^2 + 294 t u + 391 u^2) 
\nonumber\\[1.2ex]
&&               + m_Z^2 (t + u)^2 
                (3 t^4 - 70 t^3 u + 1230 t^2 u^2 - 70 t u^3 + 
                  3 u^4) + 
               8 m_Z^4 (3 t^5 + 111 t^4 u 
\nonumber\\[1.2ex]
&&                  - 122 t^3 u^2 - 
                  122 t^2 u^3 + 111 t u^4 + 3 u^5))))
\nonumber\\[1.2ex]
V_{12}&=& {1 \over  (3 (m^2 - t) t^2 (-m_Z^2 + t) (m^2 - u) u^2 (-m_Z^2 + u)
          (m_Z^2 -s)^2 (4 m^2 m_Z^2 - (t + u)^2))}
\nonumber\\[1.2ex]
&&       \times ((-12 m^{16} m_Z^2 t u 
             (22 m_Z^4 + 20 t u - 21 m_Z^2 (t + u)) + 
            3 m^{14} (20 t^2 u^2 (t + u)^2 
\nonumber\\[1.2ex]
&&             + 44 m_Z^8 (t^2 - 12 t u + u^2) - 
               10 m_Z^4 t u (29 t^2 + 118 t u + 29 u^2) + 
               m_Z^6 (-44 t^3 
\nonumber\\[1.2ex]
&&               + 852 t^2 u + 852 t u^2 - 44 u^3) - 
               21 m_Z^2 t u (t^3 - 13 t^2 u - 13 t u^2 + u^3)) 
\nonumber\\[1.2ex]
&&             + 4 t^3 u^3 (t + u)^2 
             (64 m_Z^{10} - 176 m_Z^8 (t + u) + 
               8 m_Z^6 (23 t^2 + 49 t u + 23 u^2) 
\nonumber\\[1.2ex]
&&              - 3 t u (7 t^3 + 5 t^2 u + 5 t u^2 + 7 u^3) - 
               6 m_Z^4 (16 t^3 + 51 t^2 u + 51 t u^2 + 16 u^3) 
\nonumber\\[1.2ex]
&&              + m_Z^2 (24 t^4 + 111 t^3 u + 134 t^2 u^2 + 
                  111 t u^3 + 24 u^4)) - 
            3 m^{12} (792 m_Z^{10} t u 
\nonumber\\[1.2ex]
&&           + 4 m_Z^8 (33 t^3 - 601 t^2 u - 601 t u^2 + 
                  33 u^3) + 
               t^2 u^2 (79 t^3 + 257 t^2 u + 257 t u^2 
\nonumber\\[1.2ex]
&&               + 79 u^3) + 
               m_Z^6 (-121 t^4 + 2026 t^3 u + 5262 t^2 u^2 + 
                  2026 t u^3 - 121 u^4) 
\nonumber\\[1.2ex]
&&              + m_Z^2 t u (-73 t^4 + 142 t^3 u + 574 t^2 u^2 + 
                  142 t u^3 - 73 u^4) - 
               m_Z^4 (11 t^5 
\nonumber\\[1.2ex]
&&             + 326 t^4 u + 3015 t^3 u^2 + 
                  3015 t^2 u^3 + 326 t u^4 + 11 u^5)) - 
            3 m^{10} (352 m_Z^{12} t u 
\nonumber\\[1.2ex]
&&           - 2428 m_Z^{10} t u (t + u) - 
               t^2 u^2 (t + u)^2 (127 t^2 + 386 t u + 127 u^2) 
\nonumber\\[1.2ex]
&&               - 22 m_Z^8 (6 t^4 - 167 t^3 u - 430 t^2 u^2 - 
                  167 t u^3 + 6 u^4) + 
               m_Z^6 (99 t^5 - 1796 t^4 u 
\nonumber\\[1.2ex]
&&                 - 9887 t^3 u^2 - 9887 t^2 u^3 
             - 1796 t u^4 + 99 u^5) + 
               m_Z^2 t u (104 t^5 + 521 t^4 u 
\nonumber\\[1.2ex]
&&             + 707 t^3 u^2 + 707 t^2 u^3 + 521 t u^4 + 104 u^5) + 
               m_Z^4 (33 t^6 + 87 t^5 u + 2369 t^4 u^2 
\nonumber\\[1.2ex]
&&             + 6046 t^3 u^3 + 2369 t^2 u^4 + 87 t u^5 + 
                  33 u^6)) + 
            m^8 (2112 m_Z^{12} t u (t + u) 
\nonumber\\[1.2ex]
&&            - 16 m_Z^{10} t u (453 t^2 + 1241 t u + 453 u^2) + 
               m_Z^8 (-132 t^5 + 6321 t^4 u 
\nonumber\\[1.2ex]
&&              + 36839 t^3 u^2 + 36839 t^2 u^3 
                 + 6321 t u^4 - 132 u^5) - 
               3 t^2 u^2 (107 t^5 + 809 t^4 u 
\nonumber\\[1.2ex]
&&               + 1912 t^3 u^2 + 
                  1912 t^2 u^3 + 809 t u^4 + 107 u^5) + 
               m_Z^6 (33 t^6 - 1209 t^5 u 
\nonumber\\[1.2ex]
&&               - 20115 t^4 u^2 - 
                  42434 t^3 u^3 - 20115 t^2 u^4 - 1209 t u^5 + 
                  33 u^6) + 
               3 m_Z^2 t u (84 t^6 
\nonumber\\[1.2ex]
&&                 + 943 t^5 u + 3098 t^4 u^2 + 
                  4694 t^3 u^3 + 3098 t^2 u^4 + 943 t u^5 + 
                  84 u^6) + 
               3 m_Z^4 (33 t^7 
\nonumber\\[1.2ex]
&&              - 77 t^6 u - 544 t^5 u^2 + 
                  1792 t^4 u^3 + 1792 t^3 u^4 - 544 t^2 u^5 - 
                  77 t u^6 + 33 u^7)) 
\nonumber\\[1.2ex]
&&               + m^2 t^2 u^2 (-1024 m_Z^{12} t u - 
               m_Z^6 (t + u)^3 (1033 t^2 + 3430 t u + 1033 u^2) 
\nonumber\\[1.2ex]
&&             - 32 m_Z^{10} (14 t^3 - 37 t^2 u - 37 t u^2 + 
                  14 u^3) + 
               3 t u (t + u)^2 
                (41 t^4 + 170 t^3 u 
\nonumber\\[1.2ex]
&&            + 194 t^2 u^2 + 170 t u^3 + 
                  41 u^4) + 
               8 m_Z^8 (142 t^4 + 405 t^3 u + 478 t^2 u^2 + 
                  405 t u^3 
\nonumber\\[1.2ex]
&&               + 142 u^4) + 
               m_Z^4 (483 t^6 + 4326 t^5 u + 13853 t^4 u^2 + 
                  20660 t^3 u^3 + 13853 t^2 u^4 
\nonumber\\[1.2ex]
&&               + 4326 t u^5 + 
                  483 u^6) - 
               2 m_Z^2 (69 t^7 + 660 t^6 u + 2578 t^5 u^2 
             + 5301 t^4 u^3 
\nonumber\\[1.2ex]
&&              + 5301 t^3 u^4 + 2578 t^2 u^5 + 
                  660 t u^6 + 69 u^7)) + 
            m^6 (-32 m_Z^{12} t u (33 t^2 
\nonumber\\[1.2ex]
&&            + 131 t u + 33 u^2) + 
               4 m_Z^{10} t u 
                (537 t^3 + 4225 t^2 u + 4225 t u^2 + 537 u^3) 
\nonumber\\[1.2ex]
&&             + 3 t^2 u^2 (t + u)^2 
                (49 t^4 + 469 t^3 u + 862 t^2 u^2 + 469 t u^3 + 
                  49 u^4) - 
               4 m_Z^8 t u (126 t^4 
\nonumber\\[1.2ex]
&&             + 4141 t^3 u + 9014 t^2 u^2 + 
                  4141 t u^3 + 126 u^4) + 
               m_Z^6 (33 t^7 - 630 t^6 u + 775 t^5 u^2 
\nonumber\\[1.2ex]
&&             + 14222 t^4 u^3 + 14222 t^3 u^4 + 775 t^2 u^5 - 
                  630 t u^6 + 33 u^7) - 
               3 m_Z^2 t u (43 t^7 
\nonumber\\[1.2ex]
&&              + 686 t^6 u + 3478 t^5 u^2 + 
                  7861 t^4 u^3 + 7861 t^3 u^4 + 3478 t^2 u^5 + 
                  686 t u^6 + 43 u^7) 
\nonumber\\[1.2ex]
&&                  + m_Z^4 (-33 t^8 + 171 t^7 u + 5013 t^6 u^2 + 
                  15053 t^5 u^3 + 18824 t^4 u^4 + 
                  15053 t^3 u^5 
\nonumber\\[1.2ex]
&&                + 5013 t^2 u^6 + 171 t u^7 - 
                  33 u^8)) + 
            m^4 t u (2080 m_Z^{12} t u (t + u) 
\nonumber\\[1.2ex]
&&           - 6 t u (t + u)^3 
                (5 t^4 + 96 t^3 u + 172 t^2 u^2 + 96 t u^3 + 
                  5 u^4) + 
               8 m_Z^{10} (24 t^4 
\nonumber\\[1.2ex]
&&              - 485 t^3 u 
              - 1382 t^2 u^2 - 
                  485 t u^3 + 24 u^4) - 
               m_Z^8 (423 t^5 + 375 t^4 u - 8054 t^3 u^2 
\nonumber\\[1.2ex]
&&                  - 8054 t^2 u^3 + 375 t u^4 + 423 u^5) + 
               m_Z^6 (237 t^6 + 3942 t^5 u + 10563 t^4 u^2 
\nonumber\\[1.2ex]
&&               + 11668 t^3 u^3 + 10563 t^2 u^4 + 3942 t u^5 + 
                  237 u^6) - 
               m_Z^4 (39 t^7 + 2514 t^6 u 
\nonumber\\[1.2ex]
&&               + 14408 t^5 u^2 + 
                  29959 t^4 u^3 + 29959 t^3 u^4 + 
                  14408 t^2 u^5 + 2514 t u^6 + 39 u^7) 
\nonumber\\[1.2ex]
&&                 + m_Z^2 (33 t^8 + 777 t^7 u + 5388 t^6 u^2 + 
                  16631 t^5 u^3 + 24742 t^4 u^4 + 
                  16631 t^3 u^5 
\nonumber\\[1.2ex]
&&                 + 5388 t^2 u^6 + 777 t u^7 + 
                  33 u^8))))
\nonumber\\[1.2ex]
V_{13}&=& {1 \over ((m_Z^2 - t)^2 t^2 (m_Z^2 - u)^2 u^2 (-m^2 + t + u)
          (-4 m^2 m_Z^2 + (t + u)^2)^2)} 
\nonumber\\[1.2ex]
&& \times       ( m_Z^2 (-16 m^{12} m_Z^4 t u 
             (6 m_Z^6 - 8 m_Z^4 (t + u) - 2 t u (t + u) + 
               m_Z^2 (3 t^2 + 8 t u 
\nonumber\\[1.2ex]
&&             + 3 u^2)) + 
            8 m^{10} m_Z^2 (6 m_Z^{10} (t^2 - 12 t u + u^2) + 
               m_Z^8 (-12 t^3 + 149 t^2 u + 149 t u^2 
\nonumber\\[1.2ex]
&&               - 12 u^3) + 
               3 m_Z^4 t u (t^3 + 45 t^2 u + 45 t u^2 + u^3) - 
               t^2 u^2 (2 t^3 + t^2 u + t u^2 + 2 u^3) 
\nonumber\\[1.2ex]
&&              + m_Z^2 t u (3 t^4 - 2 t^3 u - 54 t^2 u^2 - 
                  2 t u^3 + 3 u^4) + 
               m_Z^6 (6 t^4 
          - 80 t^3 u - 284 t^2 u^2 
\nonumber\\[1.2ex]
&&           - 80 t u^3 + 6 u^4)) + 
            4 t^2 u^2 (t + u)^3 
             (6 m_Z^{12} - 28 m_Z^{10} (t + u) + 
               2 t u (t + u)^2 (t^2 
\nonumber\\[1.2ex]
&&           + 3 t u + u^2) + 
               m_Z^8 (49 t^2 + 110 t u + 49 u^2) - 
               m_Z^6 (41 t^3 + 155 t^2 u + 155 t u^2 
\nonumber\\[1.2ex]
&&              + 41 u^3) + 
               m_Z^4 (17 t^4 + 97 t^3 u + 166 t^2 u^2 + 
                  97 t u^3 + 17 u^4) - 
               m_Z^2 (3 t^5 + 26 t^4 u 
\nonumber\\[1.2ex]
&&              + 71 t^3 u^2 + 
                  71 t^2 u^3 + 26 t u^4 + 3 u^5)) - 
            m^8 (864 m_Z^{14} t u + 
               24 m_Z^{12} (2 t^3 - 117 t^2 u 
\nonumber\\[1.2ex]
&&            - 117 t u^2 + 
                  2 u^3) - 
               4 m_Z^8 t u (293 t^3 + 1581 t^2 u + 1581 t u^2 + 
                  293 u^3) 
\nonumber\\[1.2ex]
&&               + m_Z^{10} (-72 t^4 + 3012 t^3 u + 7672 t^2 u^2 + 
                  3012 t u^3 - 72 u^4) - 
               2 t^2 u^2 (t^5 + 15 t^4 u 
\nonumber\\[1.2ex]
&&             - 20 t^3 u^2 - 
                  20 t^2 u^3 + 15 t u^4 + u^5) - 
               4 m_Z^4 t u (t^5 + 19 t^4 u - 54 t^3 u^2 
\nonumber\\[1.2ex]
&&               - 54 t^2 u^3 + 19 t u^4 + u^5) + 
               m_Z^2 t u (3 t^6 + 36 t^5 u - 23 t^4 u^2 - 
                  688 t^3 u^3 
\nonumber\\[1.2ex]
&&                - 23 t^2 u^4 + 36 t u^5 + 3 u^6)
                + 2 m_Z^6 (12 t^6 + 57 t^5 u 
               + 812 t^4 u^2 + 
                  1470 t^3 u^3 
\nonumber\\[1.2ex]
&&               + 812 t^2 u^4 + 57 t u^5 + 12 u^6
                  )) - m^2 t u 
             (4 m_Z^{12} (t + u)^2 (23 t^2 - 352 t u + 23 u^2) 
\nonumber\\[1.2ex]
&&              - 24 m_Z^{14} (t^3 - 13 t^2 u - 13 t u^2 + u^3) + 
               2 t u (t + u)^4 
                (t^4 + 13 t^3 u + 38 t^2 u^2 
\nonumber\\[1.2ex]
&&                + 13 t u^3 + u^4)
                + 2 m_Z^8 (t + u)^2 
                (59 t^4 - 542 t^3 u - 2218 t^2 u^2 - 542 t u^3 + 
                  59 u^4) 
\nonumber\\[1.2ex]
&&                 - 2 m_Z^{10} (73 t^5 - 845 t^4 u - 3284 t^3 u^2 - 
                  3284 t^2 u^3 - 845 t u^4 + 73 u^5) 
\nonumber\\[1.2ex]
&&               - m_Z^2 (t + u)^3 
                (3 t^6 + 48 t^5 u + 237 t^4 u^2 + 236 t^3 u^3 + 
                  237 t^2 u^4 + 48 t u^5 + 3 u^6) 
\nonumber\\[1.2ex]
&&                + 2 m_Z^4 (t + u)^2 
                (6 t^6 + 87 t^5 u + 104 t^4 u^2 - 238 t^3 u^3 + 
                  104 t^2 u^4 + 87 t u^5 + 6 u^6) 
\nonumber\\[1.2ex]
&&                + m_Z^6 (-49 t^7 - 73 t^6 u + 2017 t^5 u^2 + 
                  6457 t^4 u^3 + 6457 t^3 u^4 + 2017 t^2 u^5 
\nonumber\\[1.2ex]
&&               - 73 t u^6 - 49 u^7)) + 
            m^6 (-384 m_Z^{16} t u + 1936 m_Z^{14} t u (t + u) - 
               4 m_Z^{12} t u (859 t^2 
\nonumber\\[1.2ex]
&&            + 1806 t u + 859 u^2) - 
               t^2 u^2 (t + u)^2 
                (3 t^4 + 64 t^3 u - 130 t^2 u^2 + 64 t u^3 + 
                  3 u^4) 
\nonumber\\[1.2ex]
&&               + 8 m_Z^{10} (3 t^5 + 347 t^4 u + 1095 t^3 u^2 + 
                  1095 t^2 u^3 + 347 t u^4 + 3 u^5) - 
               m_Z^8 (45 t^6 
\nonumber\\[1.2ex]
&&             + 1146 t^5 u + 3887 t^4 u^2 + 
                  3460 t^3 u^3 + 3887 t^2 u^4 + 1146 t u^5 + 
                  45 u^6) + 
               m_Z^2 t u (3 t^7 
\nonumber\\[1.2ex]
&&               + 115 t^6 u + 165 t^5 u^2 - 
                  1811 t^4 u^3 - 1811 t^3 u^4 + 165 t^2 u^5 + 
                  115 t u^6 + 3 u^7) 
\nonumber\\[1.2ex]
&&                 + 2 m_Z^6 (9 t^7 + 142 t^6 u + 348 t^5 u^2 - 
                  1971 t^4 u^3 - 1971 t^3 u^4 + 348 t^2 u^5 + 
                  142 t u^6 
\nonumber\\[1.2ex]
&&                 + 9 u^7) + 
               m_Z^4 (3 t^8 - 32 t^7 u - 340 t^6 u^2 + 
                  2156 t^5 u^3 + 6490 t^4 u^4 + 2156 t^3 u^5 
\nonumber\\[1.2ex]
&&              - 340 t^2 u^6 - 32 t u^7 + 3 u^8)) + 
            m^4 (192 m_Z^{16} t u (t + u) - 
               8 m_Z^{14} t u (101 t^2 
\nonumber\\[1.2ex]
&&             + 138 t u + 101 u^2) + 
               4 m_Z^{12} t u 
                (333 t^3 + 379 t^2 u + 379 t u^2 + 333 u^3) 
\nonumber\\[1.2ex]
&&               + 2 m_Z^{10} t u 
                (-581 t^4 + 30 t^3 u + 2118 t^2 u^2 + 30 t u^3 - 
                  581 u^4) + 
               3 t^2 u^2 (t + u)^3 
                (t^4 
\nonumber\\[1.2ex]
&&               + 18 t^3 u - 2 t^2 u^2 + 18 t u^3 + u^4) - 
               m_Z^2 t u (t + u)^2 
                (3 t^6 + 96 t^5 u + 209 t^4 u^2 
\nonumber\\[1.2ex]
&&                - 1132 t^3 u^3 + 209 t^2 u^4 + 96 t u^5 + 3 u^6)
                 - m_Z^8 (3 t^7 - 613 t^6 u + 921 t^5 u^2 
\nonumber\\[1.2ex]
&&                 + 11961 t^4 u^3 + 11961 t^3 u^4 + 921 t^2 u^5 - 
                  613 t u^6 + 3 u^7) + 
               m_Z^6 (6 t^8 - 177 t^7 u 
\nonumber\\[1.2ex]
&&               + 12 t^6 u^2 + 
                  8745 t^5 u^3 + 18124 t^4 u^4 + 8745 t^3 u^5 + 
                  12 t^2 u^6 - 177 t u^7 + 6 u^8) 
\nonumber\\[1.2ex]
&&                 - m_Z^4 (3 t^9 - 13 t^8 u - 338 t^7 u^2 + 
                  1474 t^6 u^3 + 8410 t^5 u^4 + 8410 t^4 u^5 
\nonumber\\[1.2ex]
&&                 + 1474 t^3 u^6 - 338 t^2 u^7 - 13 t u^8 + 3 u^9)
               )) )
\nonumber\\[1.2ex]
V_{14}&=&          {1 \over  ((m^2 - t)^2 t^2 (m^2 - u)^2 (m^2 - t - u) u^2
          (-4 m^2 m_Z^2 + (t + u)^2)^2)} 
\nonumber\\[1.2ex]
&&       (m^2 (96 m^{18} m_Z^4 t u - 
            4 m^{16} m_Z^2 (158 m_Z^2 t u (t + u) + 
               t u (-3 t^2 + 34 t u - 3 u^2) 
\nonumber\\[1.2ex]
&&            + 12 m_Z^4 (t^2 - 12 t u + u^2)) - 
            4 t^3 u^3 (t + u)^2 
             (12 m_Z^6 t u + 6 t u (t + u)^3 
\nonumber\\[1.2ex]
&&             + m_Z^2 (t + u)^2 (t^2 - 11 t u + u^2) - 
               m_Z^4 (t^3 + 7 t^2 u + 7 t u^2 + u^3)) 
\nonumber\\[1.2ex]
&&             + 2 m^{14} (432 m_Z^8 t u + 
               t u (t + u)^2 (3 t^2 - 4 t u + 3 u^2) - 
               4 m_Z^2 t u (8 t^3 - 113 t^2 u 
\nonumber\\[1.2ex]
&&               - 113 t u^2 + 
                  8 u^3) + 
               4 m_Z^6 (18 t^3 - 281 t^2 u - 281 t u^2 + 
                  18 u^3) + 
               2 m_Z^4 (6 t^4 
\nonumber\\[1.2ex]
&&               + 411 t^3 u + 578 t^2 u^2 + 
                  411 t u^3 + 6 u^4)) + 
            m^2 t^2 u^2 (48 m_Z^8 t u (3 t^2 - 2 t u + 3 u^2) 
\nonumber\\[1.2ex]
&&             + 2 (t + u)^4 
                (3 t^4 + 29 t^3 u + 72 t^2 u^2 + 29 t u^3 + 
                  3 u^4) - 
               m_Z^2 (t + u)^3 
                (5 t^4 
\nonumber\\[1.2ex]
&&               + 124 t^3 u - 102 t^2 u^2 + 124 t u^3 + 
                  5 u^4) + 
               2 m_Z^4 (t + u)^2 
                (7 t^4 + 86 t^3 u - 522 t^2 u^2 
\nonumber\\[1.2ex]
&&               + 86 t u^3 + 
                  7 u^4) - 
               4 m_Z^6 (3 t^5 + 61 t^4 u - 164 t^3 u^2 - 
                  164 t^2 u^3 + 61 t u^4 + 3 u^5)) 
\nonumber\\[1.2ex]
&&                + m^{12} (384 m_Z^{10} t u - 2288 m_Z^8 t u (t + u) - 
               4 t u (t + u)^3 (6 t^2 + t u + 6 u^2) - 
\nonumber\\[1.2ex]
&&               16 m_Z^6 (9 t^4 - 207 t^3 u - 416 t^2 u^2 - 
                  207 t u^3 + 9 u^4) - 
               8 m_Z^4 (9 t^5 + 284 t^4 u 
\nonumber\\[1.2ex]
&&               + 356 t^3 u^2 + 
                  356 t^2 u^3 + 284 t u^4 + 9 u^5) - 
               m_Z^2 (3 t^6 - 126 t^5 u + 2089 t^4 u^2 
\nonumber\\[1.2ex]
&&             + 4756 t^3 u^3 + 2089 t^2 u^4 - 126 t u^5 + 
                  3 u^6)) + 
            m^{10} (-576 m_Z^{10} t u (t + u) 
\nonumber\\[1.2ex]
&&            + 672 m_Z^8 t u (3 t^2 + 7 t u + 3 u^2) + 
               t u (t + u)^2 
                (39 t^4 + 170 t^3 u + 222 t^2 u^2 
\nonumber\\[1.2ex]
&&               + 170 t u^3 + 
                  39 u^4) + 
               16 m_Z^6 (3 t^5 - 147 t^4 u - 424 t^3 u^2 - 
                  424 t^2 u^3 - 147 t u^4 
\nonumber\\[1.2ex]
&&                 + 3 u^5) + 
               2 m_Z^4 (36 t^6 + 915 t^5 u + 780 t^4 u^2 - 
                  1286 t^3 u^3 + 780 t^2 u^4 + 915 t u^5 
\nonumber\\[1.2ex]
&&                 + 36 u^6) + 
               m_Z^2 (9 t^7 - 109 t^6 u + 2141 t^5 u^2 + 
                  8903 t^4 u^3 + 8903 t^3 u^4 + 2141 t^2 u^5 
\nonumber\\[1.2ex]
&&              - 109 t u^6 + 9 u^7)) - 
            m^4 t u (48 m_Z^8 t u 
                (4 t^3 + 3 t^2 u + 3 t u^2 + 4 u^3) 
\nonumber\\[1.2ex]
&&             - m_Z^2 t u (t + u)^2 
                (85 t^4 - 52 t^3 u - 1286 t^2 u^2 - 52 t u^3 + 
                  85 u^4) + 
               (t + u)^3 (3 t^6 
\nonumber\\[1.2ex]
&&              + 58 t^5 u + 303 t^4 u^2 + 
                  492 t^3 u^3 + 303 t^2 u^4 + 58 t u^5 + 3 u^6)
                - 4 m_Z^6 (3 t^6 
\nonumber\\[1.2ex]
&&               + 108 t^5 u - 55 t^4 u^2 - 
                  608 t^3 u^3 - 55 t^2 u^4 + 108 t u^5 + 3 u^6)
                + m_Z^4 (15 t^7 
\nonumber\\[1.2ex]
&&               + 353 t^6 u - 501 t^5 u^2 - 
                  5475 t^4 u^3 - 5475 t^3 u^4 - 501 t^2 u^5 + 
                  353 t u^6 + 15 u^7)) 
\nonumber\\[1.2ex]
&&                 + m^8 (192 m_Z^{10} t u (t^2 + 4 t u + u^2) - 
               32 m_Z^8 t u 
                (20 t^3 + 97 t^2 u + 97 t u^2 + 20 u^3) 
\nonumber\\[1.2ex]
&&                - 3 t u (t + u)^3 
                (11 t^4 + 80 t^3 u + 122 t^2 u^2 + 80 t u^3 + 
                  11 u^4) + 
               32 m_Z^6 t u 
                (27 t^4 
\nonumber\\[1.2ex]
&&          + 104 t^3 u + 77 t^2 u^2 + 104 t u^3 + 
                  27 u^4) - 
               2 m_Z^4 (12 t^7 + 423 t^6 u + 483 t^5 u^2 
\nonumber\\[1.2ex]
&&               - 3970 t^4 u^3 - 3970 t^3 u^4 + 483 t^2 u^5 + 
                  423 t u^6 + 12 u^7) - 
               m_Z^2 (9 t^8 - 36 t^7 u 
\nonumber\\[1.2ex]
&&               + 880 t^6 u^2 + 
                  7376 t^5 u^3 + 13062 t^4 u^4 + 7376 t^3 u^5 + 
                  880 t^2 u^6 - 36 t u^7 + 9 u^8)) 
\nonumber\\[1.2ex]
&&                + m^6 (-192 m_Z^{10} t^2 u^2 (t + u) + 
               16 m_Z^8 t u 
                (3 t^4 + 58 t^3 u + 62 t^2 u^2 + 58 t u^3 
\nonumber\\[1.2ex]
&&                 + 3 u^4) - 
               4 m_Z^6 t u (41 t^5 + 337 t^4 u - 308 t^3 u^2 - 
                  308 t^2 u^3 + 337 t u^4 + 41 u^5) 
\nonumber\\[1.2ex]
&&               + t u (t + u)^2 
                (15 t^6 + 194 t^5 u + 697 t^4 u^2 + 
                  1016 t^3 u^3 + 697 t^2 u^4 + 194 t u^5 
\nonumber\\[1.2ex]
&&                 + 15 u^6) + m_Z^4 t u (195 t^6 + 916 t^5 u - 4815 t^4 u^2 - 
                  12032 t^3 u^3 - 4815 t^2 u^4 
\nonumber\\[1.2ex]
&&                + 916 t u^5 + 195 u^6) + 
               m_Z^2 (3 t^9 - t^8 u - 20 t^7 u^2 + 
                  2332 t^6 u^3 + 7838 t^5 u^4 
\nonumber\\[1.2ex]
&&              + 7838 t^4 u^5 + 
                  2332 t^3 u^6 - 20 t^2 u^7 - t u^8 + 3 u^9))))
\nonumber\\[1.2ex]
V_{15}&=&         {1 \over (6 t^2 u^2 (-m^2 + t + u)^2)} 
       ((144 m^{10} t u - 
            6 m^8 (97 t u (t + u) + 
               12 m_Z^2 (t^2 - 12 t u 
\nonumber\\[1.2ex]
&&            + u^2)) + 
            3 m^6 (432 m_Z^4 t u + 
               t u (353 t^2 + 700 t u + 353 u^2) + 
               m_Z^2 (48 t^3 - 746 t^2 u 
\nonumber\\[1.2ex]
&&            - 746 t u^2 + 48 u^3)) + 
            12 m^4 (48 m_Z^6 t u - 208 m_Z^4 t u (t + u) - 
               t u (89 t^3 + 249 t^2 u 
\nonumber\\[1.2ex]
&&       + 249 t u^2 + 89 u^3) + 
               m_Z^2 (-6 t^4 + 199 t^3 u + 482 t^2 u^2 + 
                  199 t u^3 - 6 u^4)) 
\nonumber\\[1.2ex]
&&               + m^2 t u (-696 m_Z^6 (t + u) + 
               40 m_Z^4 (39 t^2 + 107 t u + 39 u^2) - 
               6 m_Z^2 (229 t^3 
\nonumber\\[1.2ex]
&&          + 817 t^2 u + 817 t u^2 + 
                  229 u^3) + 
               21 (27 t^4 + 98 t^3 u + 126 t^2 u^2 + 98 t u^3 + 
                  27 u^4)) 
\nonumber\\[1.2ex]
&&             - 4 t u (-2 m_Z^6 (15 t^2 + 86 t u + 15 u^2) + 
               m_Z^4 (90 t^3 + 406 t^2 u + 406 t u^2 + 90 u^3) 
\nonumber\\[1.2ex]
&&             - m_Z^2 (90 t^4 + 381 t^3 u + 512 t^2 u^2 + 
                  381 t u^3 + 90 u^4) + 
               3 (10 t^5 + 49 t^4 u + 69 t^3 u^2 
\nonumber\\[1.2ex]
&&             + 69 t^2 u^3 + 
                  49 t u^4 + 10 u^5))) )
\end{eqnarray}

%
%


\section{$B_0$ Integrals}

\begin{eqnarray}
B_0\left({\cal{P}}\right) = \frac{i}{(4\pi)^2}\left[-\frac{2}{\epsilon} +2 -\gamma_E - f\left({\cal{P}}\right)\right] 
\end{eqnarray}
where ${\cal{P}}\ \in\ \{p_3,p_4,p_5,k,q\}$,  
$k$ is the momentum of vector boson,
$q$, the momentum of graviton, $p_3 = p_1+k, p_4 = p_2+k , p_5 = p_1+p_2$
 and,
\begin{eqnarray}
f\left({\cal{P}}\right) =  \left\{ \begin{array}{ll}
                        \ln \left(\frac{-{\cal{P}}^2}{4 \pi \mu_r^2}\right) & \mbox{for}\ \ {\cal{P}} = p_3,p_4 \\
                        \ln \left(\frac{{\cal{P}}^2}{4 \pi \mu_r^2}\right) - i \pi  & \mbox{for}\ \ {\cal{P}} = p_5,k,q \\
                                   \end{array}   \right.       
\end{eqnarray}

\section{$C_0$ Integrals}

\begin{eqnarray}
C_0\left({\cal{P'}},{\cal{P''}}\right) = \!\!\!\!\!\!\!\!\!\!&&
                       \frac{-i}{(4\pi)^2}\ \frac{1}{[({\cal{P'}}-{\cal{P''}})^2-{\cal{P''}}^2]}
                       \left[-\frac{2}{\epsilon}
                       {\left\{\ln \left(\frac{-({\cal{P'}}-{\cal{P''}})^2}{{\cal{P''}}^2}\right)+i \pi
                       \right\}}\ +\ \right.  \nonumber \\
                       &&\left.\frac{1}{2}\left\{{\left(\gamma_E + \ln
                        \left(\frac{{\cal{P''}}^2}{4\pi\mu_r^2}\right)-i\pi\right)^2-  
                      {\left(\gamma_E + \ln\left(\frac{-({\cal{P'}}-{\cal{P''}})^2}{4\pi\mu_r^2}\right)\right)}^2}\right\} \right]
\end{eqnarray}
where \ ${\cal{P}'}\ \in\ \{p_1,p_2\}$\quad  and,\ ${\cal{P}''}\ \in\ \{k,q\}$
\begin{eqnarray}
C_0\left(p_1,p_2\right) = \!\!\!\!\!\!\!\!\!\!&&
                       \frac{-i}{(4\pi)^2}\ \frac{1}{s}
                       \left[-\frac{4}{\epsilon^2}-\frac{2}{\epsilon}
                       {\left\{\gamma_E + \ln \left(\frac{s}{4\pi\mu_r^2}\right)-i \pi
                       \right\}}\ +\ \right.  \nonumber \\
                       &&\left.\frac{1}{2}\left\{{\frac{\pi^2}{6}-
                       {\left(\gamma_E + \ln\left(\frac{s}{4\pi\mu_r^2}\right)-i\pi\right)}^2}\right\} \right] \\
\!\!\!\!\!\!\!\! C_0\left(k,q\right) = \!\!\!\!\!\!\!\!\!\!&&
                          \frac{-i}{(4\pi)^2}\ \frac{1}{s\beta}
                          \left[2 {Li}_2 \left(\frac{2}{1-\alpha+\beta}\right)
                              -2 {Li}_2 \left(\frac{2}{1-\alpha-\beta}\right) -  \right. \nonumber \\
                         \!\!\!\!\!\!\!\!\!\!\!\!\!&&
                          \left.\ln \left(\frac{(1-\alpha)^2-\beta^2}{4}\right) \!\!
                         \left\{\ln \left(\frac{\alpha-\beta+1}{\alpha-\beta-1}\right)
                         - \ln \left(\frac{\alpha+\beta+1}{\alpha+\beta-1}\right)\right\} \right]                  
\end{eqnarray}
\\ where $\alpha = \frac{m^2-m_z^2}{s}$  and $\beta = \frac{1}{s}\sqrt{(t+u)^2-4 m_z^2 m^2}$
\section{$D_0$ Integrals}

\begin{eqnarray}
D_0\left(p_1,k,q\right) = \!\!\!\!\!\!\!\!\!\!&&
                       \frac{i}{(4\pi)^2}\ \frac{1}{st}
                       \left[\frac{4}{\epsilon^2}+\frac{2}{\epsilon}
                       {\left\{\gamma_E + \ln \left(\frac{-t}{4\pi\mu_r^2}\right)+\ln \left(\frac{s}{m_z^2}\right)
                       +\ln \left(\frac{-t}{m^2}\right)+i \pi
                       \right\}} \right.  \nonumber \\
                       \!\!\!\!\!\!\!\!\!\!&&
                       +\ {\left(\gamma_E + \ln \left(\frac{s}{4\pi\mu_r^2}\right) - i \pi \right)}^2
                       + {\left(\gamma_E + \ln \left(\frac{-t}{4\pi\mu_r^2}\right) \right)}^2 \nonumber \\
                       \!\!\!\!\!\!\!\!\!\!&&
                       -\ {\left(\gamma_E + \ln \left(\frac{m_z^2}{4\pi\mu_r^2}\right) - i \pi \right)}^2
                       - {\left(\gamma_E + \ln \left(\frac{m^2}{4\pi\mu_r^2}\right) - i \pi \right)}^2 \nonumber \\
                       \!\!\!\!\!\!\!\!\!\!&&
                       +\ \frac{1}{2}{\left(\gamma_E + \ln \left(\frac{m_z^2}{s}\right) 
                       + \ln \left(\frac{m^2}{4\pi\mu_r^2}\right) - i \pi \right)}^2 - \frac{\pi^2}{12} \nonumber \\
                       \!\!\!\!\!\!\!\!\!\!&&
                       +\ \frac{1}{3} \left(-3 \ln ^2\left(1-\frac{t}{m^2}\right)-3 \ln
                      ^2\left(\frac{{m_z}^2-t}{s}\right)-\pi ^2\right)-2
                      {Li}_2\left(\frac{t}{m^2}\right) \nonumber \\
                      \!\!\!\!\!\!\!\!\!\!&&
                      +\ \ln^2\left(1-\frac{m^2}{t}\right)-2 i \pi  \ln
                      \left(1-\frac{m^2}{t}\right)-2
                      {Li}_2\left(\frac{t}{{m_z}^2}\right) \nonumber \\
                      \!\!\!\!\!\!\!\!\!\!&&
                      +\ 2 \ln \left(1-\frac{{m_z}^2}{s}\right) \left(\ln
                      \left(1-\frac{{m_z}^2}{t}\right)-i \pi \right) \nonumber \\
                      \!\!\!\!\!\!\!\!\!\!&&
                      +\ \left(\ln
                      \left(1-\frac{{m_z}^2}{t}\right)-\ln
                      \left(\frac{{m_z}^2-s}{t}\right)\right) \times \nonumber \\
                      \!\!\!\!\!\!\!\!\!\!&&
                      \quad\  \left(\ln
                      \left(\frac{{m_z}^2-s}{t}\right)+\log
                      \left(1-\frac{{m_z}^2}{t}\right)-2 i \pi \right) \nonumber \\
                      \!\!\!\!\!\!\!\!\!\!&&
                      -\ 2 \ln
                      \left(\frac{s}{{m_z}^2}-1\right) \ln
                      \left(1-\frac{t}{{m_z}^2}\right)+\ln
                      ^2\left(1-\frac{{m_z}^2}{s}\right) \nonumber \\
                       \!\!\!\!\!\!\!\!\!\!&&
                      \left.-\ \ln^2\left(\frac{s}{{m_z}^2}\right)+2 \ln
                      \left(\frac{s}{{m_z}^2}\right) \ln
                      \left(\frac{s}{{m_z}^2}-1\right) \right]
\end{eqnarray}
$D_0(p_2,k,q)$ can be readily obtained by replacing \textquoteleft$t$\textquoteright\ 
by \textquoteleft$u$\textquoteright\ in the above $D_0(p_1,k,q)$ expression.
\begin{eqnarray}
D_0\left(k,p_2,q\right) = \!\!\!\!\!\!\!\!\!\!&&
                       \frac{i}{(4\pi)^2}\ \frac{1}{(tu-m_z^2m^2)}
                       \left[\frac{4}{\epsilon}
                       {\left\{ \ln \left(\frac{-t}{m_z^2}\right) + \ln \left(\frac{-u}{m^2}\right)
                        +2 i \pi
                       \right\}} \right.  \nonumber \\
                       \!\!\!\!\!\!\!\!\!\!&&
                       -\ {\left(\gamma_E + \ln \left(\frac{m_z^2}{4\pi\mu_r^2}\right) - i \pi \right)}^2
                       - {\left(\gamma_E + \ln \left(\frac{m^2}{4\pi\mu_r^2}\right) - i \pi \right)}^2 \nonumber \\
                        \!\!\!\!\!\!\!\!\!\!&&
                       +\ {\left(\gamma_E + \ln \left(\frac{-t}{4\pi\mu_r^2}\right) \right)}^2
                       + {\left(\gamma_E + \ln \left(\frac{-u}{4\pi\mu_r^2}\right) \right)}^2 -\frac{4 \pi ^2}{3} \nonumber \\
                       \!\!\!\!\!\!\!\!\!\!&&
                       +\ 2 {Li}_2\left(\frac{\left(m^2-t\right)
                      \left({m_z}^2-t\right)}{m^2 {m_z}^2-t u}\right)+2
                      {Li}_2\left(\frac{\left(m^2-u\right)
                      \left({m_z}^2-u\right)}{m^2 {m_z}^2-t u}\right) \nonumber \\
                       \!\!\!\!\!\!\!\!\!\!&&
                      +\ 2 {Li}_2\left(\frac{t u-m^2
                      {m_z}^2}{\left(m^2-t\right) \left(m^2-u\right)}\right)+2
                      {Li}_2\left(\frac{t u-m^2
                      {m_z}^2}{\left({m_z}^2-t\right)
                      \left({m_z}^2-u\right)}\right) \nonumber \\
                      \!\!\!\!\!\!\!\!\!\!&&
                      +\ \ln^2\left(\frac{\left(m^2-t\right) \left(m^2-u\right)}{t u-m^2
                      {m_z}^2}\right)+\ln
                      ^2\left(\frac{\left({m_z}^2-t\right)
                      \left({m_z}^2-u\right)}{t u-m^2 {m_z}^2}\right) \nonumber \\
                      \!\!\!\!\!\!\!\!\!\!&&
                      -\ 2 i \pi  \left(\ln \left(\frac{\left(m^2-t\right)
                      \left(m^2-u\right)}{t u-m^2 {m_z}^2}\right)+ \ln
                      \left(\frac{\left({m_z}^2-t\right)
                      \left({m_z}^2-u\right)}{t u-m^2
                      {m_z}^2}\right)\right)
\end{eqnarray}


\end{document}